\documentclass[draft,tightenlines,nofootinbib,preprint,aps,eqsecnum,amsmath,amssymb]{revtex4}
\begin{document}

\newcommand{\beq}{\begin{equation}}
\newcommand{\eeq}{\end{equation}}
\newcommand{\bea}{\begin{eqnarray}}
\newcommand{\eea}{\end{eqnarray}}
\newcommand{\cir}{{\buildrel \circ \over =}}

\title{New Directions in Non-Relativistic and Relativistic Rotational
and Multipole Kinematics for N-Body and Continuous Systems.}

\author{David Alba}

\affiliation
 {Dipartimento di Fisica\\ Universita' di Firenze\\
Polo Scientifico, via Sansone 1\\
 50019 Sesto Fiorentino, Italy\\
 E-mail: ALBA@FI.INFN.IT}

\author{Luca Lusanna}

\affiliation{Sezione INFN di Firenze\\Polo Scientifico, via Sansone 1\\
 50019 Sesto Fiorentino, Italy\\
 E-mail LUSANNA@FI.INFN.IT}

\author{Massimo Pauri}

\affiliation
{Dipartimento di Fisica\\ Universita' di Parma\\
 Parco Area Scienze 7/A\\
  43100 Parma, Italy\\
   E-mail: PAURI@PR.INFN.IT}

\begin{abstract}

In non-relativistic mechanics the center of mass of an isolated
system is easily separated out from the relative variables. For a
N-body system these latter are usually described by a set of
Jacobi normal coordinates, based on the clustering of the centers
of mass of sub-clusters. The Jacobi variables are then the
starting point for separating {\it orientational} variables,
connected with the angular momentum constants of motion, from {\it
shape} (or {\it vibrational}) variables. Jacobi variables,
however, cannot be extended to special relativity. We show by
group-theoretical methods that two new sets of relative variables
can be defined in terms of a {\it clustering of the angular
momenta of sub-clusters} and directly related to the so-called
{\it dynamical body frames} and {\it canonical spin bases}. The
underlying group-theoretical structure allows a direct extension
of such notions from a non-relativistic to a special-relativistic
context if one exploits the {\it rest-frame instant form of
dynamics}. The various known definitions of relativistic center of
mass are recovered. The separation of suitable relative variables
from the so-called {\it canonical internal} center of mass leads
to the correct kinematical framework for the relativistic theory
of the orbits for a N-body system with action-at-a-distance
interactions. The rest-frame instant form is also shown to be the
correct kinematical framework for introducing the Dixon
multi-poles for closed and open N-body systems, as well as for
continuous systems, exemplified here by the configurations of the
Klein-Gordon field that are compatible with the previous notions
of center of mass.

\vskip 1truecm

Invited contribution for the book {\it Atomic and Molecular
Clusters: New Research} (Nova Science).

\vskip 1truecm

 \today

\end{abstract}

\maketitle

\vfill\eject

\section{Introduction.}

In nuclear, atomic and molecular physics, as well as in celestial
mechanics, an old basic problem is to exploit the translational
and rotational invariance of a N-body system for eliminating as
many global variables as possible to get a reduced system
described by a well defined set of {\it relative degrees of
freedom}. In molecular dynamics, for instance, this reduction is
instrumental for the definition of molecular vibrations and
rotations. A similar old problem exists for isolated continuous
deformable bodies: in this case the usual Euler kinematics of
rigid bodies in the body frame must be generalized to understand
phenomena like the {\it falling cat} and the {\it diver}. We
suggest to see the review part of Ref.\cite{1} for an essential
selection of the huge bibliography on such issues.

\medskip

Since most of the applications concerning these problems are
typically non-relativistic, nearly all treatments deal with
non-relativistic systems of the {\it kinetic-plus-potential type}.
However, developments in particle physics, astrophysics and
general relativity suggest to extend the treatment from the
absolute Galilei space plus time to Minkowski space-time and then
to Einstein space-times. Furthermore, the case of the general
relativistic N-body problem, in particular, forces us to simulate
the issue of the divergences in the self-energies with a
multipolar expansion. We need, therefore, a translation of every
result in the language of such kind of expansions.

\medskip

In this article we  show a new method for the treatment of
relative variables for a many-body system {\it both
non-relativistic and relativistic}. Our proposal is based upon a
conjunction of Hamiltonian and group-theoretical methods leading
to a systematic generalization, valid for generic deformable
systems, of the standard concept of {\it body frame} for rigid
systems. We also show that our procedure constitutes the proper
kinematical framework for introducing multipolar expansions for
closed and open systems.

\bigskip

In Newton mechanics isolated systems of  N particles possess 3N
degrees of freedom in configuration space and 6N in phase space.
The Abelian nature of the overall translational invariance, with
its associated three commuting Noether constants of the motion,
makes possible the decoupling and, therefore, the elimination of
either three configurational variables or three pairs of canonical
variables, respectively ({\it separation of the center-of-mass
motion}). In this way one is left with either 3N-3 relative
coordinates ${\vec \rho}_a$ or 6N-6 relative phase space variables
${\vec \rho}_a$, ${\vec p}_a$, $a=1,.., N-1$ while the
center-of-mass angular momentum or spin is $\vec S=
\sum_{a=1}^{N-1}{\vec \rho}_a \times {\vec p}_a$. Most of the
calculations of the non-relativistic theory employ the sets of
$3N-3$ {\it Jacobi normal relative coordinates} ${\vec s}_a$ (see
Ref.\cite{1}) that diagonalize the quadratic form associated with
the relative kinetic energy (the spin becomes $\vec
S=\sum_{a=1}^{N-1}{\vec s}_a \times {\vec \pi}_{sa}$, with ${\vec
\pi}_{sa}$  momenta conjugated to the ${\vec s}_a$'s). Each set of
Jacobi normal coordinates ${\vec s}_a$ orders the N particles into
a {\it hierarchy of clusters}, in which each cluster of two or
more particles has a mass given by an eigenvalue ({\it reduced
mass}) of the quadratic form; the Jacobi normal coordinates join
the centers of mass of cluster pairs.

\medskip

On the other hand, the non-Abelian nature of the overall
rotational invariance entails that an analogous intrinsic
separation of {\it rotational} (or {\it orientational})
configurational variables from others which could be called {\it
shape} or {\it vibrational} be impossible. As a matter of fact,
this is one of the main concerns of molecular physics and of
advanced mechanics of deformable bodies. In fact, in the theory of
deformable bodies one looses any intrinsic notion of {\it body
frame}, which is the fundamental tool for the description of rigid
bodies and their associated Euler equations. A priori, given any
configuration of a non-relativistic continuous body (in particular
a N-body system), any barycentric orthogonal frame could be named
{\it body frame} of the system with its associated notion of {\it
vibrations}.

\medskip

This state of affairs suggested \cite{1} to replace the
kinematically accessible region of the non-singular configurations
\cite{2} in the (3N-3)-dimensional relative configuration space by
a SO(3) principal fiber bundle over a (3N-6)-dimensional base
manifold, called {\it shape} space. The SO(3) fiber on each shape
configuration carries the {\it orientational} variables (e.g. the
usual Euler angles) referred to the chosen {\it body frame}. Then,
a local cross section of the principal fiber bundle selects just
one orientation of a generic N-body configuration in each fiber
(SO(3) orbit). This is in fact equivalent to a {\it gauge
convention} namely, after a preliminary choice of the {\it shape}
variables, to a possible definition of a {\it body frame} ({\it
reference orientation}). It turns out that this principal bundle
is trivial only for N=3, so that in this case global cross
sections exist, and in particular the identity cross section may
be identified with the {\it space frame}. Any global cross section
is a copy of the 3-body shape space and its coordinatization
provides a description of the {\it internal vibrational} motions
associated with the chosen gauge convention for the reference
orientation. For $N\geq 4$, however, global cross sections do not
exist \cite{4} and the definition of the reference orientation
({\it body frame}) can be given only locally. This means that the
{\it shape space cannot be identified with a (3N-6)-dimensional
sub-manifold} of the (3N-3)-dimensional relative configuration
space. The {\it gauge convention} about the reference orientation
and the consequent individuation of the internal {\it vibrational}
degrees of freedom requires the choice of a connection $\Gamma$ on
the SO(3) principal bundle (i.e. a concept of {\it
horizontality}), which leads in turn to the introduction of a
SO(3) gauge potential on the base manifold. In this way a natural
gauge invariant concept exists of {\it purely rotational} N-body
configurations ({\it vertical} velocity vector field, i.e. {\it
null shape velocities}). Clearly, a gauge fixing is needed in
order to select a particular local cross section and the
correlated gauge potential on the shape space. Obviously, {\it
physical quantities like the rotational or vibrational kinetic
energies and, in general, any observable feature of the system
must be gauge invariant}. On the other hand, in the {\it
orientation-shape} bundle approach, both the space frame and the
body frame components of the angular velocity are gauge quantities
so that their definition depends upon the gauge convention. See
Ref.\cite{1} for a review of the gauge fixings used in molecular
physics' literature and, in particular, for the virtues of a
special connection C corresponding to the shape configurations
with vanishing center-of-mass angular momentum $\vec S$. The C
connection is defined by introducing local cross sections
orthogonal to the fibers with respect to the Riemannian metric
dictated by the kinetic energy.

The {\it orientation-shape} approach replaces the usual Euler
kinematics of rigid bodies and implies in general a coupling
between the internal {\it shape} variables and some of the {\it
orientational} degrees of freedom. In Ref.\cite{1} it is
interestingly shown that the non-triviality of the SO(3) principal
bundle, when extended to continuous deformable bodies, is the core
of the physical explanation of the {\it falling cat} and the {\it
diver}. A characteristic role of SO(3) gauge potentials in this
case is to {\it generate rotations by changing the shape}. This
approach is defined in configuration space and leads to {\it
momentum-independent shape variables}, so that it can be easily
extended to the quantum level (the N-body Schr\"odinger equation).
In Ref.\cite{1} the Hamiltonian formulation of this framework is
also given, but no explicit procedure is worked out for the
construction of a canonical Darboux basis for the orientational
and shape variables. See Refs.\cite{1,3} for the existing sets of
shape variables for $N=3,4$ and for the determination of their
physical domain.

\bigskip

Consider now the description of a N-body system in a special
relativistic context. Unlike the Newtonian case, manifest Lorentz
covariance requires the introduction of 4N degrees of freedom in
configuration space and 8N in phase space. In phase space there
are N mass-shell first-class constraints ($\epsilon\, p_i^2 -
m^2_i \approx 0$ in the free case \cite{5}). They determine the
energies $p_i^o \approx \pm \sqrt{m_i^2 + {\vec p}_i^2}$ and
entail that the time variables $x^o_i$ be gauge variables. Such
gauge variables can be replaced by a {\it center-of-mass time}
(arbitrariness in the choice of the rate of the center-of-mass
clock) and $N - 1$ {\it relative times} (arbitrariness in the
synchronization of the clocks associated to each body). The
problem of relative times has been a big obstacle to the
development of relativistic mechanics (see Refs.\cite{6} for the
essential bibliography). The solution of this problem led to the
development of parametrized Minkowski theories \cite{6}, to the
{\it Wigner-covariant rest-frame instant form of dynamics}
referred to the intrinsic inertial rest-frame of the N-body system
\cite{6,7} (see Ref.\cite{8} for Dirac's forms of the dynamics),
and to the development of generalized radar coordinates for the
synchronization of distant clocks in arbitrary non-inertial frames
\cite{9}. A consistent elimination of relative times (i.e. the
choice of a convention for the synchronization of clocks,
different in general from Einstein's convention), together with a
choice of the center-of-mass time, reduce the variables of the
N-body system to the same number as in the non-relativistic case.
In special relativity the rest-frame instant form of dynamics,
which corresponds to Einstein's synchronization convention due to
the inertial nature of the rest-frame, implements the {\it
relativistic separation of the center of mass}, so that only
$3N-3$ relative coordinates (or $6N-6$ relative phase space
variables) survive in the rest-frame.

\medskip

As it will be shown, this separation is well defined in the rest
frame, which is intrinsically defined by the Wigner hyper-planes
orthogonal to the conserved 4-momentum of the isolated system. It
is important to stress that {\it each such hyperplane is an
instantaneous Euclidean and Wigner-covariant 3-space}. This leads
to the characterization of {\it two} distinct realizations of the
Poincare' group, referred to an {\it abstract observer} sitting
outside or inside the Wigner hyper-planes, namely the {\it
external} and the {\it internal} realization, respectively. Also,
we get the characterization of the relevant notions of
relativistic external and internal 4- and 3- centers of mass, as
well as of a final set of canonical relative variables with
respect to the internal canonical 3-center of mass. In absence of
interactions, such relative variables are identified by a
canonical transformation, which, unlike the non-relativistic case,
is {\it point only in the momenta}. However, in presence of
action-at-a-distance interactions among the particles, the
canonical transformation identifying the relative variables
becomes interaction-dependent. In any case the use of the
non-interacting internal 3-centers of mass and relative variables
shows that the Wigner hyper-planes constitute the natural
framework for the {\it relativistic theory of orbits}. It is also
argued that a future relativistic theory of orbits will be a
non-trivial extension of the non-relativistic theory \cite{10},
for, in the instant form of relativistic dynamics, the potentials
appear both in the Hamiltonian and in the Lorentz boosts.

\bigskip

An important point is that, as shown in Ref.\cite{11}, Jacobi
normal coordinates, as well as notions like reduced masses and
inertia tensors, do not survive in special relativity. Some of
such notions can be recovered (and still in a non-unique way)
\cite{12} only by replacing the N-body system with a multi-polar
expansion. A different strategy must be consequently devised {\it
already at the non-relativistic level} \cite{13}.

\medskip

Due to the presence of the mass-shell first-class constraints, the
description of N-body systems in the rest-frame instant form make
use of a special class of canonical transformations, of the
Shanmugadhasan type \cite{14} and \cite{7}. Such transformations
are simultaneously adapted to: i) the Dirac first-class
constraints appearing in the Hamiltonian formulation of
relativistic models (the transformations have the effect that the
constraint equations are replaced by the vanishing of an equal
number of new momenta, whose conjugate variables are the
Abelianized gauge variables of the system); and to ii) the
time-like Poincar\'e orbits associated with most of their
configurations. In the Darboux bases one of the final canonical
variables is the square root of the Poincar\'e invariant $P^2$
(where $P_{\mu}$ is the conserved time-like four-momentum of the
isolated system).

\medskip

By exploiting the constructive theory of the canonical
realizations of Lie groups \cite{15,16,17,18,19}, a new family of
canonical transformations was introduced in Ref.\cite{20}. This
family of transformations leads to the definition of the so-called
{\it canonical spin bases}, in which also the Pauli-Lubanski
Poincar\'e invariant $W^2=-P^2 {\vec S}^2_T$ for the time-like
Poincar\'e orbits becomes one of the final canonical variables
(provided the rest-frame Thomas spin ${\vec S}_T =
\sum_{a=1}^{N-1}\, {\vec S}_a$ is different from zero). The
essential point is that the construction of the spin bases {\it
exploits the clustering of spins instead of the Jacobi clustering
of centers of mass, which is an ill defined notion at the
relativistic level}.
\medskip

Note that, in spite of its genesis in a relativistic context, the
technique used in the determination  of the spin bases, related to
a {\it typical form} \cite{15} of the canonical realizations of
the E(3) group, {\it can be easily adapted to the non-relativistic
case}, where $W^2$ is replaced by the invariant ${\vec S}^2$ of
the extended Galilei group. The fact that the traditional Jacobi
clustering of the centers of mass of the sub-clusters is replaced
by the clustering of the spins ${\vec S}_a$, ($a=1,..,N-1$) of the
sub-clusters, as in the composition of quantum mechanical angular
momenta, is {\it the basic trick that makes the treatment of the
non-relativistic N-body problem directly extendible to the
relativistic case}. The clustering can be achieved by means of
suitable canonical transformations, which {\it in general are
non-point both in the coordinates and the momenta}. This entails
that the quantum implementations of such canonical transformations
as unitary transformations be non-trivial. The extension of our
formalism to quantum mechanics remains an open problem.

\medskip

Our aim at this point the construction of a canonical Darboux
basis adapted to the non-Abelian SO(3) symmetry, both at the
non-relativistic and the relativistic level. The three non-Abelian
Noether constants of motion $\vec S = S^r\, {\hat f}_r$ (${\hat
f}_r$ are the axes of the inertial {\it laboratory or space
frame}) are arranged in these canonical Darboux bases as an array
containing the canonical pair $S^3$, $\beta = tg^{-1}{{S^2}\over
{S^1}}$ and the unpaired variable $S=|\vec S|$ ({\it scheme A} of
the canonical realization of SO(3) \cite{16}; the configurations
with $\vec S = 0$ are singular and have to be treated separately).
The angle canonically conjugated to $S$, say $\alpha$, is an {\it
orientational} variable, which, being coupled to the internal {\it
shape} degrees of freedom, cannot be a constant of motion.
However, being conjugated to a constant of the motion, it is an
{\it ignorable} variable in the Hamiltonian formalism, so that its
equation of motion can be solved by quadratures after the solution
of the other equations. In conclusion, in this non-Abelian case
one has only two (instead of three  as in the Abelian case)
commuting constants of motion, namely $S$ and $S^3$ (like in
quantum mechanics). This is also the outcome of the momentum map
canonical reduction \cite{21,22} by means of adapted coordinates.
Let us stress that $\alpha$, $S^3$, $\beta$ are a local
coordinatization of any co-adjoint orbit of SO(3) contained in the
N-body phase space. Each co-adjoint orbit is a 3-dimensional
embedded sub-manifold and is endowed with a Poisson structure
whose neutral element is $\alpha$. By fixing non-zero values of
the variables $S^3$, $\beta = tg^{-1}{{S^2}\over {S^1}}$ through
second-class constraints, one can define a (6N-8)-dimensional
reduced phase space. {\it The canonical reduction cannot be
furthered by eliminating $S$, just because $\alpha$ is not a
constant of motion}. Yet, the angle $\alpha$ allows us to
construct a unit vector $\hat R$, orthogonal to $\vec S$, such
that $\hat S$, $\hat R$, $\hat S \times \hat R$  (the notation
$\hat {}$ means unit vector) is an orthonormal frame that we call
{\it spin frame}.

\bigskip

The final lacking ingredient for our construction of body frames
comes from the group-theoretical treatment of rigid bodies
\cite{22} (Chapter IV, Section 10). Such treatment is based on the
existence of realizations of the (free and transitive) {\it left}
and {\it right} Hamiltonian actions of the SO(3) rotation group on
either the tangent or cotangent bundle over their configuration
space. The generators of the {\it left} Hamiltonian action
\cite{23}, which is a {\it symmetry action}, are the above
non-Abelian constants of motion $S^1$, $S^2$, $S^3$, [$\{ S^r,S^s
\} = \epsilon^{rsu} S^u$].
\medskip

At this point let us stress that, in the approach of Ref.\cite{1}
the SO(3) principal bundle is built starting from the {\it
relative configuration space} and, upon the choice of a body-frame
convention, a gauge-dependent SO(3) {\it right} action is
introduced. The corresponding task in our case is the following:
taking into account the {\it relative phase space} of any isolated
system, we have to find out whether one or more SO(3) {\it right}
Hamiltonian actions could be implemented besides the global SO(3)
{\it left} Hamiltonian action. In other words, we have to look for
solutions ${\check S}^r$, r=1,2,3, [with $\sum_r ({\check S}^r)^2
= \sum_r (S^r)^2 = S^2$], of the partial differential equations
$\{ S^r, {\check S}^s \} =0$, $\{ {\check S}^r, {\check S}^s \} =-
\epsilon^{rsu} {\check S}^u$ and then build corresponding {\it
left} invariant Hamiltonian vector fields. Alternatively, one may
search for the existence of a pair ${\check S}^3$, $\gamma =
tg^{-1} {{{\check S}^2}\over {{\check S}^1}}$, of canonical
variables satisfying $\{ \gamma ,{\check S}^3 \} =-1$, $\{ \gamma
,S^r \} = \{ {\check S}^3, S^r \} =0$ and also $\{ \gamma ,\alpha
\} = \{ {\check S}^3, \alpha \} =0$. Local theorems given in
Refs.\cite{15,16} guarantee that this is always possible provided
$N \geq 3$. Clearly, the functions ${\check S}^r$, which are not
constants of the motion, do not generate symmetry actions. What
matters here is that each explicitly given right action leads to
the characterization of the following two structures: i) a
dynamical reference frame (say $\hat N$, $\hat \chi$, $\hat N
\times \hat \chi$), that we call {\it dynamical body frame}; ii) a
{\it canonical spin basis} including both the {\it orientational}
and the {\it shape} variables.

\bigskip

In conclusion, we show that, after the center-of-mass separation,
by exploiting the new notions of {\it dynamical body frames} and
{\it canonical spin bases}, it is possible to build a geometrical
and group-theoretical procedure for the common characterization of
the rotational kinematics of non-relativistic and relativistic
N-body systems. The two cases are treated in Sections II and III,
respectively.

\bigskip

In the last Subsection of Section III we show that the
relativistic separation of the center of mass, realized by the
rest-frame instant form, can be extended to continuous deformable
relativistic isolated systems, namely relativistic field
configurations, strings and fluids. The action principle of such
systems can be transformed into a parametrized Minkowski theory on
space-like hyper-surfaces whose embeddings in Minkowski space-time
are the {\it gauge} variables connected with the arbitrariness in
the choice of the convention for clock synchronization (namely the
{\it choice of the equal-time Cauchy surfaces} for the field
equations). Then the rest-frame instant form on Wigner
hyper-planes emerges in a natural way also for fields (ADM
canonical metric \cite{24} and tetrad gravity \cite{7} naturally
deparametrize to it). All the notions of external and internal 4-
and 3-centers of mass can be extended to field configurations
under the condition that a collective 4-vector \cite{25},
canonically conjugate to the configuration conserved 4-momentum,
be definable.

The construction of such a collective variable, with respect to
which the energy-momentum distribution of the configuration itself
is peaked, is exemplified for a classical real Klein-Gordon field
\cite{26}. New features, like an {\it internal time variable},
absent in the particle case, emerge for fields and entail that
each constant energy surface of the configuration be a disjoint
union of symplectic manifolds. These results can be extended to
relativistic perfect fluids \cite{27}.

\bigskip

The next issue has to do with the simulation of an extended system
by means of as few as possible global parameters, sufficient to
maintain an acceptable phenomenological description of the system.
This can be achieved by replacing the extended system with a
multipolar expansion around a world-line describing its mean
motion. After many attempts, a general approach to this problem
has been given by Dixon in Ref.\cite{28} for special relativity
and in Ref.\cite{29} (see also Refs.\cite{30,31}) for general
relativity, after a treatment of the non-relativistic case
\cite{29}.

\medskip

With this in view, we show in Section IV that the rest-frame
instant-form of the dynamics is the natural framework (extendible
to general relativity \cite{7}) for the formulation of
relativistic multipolar expansions \cite{12} exhibiting a clear
identification of the internal canonical 3-center of mass (with
associated spin dipole and higher multi-poles) as a preferred
center of motion. Besides a system of N free relativistic
particles, we also discuss an {\it open system} defined by cluster
of $n < N$ charged particles inside an isolated system of N
charged particles, plus the electro-magnetic field in the
radiation gauge \cite{6}. Finally, as a prerequisite to the
treatment of relativistic perfect fluids \cite{27}, we give some
results concerning a configuration of a classical Klein-Gordon
field \cite{26}, where the collective variable allows identifying
a natural center of motion together with the associated Dixon
multi-poles.

\section{The non-relativistic canonical spin bases and dynamical
body frames.}

Let us consider N free non-relativistic particles of masses $m_i$,
$i=1,..,N$, described by the configuration variables ${\vec
\eta}_i$ and by the momenta ${\vec \kappa}_i$. The Hamiltonian $H
= \sum_{i=1}^N\, {{{\vec \kappa}_i^2}\over {2 m_i}}$ is restricted
by the three first-class constraints ${\vec \kappa}_+ =
\sum_{i=1}^N\, {\vec \kappa}_i \approx 0$ defining the
center-of-mass or rest frame.

Let us introduce the following family of point canonical
transformations ($m = \sum_{i=1}^N\, m_i$) realizing the
separation of the center of mass from arbitrary relative variables
(for instance some set of Jacobi normal coordinates)

\begin{equation}
\begin{minipage}[t]{4cm}
\begin{tabular}{|l|} \hline
${\vec \eta}_i$ \\  \hline
 ${\vec \kappa}_i$ \\ \hline
\end{tabular}
\end{minipage} \ {\longrightarrow \hspace{.2cm}} \
\begin{minipage}[t]{4 cm}
\begin{tabular}{|l|l|} \hline
${\vec q}_{nr}$   & ${\vec \rho}_a$   \\ \hline
 ${\vec \kappa}_{+}$&${\vec \pi}_{a}$ \\ \hline
\end{tabular}
\end{minipage}
\label{2.1}
\end{equation}

\noindent defined by:

\begin{eqnarray*}
{\vec \eta}_i&=&
 {\vec q}_{nr} + {1\over {\sqrt{N}}} \sum_{a=1}^{N-1} \Gamma_{ai} {\vec
\rho}_{a},\qquad {\vec \kappa}_i =
 {{m_i}\over m} {\vec \kappa}_{+} + \sqrt{N} \sum_{a=1}^{N-1}
\gamma_{ai} {\vec \pi}_{a},
 \end{eqnarray*}

\begin{eqnarray*}
 {\vec q}_{nr}&=&\sum_{i=1}^N{{m_i}\over m} {\vec
 \eta}_i,\qquad
  {\vec \kappa}_{+} = \sum_{i=1}^N {\vec \kappa}_i,\nonumber \\
  {\vec \rho}_a&=&\sqrt{N} \sum_{i=1}^N
\gamma_{ai} {\vec \eta}_i,\qquad
  {\vec \pi}_{a} = {1\over {\sqrt{N}}} \sum_{i=1}^N \Gamma_{ai} {\vec
 \kappa}_i,\qquad
 \vec S =  \sum_{a=1}^{N-1}{\vec \rho}_a\times {\vec
\pi}_{a},
 \end{eqnarray*}

\begin{eqnarray}
 \Gamma_{ai}&=& \gamma_{ai} - \sum_{k=1}^N {{m_k}\over m} \gamma_{ak},\quad\quad
\gamma_{ai}=\Gamma_{ai}-{1\over N} \sum_{k=1}^N
\Gamma_{ak},\nonumber \\ &&\sum_{i=1}^N\gamma_{ai}=0,\quad\quad
\sum_{i=1}^N{{m_i}\over m}\Gamma_{ai}=0,\nonumber \\
&&\sum_{i=1}^N\gamma_{ai}\gamma_{bi}=\delta_{ab},\quad\quad
\sum_{i=1}^N\gamma_{ai} \Gamma_{bi}=\delta_{ab},\nonumber \\
&&\sum_{a=1}^{N-1}\gamma_{ai}\gamma_{aj}=\delta_{ij}-{1\over
N},\quad\quad
\sum_{a=1}^{N-1}\gamma_{ai}\Gamma_{aj}=\delta_{ij}-{{m_i}\over m}.
\label{2.2}
\end{eqnarray}

Here, the $\gamma_{ai}$'s and the $\Gamma_{ai}$'s are numerical
parameters depending on ${1\over 2}(N-1)(N-2)$ free parameters
\cite{6}.

It can be shown \cite{13} that the relative motion is described by
the following Lagrangian and Hamiltonian

\begin{eqnarray}
  L_{rel}(t) &=& {1\over 2}
\sum_{a,b}^{1..N-1} k_{ab}[m_i,\Gamma_{ai}]\, {\dot {\vec
\rho}}_{a}(t)\cdot {\dot {\vec \rho}}_{b}(t),\nonumber \\
 &&{}\nonumber \\
 &&k_{ab}[m_i,\Gamma_{ci}]=k_{ba}[m_i,\Gamma_{ci}]={1\over N}\sum_{i=1}^N
 m_i\Gamma_{ai}\Gamma_{bi}, \nonumber \\
 &&k^{-1}_{ab}[m_i,\Gamma_{ci}] = N \sum_{i=1}^N {{\gamma_{ai}\gamma_{bi}}\over {m_i}},
 \nonumber \\
 &&{}\nonumber \\
 &&\Downarrow \nonumber \\
 {\vec \pi}_{a}(t)&=&\sum_{b=1}^{N-1} k_{ab}[m_i,\Gamma_{ci}]\,
  {\dot {\vec \rho}}_{b}(t),\nonumber \\
 &&{}\nonumber \\
\Rightarrow&& H_{rel}={1\over 2} \sum_{ab}^{1..N-1}
k^{-1}_{ab}[m_i,\Gamma_{ai}]\,\,
 {\vec \pi}_{a}(t)\cdot {\vec \pi}_{b}(t).
\label{2.3}
\end{eqnarray}

\noindent If we add the gauge fixings ${\vec q}_{nr}\approx 0$ and
we go to Dirac brackets with respect to the second-class
constraints ${\vec \kappa}_{+}\approx 0$, ${\vec q}_{nr}\approx
0$, we get a (6N-6)-dimensional reduced phase space spanned by
${\vec \rho}_{a}$, ${\vec \pi}_{a}$ and with $\vec S\equiv
\sum_{a=1}^{N-1} {\vec \rho}_a\times {\vec \pi}_{a} =
\sum_{a=1}^{N-1}\, {\vec S}_a$. At the non-relativistic level
\cite{1} the next problem of the standard approach for each N is
the diagonalization of the matrix $k_{ab}[m_i,\Gamma_{ai}]$. The
off-diagonal terms of the matrix $k_{ab}[m_i,\Gamma_{ai}]$ are
called {\it mass polarization terms}, while its eigenvalues are
the {\it reduced masses} (see for instance Ref.\cite{32}). In this
way the Jacobi normal coordinates ${\vec \rho}_a = {\vec s}_a$,
with conjugate momenta ${\vec \pi}_a = {\vec \pi}_{sa}$, are
introduced.

\medskip

In the rest frame the 11 generators of the extended Galilei group
(the total mass $m$ is a central charge) \cite{17} are

\bea
 m &=& \sum_{i=1}^N\, m_i,\qquad
 E = \sum_{i=1}^N\, {{{\vec \kappa}_i^2}\over {2 m_i}}
 \approx H_{rel},\qquad
 {\vec \kappa}_+ = \sum_{i=1}^N\, {\vec \kappa}_i \approx
 0,\nonumber \\
 \vec J &=& \sum_{i=1}^N\, {\vec \eta}_i \times {\vec \kappa}_i
 \approx \vec S = \sum_{a=1}^{N-1}\, {\vec S}_a =
 \sum_{a=1}^{N-1}\, {\vec \rho}_a \times {\vec \pi}_a,\nonumber \\
 \vec K &=& - \sum_{i=1}^N\, m_i\, {\vec \eta}_i + {\vec
 \kappa}_+\, t \approx - m\, {\vec q}_{nr},
 \label{2.4}
 \eea

\noindent and the gauge fixings ${\vec q}_{nr} \approx 0$ are
equivalent to $\vec K \approx 0$.

\bigskip

Following the strategy delineated in the Introduction, we search
for a geometrical and group-theoretical characterization of
further canonical transformations leading to a privileged class of
canonical Darboux bases for the N-body. Specifically, they must be
adapted to the SO(3) subgroup \cite{15,16} of the extended Galilei
group and contain one of its invariants, namely the modulus of the
spin:
\medskip

1) Every such basis must be a {\it scheme B} (i.e. a canonical
completion of {\it scheme A}, according to the language of
Ref.\cite{15,16,17,20}) for the canonical realization of the
rotation group SO(3), viz. it must contain its invariant $S$ and
the canonical pair $S^3$, $\beta = tg^{-1}{{S^2}\over {S^1}}$.
This entails that, except for $\alpha$, all the remaining
variables in the canonical basis be SO(3) scalars.

2) The existence of the angle $\alpha$  satisfying $\{ \alpha , S
\} =1$ and $\{ \alpha ,S^3 \} = \{ \alpha ,\beta \} =0$ leads to
the geometrical identification of a unit vector $\hat R$
orthogonal to $\vec S$ and, therefore, of an orthonormal frame,
the {\it spin frame} $\hat S$, $\hat R$, $\hat S\times \hat R$.

3) From the equations ${\hat R}^2=1$ and $\{ \vec S\cdot {\hat R},
{\hat R}^i \} =0$ it follows the symplectic result $\{ {\hat R}^i,
{\hat R}^j \} =0$. As a byproduct, we get a canonical realization
of a Euclidean group E(3) with generators $\vec S$, $\hat R\,\,$
[$\{ {\hat R}^i,{\hat R}^j \} =0$, $\{ {\hat R}^i, S^j \} =
\epsilon^{ijk} {\hat R}^k$] and fixed values of its invariants
${\hat R}^2=1$, $\vec S\cdot \hat R=0$ (non-irreducible {\it type
3} realization according to Ref.\cite{20}).

4) In order to implement a SO(3) {\it Hamiltonian right action} in
analogy with  the rigid body theory \cite{22}, we must construct
an orthonormal triad or {\it body frame} $\hat N$, $\hat \chi$,
$\hat N\times \hat \chi$. The decomposition

\begin{equation}
 \vec S = {\check S}^1\hat \chi +{\check S}^2 \hat N\times
\hat \chi + {\check S}^3 \hat N \,\,\, {\buildrel {def}\over
=}\,\,\, {\check S}^r {\hat e}_r,
 \label{2.5}
  \end{equation}

\noindent identifies the SO(3) scalar generators ${\check S}^r$ of
the {\it right action} provided they satisfy $\{ {\check
S}^r,{\check S}^s \} = -\epsilon^{rsu} {\check S}^u$. This latter
condition together with the obvious requirement that $\hat N$,
$\hat \chi$, $\hat N\times \hat \chi$ be SO(3) vectors [$\{ {\hat
N}^r,S^s \} =\epsilon^{rsu}{\hat N}^u$, $\{ {\hat \chi}^r,S^s \}
=\epsilon^{rsu}{\hat \chi}^u$, $\{ {\hat N\times \hat \chi}^r,S^s
\} =\epsilon^{rsu}{\hat N\times \hat \chi}^u$] entails the
equations \cite{33}

\begin{equation}
 \{ {\hat N}^r,{\hat N}^s \} = \{ {\hat N}^r,{\hat \chi}^s \}
= \{ {\hat \chi}^r,{\hat \chi}^s \} =0.
 \label{2.6}
  \end{equation}

Each solution of these equations identifies a couple of canonical
realizations of the E(3) group (non-irreducible, {\it type 2}):
one with generators $\vec S$, $\vec N$ and non-fixed invariants
${\check S}^3=\vec S \cdot \hat N$ and $|{\vec N}|$; another with
generators $\vec S$, $\vec \chi$ and non-fixed invariants ${\check
S}^1=\vec S\cdot \hat \chi$ and $|{\vec \chi}|$. Such realizations
contain the relevant information for constructing the new
canonical pair ${\check S}^3$, $\gamma =tg^{-1}{{{\check
S}^2}\over {{\check S}^1}}$ of SO(3) scalars. Since $\{  \alpha ,
{\check S}^3 \} = \{ \alpha , \gamma \} =0$ must hold, it follows
\cite{20} that the vector $\hat N$ necessarily belongs to the
$\hat S$-$\hat R$ plane. The three canonical pairs $S$, $\alpha$,
$S^3$, $\beta$, ${\check S}^3$, $\gamma$ will describe the {\it
orientational} variables of our Darboux basis, while $|\vec N|$
and $|\vec \chi |$ will belong to the {\it shape} variables. For
each independent right action [i.e. for each solution $\hat N$,
$\hat \chi$ of Eqs.(\ref{2.6})], we can identify a {\it canonical
spin basis} containing the above 6 orientational variables and
$6N-12$ canonical shape variables. Alternatively, an anholonomic
basis can be constructed by replacing the above six variables by
${\check S}^r$ (or $S^r$) and three uniquely determined Euler
angles $\tilde \alpha$, $\tilde \beta$, $\tilde \gamma$ (see
Ref.\cite{13}). Let us stress that the non-conservation of
${\check S}^r$ entails that {\it the dynamical body frame evolves
in a way dictated by the equations of motion}, just as it happens
in the rigid body case.

\bigskip

We can conclude that the N-body problem has {\it hidden
structures} leading to the characterization of special {\it
dynamical body frames} which, {\it being independent of gauge
conditions, are endowed with a direct physical meaning}.

\subsection{The 2-body system.}

For $N=2$, a single E(3) group can be defined: it allows the
construction of an orthonormal {\it spin frame} $\hat S$, $\hat
R$, $\hat R\times \hat S$ in terms of the measurable relative
coordinates and momenta of the particles. The relative variables
are ${\vec \rho}=\vec \rho$, ${\vec \pi}$ and the Hamiltonian is
$H_{rel}={{{\vec \pi}^2}\over {2\mu}}$, where $\mu={{m_1m_2}\over
{m_1+m_2}}$ is the reduced mass. The spin  is ${\vec S}={\vec
\rho}\times {\vec \pi}$. Let us define the following decomposition

\begin{eqnarray}
{\vec \rho}&=& \rho \hat R,\quad\quad \rho=\sqrt{{\vec
\rho}^2},\quad \quad \hat R={{{\vec \rho}}\over
{\rho}}={\hat \rho},\quad\quad {\hat R}^2=1, \nonumber \\
&&{}\nonumber \\
 {\vec \pi}&=& {\tilde \pi} \hat R -{{S}\over
{\rho}} \hat R\times {\hat S}={\tilde \pi}{\hat \rho}-{{S}\over
{\rho}}{\hat \rho}\times {\hat S},\nonumber \\
 && {\tilde \pi}={\vec \pi}\cdot \hat R={\vec \pi}\cdot {\hat \rho},\quad\quad
{\hat S}={{{\vec S}}\over {S}},\quad\quad {\hat S} \cdot \hat R
=0.
 \label{2.7}
\end{eqnarray}

As  shown in Ref.\cite{20}, it is instrumental considering the
following non-point canonical transformation adapted to the SO(3)
group, valid when ${\vec S}\not= 0$

\begin{eqnarray}
&&\begin{minipage}[t]{1cm}
\begin{tabular}{|l|} \hline
${\vec \rho}$ \\ \hline ${\vec \pi}$ \\ \hline
\end{tabular}
\end{minipage} \ {\longrightarrow \hspace{.2cm}} \
\begin{minipage}[t]{2 cm}
\begin{tabular}{|ll|l|} \hline
$\alpha$ & $\beta$ & $\rho$\\ \hline
 $S$   & $S^3$ & ${\tilde \pi}$  \\ \hline
\end{tabular}
\end{minipage}  ,\qquad \alpha = tg^{-1} {1\over {S}} \Big( {\vec \rho}\cdot {\vec
\pi}- {{(\rho)^2}\over {\rho^3}} \pi^3\Big).
 \label{2.8}
\end{eqnarray}

In the language of Ref.\cite{20}, the two pairs of canonical
variables $\alpha$, $S$, $\beta$, $S^3$ form the {\it irreducible
kernel} of the {\it scheme A} of a (non-irreducible, {\it type 3},
) canonical realization of the group E(3), generated by ${\vec
S}$, $\hat R$, with fixed values of the invariants ${\hat R}^2=1$,
$\hat R \cdot {\vec S}=0$, just as the variables $S^3$, $\beta$
and $S$ form the {\it scheme A} of the SO(3) group with invariant
$S$. Geometrically, we have: i) the angle $\alpha$ is the angle
between the plane determined by ${\vec S}$ and ${\hat f}_3$ and
the plane determined by ${\vec S}$ and $\hat R$; ii) the angle
$\beta$ is the angle between the plane ${\vec S}$ - ${\hat f}_3$
and the plane ${\hat f}_3$ - ${\hat f}_1$. Moreover, $S^1 = \sqrt{
(S)^2-(S^3)^2} cos\, \beta$, $S^2 = \sqrt{ (S)^2-(S^3)^2} sin\,
\beta$, ${\hat R}^1 = {\hat \rho}^1= sin\, \beta sin\, \alpha
-{{S^3}\over {S}} cos\, \beta cos\, \alpha$, ${\hat R}^2 = {\hat
\rho}^2= -cos\, \beta sin\, \alpha - {{S^3} \over {S}} sin\, \beta
cos\, \alpha$, ${\hat R}^3 = {\hat \rho}^3={1\over {S}}
\sqrt{(S)^2- (S^3)^2} cos\, \alpha$, $\alpha =- tg^{-1}\, {{
({\hat S} \times \hat R)^3}\over {[{\hat S} \times ( {\hat S}
\times \hat R)]^3}}$.

\medskip

In this degenerate case (N=2), the {\it dynamical} shape variables
$\rho$, ${\tilde \pi}$ coincide with the {\it static} ones and
describe the vibration of the dipole. The rest-frame Hamiltonian
for the relative motion becomes ($\check I$ is the barycentric
tensor of inertia of the dipole) $H_{rel}= {1\over 2} \Big[
{\check I}^{-1} S^2 +{{ {\tilde \pi}^2}\over {\mu}}\Big]$, $\check
I = \mu \rho^2$, while the body frame angular velocity is ${\check
\omega} = {{\partial H_{rel}}\over {\partial {\check S}}} ={{
{\check S}}\over {\check I}}$.

\subsection{The 3-body system.}

For $N=3$, where we have $\vec S = {\vec S}_1+{\vec S}_2$, a {\it
pair} of E(3) groups emerge, associated with ${\vec S}_1$ and
${\vec S}_2$, respectively. We have now {\it two} unit vectors
${\hat R}_a$ and {\it two} E(3) realizations generated by ${\vec
S}_{a}$, ${\hat R}_a$ respectively and fixed invariants ${\hat
R}^2_a=1$, ${\vec S}_{a}\cdot {\hat R}_a=0$ (non-irreducible, type
2, see Ref.\cite{20}). We shall assume ${\vec S} \not= 0$, because
the exceptional SO(3) orbit $S=0$ has to be studied separately by
adding $S \approx 0$ as a first-class constraint.
\medskip

For each value of $a=1,2$, we consider the non-point canonical
transformation (\ref{2.8})

\bea
 &&\begin{minipage}[t]{1cm}
\begin{tabular}{|l|} \hline
${\vec \rho}_{a}$ \\ \hline ${\vec \pi}_{a}$ \\ \hline
\end{tabular}
\end{minipage} \ {\longrightarrow \hspace{.2cm}} \
\begin{minipage}[t]{2 cm}
\begin{tabular}{|ll|l|} \hline
$\alpha_a$ & $\beta_a$ & $\rho_{a}$\\ \hline
 $S_{a}$   & $S^3_{a}$ & ${\tilde \pi}_{a}$  \\ \hline
\end{tabular}
\end{minipage},\nonumber \\
 &&{}\nonumber \\
 &&{}\nonumber \\
 {\vec \rho}_{a}&=& \rho_{a} {\hat R}_a,\quad\quad
\rho_{a}=\sqrt{{\vec \rho} _{a}^2},\quad\quad {\hat R}_a={{{\vec
\rho}_{a}}\over {\rho_{a}}}={\hat \rho}_{a}, \quad\quad {\hat
R}_a^2=1,\nonumber \\
 &&{}\nonumber \\
 {\vec \pi}_{a}&=&
{\tilde \pi}_{a} {\hat R}_a +{{S_{a}}\over {\rho_{a}}} {\hat
S}_{a}\times {\hat R}_a,\quad\quad {\tilde \pi}_{a}= {\vec
\pi}_{a}\cdot {\hat R}_a.
 \label{2.9}
\eea

\medskip

In this case, besides the orthonormal {\it spin frame}, an
orthonormal {\it dynamical body frame} $\hat N$, $\hat \chi$,
$\hat N\times \hat \chi$, i.e. a SO(3) Hamiltonian {\it right}
action, can be defined. The {\it simplest choice},  within the
existing arbitrariness \cite{34}, for the orthonormal vectors
$\vec N$ and $\vec \chi$ functions only of the relative
coordinates is

\begin{eqnarray}
\vec N&=& {1\over 2} ({\hat R}_1+{\hat R}_2)= {1\over 2} ({\hat
\rho}_{1}+{\hat \rho}_{2}),\quad\quad \hat N={{\vec N}\over {|\vec
N|}},\qquad |\vec N|=\sqrt{ {{1+{\hat \rho}_{1}\cdot {\hat
\rho}_{2}}\over 2} },\nonumber \\ \vec \chi &=&{1\over 2}({\hat
R}_1-{\hat R}_2)=
  {1\over 2}({\hat \rho}_{1}-{\hat \rho}_{2})
,\quad\quad \hat \chi ={{\vec \chi}\over {|\vec \chi |}},\qquad
  |\vec \chi |=\sqrt{ { {1-{\hat \rho}_{1}\cdot {\hat
\rho}_{2} }\over 2} }=\sqrt{1-{\vec N}^2},\nonumber \\
 &&{}\nonumber \\
  \vec N\times \vec \chi &=&-{1\over 2}{\hat
\rho}_{1}\times {\hat \rho}_{2},\quad \quad |\vec N\times \vec
\chi |= |\vec N| |\vec \chi |={1\over 2}\sqrt{1-({\hat
\rho}_{1}\cdot {\hat \rho}_{2})^2},\qquad
 \vec N\cdot \vec \chi = 0.
  \label{2.10}
\end{eqnarray}

\medskip

As said above, (Eq.(\ref{2.5}), this choice is equivalent to the
determination of the non-conserved generators ${\check S}^r$ of a
Hamiltonian {\it right action} of SO(3).

The realization of the E(3) group with generators ${\vec S}$,
$\vec N$ and non-fixed invariants ${\vec N}^2$, ${\vec S}\cdot
\vec N$ leads to the final canonical transformation introduced in
Ref.\cite{20}

\bea
 &&\begin{minipage}[t]{1cm}
\begin{tabular}{|l|} \hline
${\vec \rho}_{a}$ \\ \hline
 ${\vec \pi}_{a}$ \\ \hline
\end{tabular}
\end{minipage}\ {\longrightarrow \hspace{.2cm}} \
\begin{minipage}[t]{4cm}
\begin{tabular}{|ll|ll|l|} \hline
$\alpha_1$ & $\beta_1$ & $\alpha_2$ & $\beta_2$ & $\rho_{a}$\\
\hline
 $S_{1}$   & $S^3_{1}$ & $S_{2}$ & $S^3_{2}$ & ${\tilde \pi}_{a}$\\ \hline
\end{tabular}
\end{minipage}
\ {\longrightarrow \hspace{.2cm}}\
\begin{minipage}[b]{4cm}
\begin{tabular}{|lll|l|l|} \hline
$\alpha$ & $\beta$ & $\gamma$ & $|\vec N|$ & $\rho_{a}$ \\ \hline
 $S={\check S}$   & $S^3$& ${\check S}^3={\vec S}\cdot \hat
N$ & $\xi$ & ${\tilde \pi}_{a}$ \\ \hline
\end{tabular}
\end{minipage},\nonumber \\
&&{}\nonumber \\
&&{}\nonumber \\
 \xi  &=&{{ \sqrt{2} \sum_{a=1}^2(-)^{a+1}{\vec
\rho}_{a}\times {\vec \pi}_{a}\cdot ({\hat \rho}_{2}\times {\hat
\rho}_{1})}\over {[1-{\hat \rho}_{1}\cdot {\hat
\rho}_{2}]\sqrt{1+{\hat \rho}_{1} \cdot {\hat \rho}_{2}} }}.
 \label{2.11}
\eea

For N=3 the {\it dynamical shape variables}, functions of the
relative coordinates ${\vec \rho}_{a}$ only, are  $|\vec N|$ and
$\rho_{a}$, while the conjugate shape momenta  are $\xi$, ${\tilde
\pi}_{a}$.
\medskip

We can now reconstruct ${\vec S}$ and define a {\it new} unit
vector $\hat R$ orthogonal to ${\vec S}$ by adopting the
prescription given after Eq.(\ref{2.8}). The vectors ${\hat S}$,
$\hat R$, ${\hat S}\times \hat R$ build up the {\it spin frame}
for N=3. The angle $\alpha$ conjugate to $S$ is explicitly given
by $\alpha = -tg^{-1}\, {{ ({\hat S}\times \hat N)^3}\over {[{\hat
S}\times ({\hat S}\times \hat N)]^3}}=- tg^{-1}\, {{({\hat S}
\times \hat R)^3}\over {[{\hat S} \times ({\hat S} \times \hat
R)]^3}}$. The two expressions of $\alpha$ given here are
consistent with the fact that ${\hat S}$, $\hat R$ and $\hat N$
are coplanar, so that $\hat R$ and $\hat N$ differ only by a term
in ${\hat S}$.

{\it As a consequence of this definition of $\hat R$}, we get the
following expressions for the {\it dynamical body frame} $\hat N$,
$\hat \chi$, $\hat N\times \hat \chi$  in terms of the final
canonical variables

\begin{eqnarray}
 \hat N&=& cos\, \psi {\hat S}+sin\, \psi \hat R={{{\check S}^3}\over {S}}
 {\hat S}+{1\over {S}}\sqrt{(S)^2-({\check S}^3)^2}\hat R=\nonumber \\
 &=&\hat N [S,\alpha ;S^3,\beta ;{\check S}^3,\gamma ],\nonumber \\
 &&{}\nonumber \\
 \hat \chi &=&sin\, \psi cos\, \gamma {\hat S}-cos\, \psi cos\, \gamma \hat R +sin\,
 \gamma {\hat S}\times \hat R=\nonumber \\
  &=& {{{\check S}^1}\over {S}}\, {\hat S}- {{{\check S}^3}\over {S}}
  {{ {\check S}^1\, \hat R + {\check S}^2\,  {\hat S}\times \hat R}\over
{ \sqrt{({ S})^2-({\check S}^3)^2}}}=
  \hat \chi [S,\alpha ;S^3,\beta ;{\check S}^3,\gamma ],\nonumber \\
 &&{}\nonumber \\
  &&\Downarrow\nonumber \\
  {\hat S} &=& sin\, \psi cos\, \gamma \hat \chi +sin\, \psi sin\, \gamma
  \hat N \times \hat \chi + cos\, \psi \hat N\nonumber \\
  &{\buildrel {def} \over =}& {1\over {S}} \Big[ {\check S}^1 \hat \chi +{\check S}^2
  \hat N \times \hat \chi +{\check S}^3 \hat N\Big],\nonumber \\
  &&{}\nonumber \\
  \hat R &=& -cos\, \psi cos\, \gamma \hat \chi -cos\, \psi sin\, \gamma
  \hat N \times \hat \chi +sin\, \psi \hat N,\nonumber \\
  &&{}\nonumber \\
  \hat R \times {\hat S} &=& -sin\, \gamma \hat \chi +
  cos\, \gamma \hat N \times \hat \chi .
\label{2.12}
\end{eqnarray}

While $\psi$ is the angle between ${\hat S}$ and $\hat N$,
$\gamma$ is the angle between the plane $\hat N - \hat \chi$ and
the plane ${\hat S} - \hat N$. As in the case N=2, $\alpha$ is the
angle between the plane ${\hat S} - {\hat f}_3$ and the plane
${\hat S} - \hat R$, while $\beta$ is the angle between the plane
${\hat S} - {\hat f}_3$ and the plane ${\hat f}_3 - {\hat f}_1$.

\medskip

Finally, as shown in Ref.\cite{13}, we can perform a sequence of a
canonical transformation to Euler angles $\tilde \alpha$, $\tilde
\beta$, $\tilde \gamma$ (explicitly calculable in terms of ${\vec
\rho}_a$ and ${\vec \pi}_a$) with their conjugate momenta,
followed by a transition to the anholonomic basis used in the
orientation-shape bundle approach \cite{1} ($q^{\mu}= (\rho_{1},
\rho_{2}, |\vec N|)$, $p_{\mu}=({\tilde \pi}_{1}, {\tilde
\pi}_{2}, \xi )$ are the dynamical  shape variables)

\bea
\begin{minipage}[t]{5cm}
\begin{tabular}{|lll|l|l|} \hline
$\alpha$ & $\beta$ & $\gamma$& $|\vec N|$ & $\rho_{a}$ \\ \hline
 $S={\check S}$   & $S^3$& ${\check S}^3$ & $\xi$ & ${\tilde \pi}_{a}$ \\ \hline
\end{tabular}
\end{minipage}&&
\ {{\buildrel {non\,\, can.} \over \longrightarrow}
\hspace{.2cm}}\
\begin{minipage}[b]{5cm}
\begin{tabular}{|lll|l|l|} \hline
$\tilde \alpha$ & $\tilde \beta$ & $\tilde \gamma$& $|\vec N|$ &
$\rho_{a}$ \\ \hline
 ${\check S}^1$ & ${\check S}^2$ & ${\check S}^3$ & $\xi$ & ${\tilde \pi}_{a}$ \\
 \hline
 \end{tabular}
 \end{minipage} =\nonumber \\
 &&{}\nonumber \\
 &&\begin{minipage}[b]{5cm}
\begin{tabular}{|lll|l|} \hline
$\tilde \alpha$ & $\tilde \beta$ & $\tilde \gamma$& $q^{\mu}({\vec
\rho}_{a})$
\\ \hline
 ${\check S}^1$ & ${\check S}^2$ & ${\check S}^3$ &
 $p_{\mu}({\vec \rho}_{a},{\vec \pi}_{a})$ \\ \hline
 \end{tabular}
 \end{minipage}.
 \label{2.13}
\eea

The Euler angles $\tilde \alpha$, $\tilde \beta$, $\tilde \gamma$
are determined as the {\it unique set} of {\it dynamical
orientation variables}.

\bigskip

For $N=3$ we get a result similar to the orientation-shape bundle
approach, namely the following relation between the space and body
components of the relative coordinates: $\rho^r_{a} = {\cal
R}^r{}_s(\tilde \alpha ,\tilde \beta ,\tilde \gamma ) {\check
\rho}^s_{a}(q)$, with ${\check \rho}^1_{a}(q) = (-)^{a+1} \rho_{a}
\sqrt{1-{\vec N}^2}$, ${\check \rho}^2_{a}(q)=0$, ${\check
\rho}^3_{a}(q) = \rho_{a} |\vec N|$, $S^r =  {\cal R}^r{}_s(\tilde
\alpha ,\tilde \beta ,\tilde \gamma ) {\check S}^s$. It can also
be shown that for N=3 our definition of {\it dynamical body frame}
can be reinterpreted as a special global cross section ({\it xxzz
gauge}, where $x$ stays for $\hat \chi$ and $z$ for $\hat N$; an
outcome that is independent of the particular choice made for
$\vec N$ and $\vec \chi$) of the trivial SO(3) principal bundle of
Ref.\cite{1}, namely a privileged choice of body frame. While the
above {\it dynamical body frame} can be identified with the global
cross section corresponding to the {\it xxzz gauge}, all other
global cross sections cannot be interpreted as {\it dynamical body
frames} (or {\it dynamical right} actions), because the SO(3)
principal bundle of Ref.\cite{1} is built starting from the
relative configuration space and, therefore, it is a {\it static},
velocity-independent, construction. As a matter of fact, after the
choice of the shape configuration variables $q^{\mu}$ and of a
space frame in which the relative variables have components
$\rho^r_a$, the approach of Ref.\cite{1} begins with the
definition of the body-frame components ${\check
\rho}^r_a(q^{\mu})$ of the relative coordinates, in the form
$\rho^r_a=R^{rs}(\theta^{\alpha}) {\check \rho}^s_a(q^{\mu})$, and
then extends it in a {\it velocity-independent} way to the
relative velocities ${\dot \rho}^r_a\, {\buildrel {def} \over =}\,
R^{rs}(\theta^{\alpha}) {\check v}^s_a$. Here $R$ is a rotation
matrix, $\theta^{\alpha}$ are arbitrary gauge orientational
parameters and ${\check \rho}^r_a(q^{\mu})$ is {\it assumed to
depend on the shape variables only}. In our construction we get
instead $\rho^r_a=R^{rs}(\tilde \alpha ,\tilde \beta ,\tilde
\gamma ) {\check \rho}^s_a(q^{\mu})$ in the {\it xxzz gauge}, so
that in the present case (N=3)  all {\it dynamical} variables of
our construction coincide with the {\it static} variables in the
xxzz gauge. On  the other hand, in the relative phase space, the
construction of the {\it evolving dynamical body frame} is based
on non-point canonical transformation.

\subsection{The N-body system.}

For N=4, it holds $\vec S = {\vec S}_1+{\vec S}_2+{\vec S}_3$.
Since we have three relative variables ${\vec \rho}_{1}$, ${\vec
\rho}_{2}$, ${\vec \rho}_{3}$ and related momenta ${\vec
\pi}_{1}$, ${\vec \pi}_{2}$, ${\vec \pi}_{3}$, it is possible to
construct {\it three} sets of {\it spin frames} and {\it dynamical
body frames}  corresponding to the hierarchy of clusterings
$((ab)c)$ [i.e. $((12)3)$, $((23)1)$, $((31)2)$; the subscripts
$a,b,c$ denote any permutation of $1,2,3$] of the relative spins
${\vec S}_a$. The associated three canonical Darboux bases share
the three variables $S^3$, $\beta$, $S$ (viz. $\vec S$), while
{\it both the remaining three orientational variables and the
shape variables depend on the spin clustering}. This entails the
existence of three different SO(3) {\it right} actions with
non-conserved canonical generators ${\check S}^r_{(A)}$, A=1,2,3.
Consistently, one can define three anholonomic bases ${\tilde
\alpha}_{(A)}$, ${\tilde \beta}_{(A)}$, ${\tilde \gamma}_{(A)}$,
${\check S}^r_{(A)}$ and associated shape variables
$q^{\mu}_{(A)}$, $p_{(A)\mu}$, $\mu =1,..,6$, connected by
canonical transformations leaving $S^r$ fixed. The relative
variables are therefore naturally split in three different ways
into 6 dynamical rotational variables and 12 generalized dynamical
shape variables. Consequently, we get three possible definitions
of {\it dynamical vibrations}.

\bigskip

By using the explicit construction given in Appendix  C of
Ref.\cite{13}, we define the following sequence of canonical
transformations (we assume $S\not= 0$; $S_{A}\not= 0$, $A=a,b,c$)
corresponding to the {\it spin clustering} pattern $abc \mapsto
(ab) c \mapsto ((ab)c)$ [build first the spin cluster $(ab)$, then
the spin cluster $((ab)c)$]:

\begin{eqnarray}
&&\begin{minipage}[t]{3cm}
\begin{tabular}{|lll|} \hline
${\vec \rho}_{a}$ & ${\vec \rho}_{b}$ & ${\vec \rho}_{c}$\\
${\vec \pi}_{a}$ & ${\vec \pi}_{b}$ & ${\vec \pi}_{c}$ \\
\hline
\end{tabular}
\end{minipage}
\ {\longrightarrow \hspace{.2cm}}\      \nonumber \\ \
{\longrightarrow \hspace{.2cm}}\ &&\begin{minipage}[t]{7cm}
\begin{tabular}{|ll|ll|ll|lll|} \hline
$\alpha_a$ & $\beta_a$ & $\alpha_b$& $\beta_b$ & $\alpha_c$ &
$\beta_c$ & $\rho_{a}$ & $\rho_{b}$ & $\rho_{c}$ \\ \hline
 $S_{a}$   & $S^3_{a}$& $S_{b}$ & $S^3_{b}$ & $S_{c}$ & $S^3_{c}$ &
 ${\tilde \pi}_{a}$ & ${\tilde \pi}_{b}$ & ${\tilde \pi}_{c}$ \\ \hline
\end{tabular}
\end{minipage}
\ {\longrightarrow \hspace{.2cm}}\     \nonumber \\ \ { {\buildrel
{(ab)c} \over \longrightarrow} \hspace{.2cm}} \
&&\begin{minipage}[t]{12cm}
\begin{tabular}{|ccc|cc|cccc|} \hline
$\alpha_{(ab)}$ & $\beta_{(ab)}$ & $\gamma_{(ab)}$& $\alpha_c$ &
$\beta_c$ & $|{\vec N}_{(ab)}|$ & $\rho_{a}$ & $\rho_{b}$ &
$\rho_{c}$\\
 \hline
 $S_{(ab)}$ & $S^3_{(ab)}$   &  ${\check S}_{(ab)}^3=
 {\vec S}_{(ab)}\cdot {\hat N}_{(ab)}$ & $S_{c}$ & $S^3_{c}$
  & $\xi_{(ab)}$ & ${\tilde \pi}_{a}$ & ${\tilde \pi}_{b}$ & ${\tilde \pi}_{c}$\\ \hline
\end{tabular}
\end{minipage}
\ {\longrightarrow \hspace{.2cm}}\     \nonumber \\ \
{\longrightarrow \hspace{.2cm}}\ &&\begin{minipage}[t]{13cm}
\begin{tabular}{|ccc|cccccc|} \hline
$\alpha_{((ab)c)}$ & $\beta_{((ab)c)}$ & $\gamma_{((ab)c)}$&
$|{\vec N}_{((ab)c)}|$ & $\gamma_{(ab)}$& $|{\vec N}_{(ab)}|$ &
$\rho_{a}$ & $\rho_{b}$ & $\rho_{c}$\\ \hline
 $S={\check S}$   & $S^3$& ${\check S}^3={\vec S}\cdot {\hat N}_{((ab)c)}$
 & $\xi_{((ab)c)}$ & ${\vec S}_{(ab)}\cdot {\hat N}_{(ab)}$ &
 $\xi_{(ab)}$ & ${\tilde \pi}_{a}$ & ${\tilde \pi}_{b}$ & $ {\tilde \pi}_{c}$\\ \hline
\end{tabular}
\end{minipage}
\ {\rightarrow \hspace{.2cm}}\     \nonumber \\ \ {{\buildrel
{non\, can.} \over \longrightarrow} \hspace{.2cm}}\
&&\begin{minipage}[t]{12cm}
\begin{tabular}{|ccc|cccccc|} \hline
$\tilde \alpha$ & $\tilde \beta$ & $\tilde \gamma$& $|{\vec
N}_{((ab)c)}|$ & $\gamma_{(ab)}$& $|{\vec N}_{(ab)}|$ & $\rho_{a}$
& $\rho_{b}$ & $\rho_{c}$\\ \hline
 ${\check S}^1$   & ${\check S}^2$& ${\check S}^3$
 & $\xi_{((ab)c)}$ & $\Omega_{(ab)}={\vec S}_{(ab)}\cdot {\hat N}_{(ab)}$ &
 $\xi_{(ab)}$ & ${\tilde \pi}_{a}$ & ${\tilde \pi}_{b}$ & $ {\tilde \pi}_{c}$\\ \hline
\end{tabular}
\end{minipage}.  \nonumber \\
&&{} \label{2.14}
\end{eqnarray}

The first non-point canonical transformation is based on the
existence of the three unit vectors ${\hat R}_A$, $A=a,b,c$, and
of three E(3) realizations with generators ${\vec S}_{A}$, ${\hat
R}_A$ and fixed values (${\hat R}_A^2=1$, ${\vec S}_A\cdot {\hat
R}_A=0$) of the invariants.
\medskip

In the next canonical transformation the spins of the {\it
relative particles} $a$ and $b$ are coupled to form the spin
cluster $(ab)$, leaving the {\it relative particle} $c$ as a
spectator. We use Eq.(\ref{2.10}) to define ${\vec
N}_{(ab)}={1\over 2}({\hat R}_a+{\hat R}_b)$, ${\vec
\chi}_{(ab)}={1\over 2}({\hat R}_a-{\hat R}_b)$, ${\vec
S}_{(ab)}={\vec S}_{a}+{\vec S}_{b}$, ${\vec W}_{(ab)}={\vec
S}_{a}-{\vec S}_{b}$. We  get ${\vec N}_{(ab)}\cdot {\vec
\chi}_{(ab)}=0$, $\{ N^r_{(ab)},N^s_{(ab)} \} = \{ N^r_{(ab)},
\chi^s_{(ab)} \} = \{ \chi^r_{(ab)}, \chi^s_{(ab)} \} =0$ and a
new E(3) realization generated by ${\vec S}_{(ab)}$ and ${\vec
N}_{(ab)}$, with non-fixed invariants $|{\vec N}_{(ab)}|$, ${\vec
S}_{(ab)}\cdot {\hat N}_{(ab)}\, {\buildrel {def} \over =}\,
\Omega_{(ab)}$. It can be shown that it holds

\bea
 &&{\vec \rho}_{a} = \rho_{a} \Big[ |{\vec N}_{(ab)}| {\hat
N}_{(ab)} + \sqrt{1-{\vec N}^2_{(ab)}} {\hat \chi}_{(ab)}\Big]
,\nonumber \\ &&{\vec \rho}_{b} = \rho_{b} \Big[ |{\vec N}_{(ab)}|
{\hat N}_{(ab)} - \sqrt{1-{\vec N}^2_{(ab)}} {\hat
\chi}_{(ab)}\Big] ,\nonumber \\
 &&{\vec \rho}_{c} = \rho_{c} {\hat R}_c.
 \label{2.15}
 \eea

\noindent It is then possible to define $\alpha_{(ab)}$ and
$\beta_{(ab)}$ and a unit vector ${\hat R}_{(ab)}$ with ${\vec
 S}_{(ab)}\cdot {\hat R}_{(ab)}=0$, $\{ {\hat R}^r_{(ab)},{\hat R}^s_{(ab)} \} =0$.
 This unit vector identifies the {\it spin cluster} $(ab)$ in
 the same way as the unit vectors ${\hat R}_A={\hat {\vec \rho}}_{A}$ identify
 the {\it relative particles} $A$.
\medskip

The next step is the coupling of the {\it spin cluster} $(ab)$
with unit vector ${\hat R}_{(ab)}$ [described by the canonical
variables $\alpha_{(ab)}$, $S_{(ab)}$, $\beta_{(ab)}$
$S^3_{(ab)}$] to the {\it relative particle} $c$ with unit vector
${\hat R}_c$ and  described by $\alpha_{c}$, $S_{c}$, $\beta_c$,
$S^3_{c}$: this characterizes the {\it spin cluster} $((ab)c)$.

Again, Eq.(\ref{2.10}) allows to define ${\vec N}_{((ab)c)}=
{1\over 2}({\hat R}_{(ab)}+{\hat R}_c)$, ${\vec
\chi}_{((ab)c)}={1\over 2}({\hat R}_{(ab)}-{\hat R}_c)$, ${\vec S}
= {\vec S}_{((ab)c)}= {\vec S}_{(ab)}+{\vec S}_{c}$, ${\vec
W}_{((ab)c)}= {\vec S}_{(ab)}-{\vec S}_{c}$. Since we have ${\vec
N}_{((ab)c)}\cdot {\vec \chi}_{((ab)c)}=0$ and  $\{
N^r_{((ab)c)},N^s_{((ab)c)} \} = \{ N^r_{((ab)c)},
\chi^s_{((ab)c)} \} = \{ \chi^r_{((ab)c)}, \chi^s_{((ab)c)} \} =0$
due to $\{ {\hat R}^r_{(ab)}, {\hat R}_c^s \} =0$, a new E(3)
realization generated by ${\vec S}$ and ${\vec N}_{((ab)c)}$ with
non-fixed invariants $|{\vec N}_{((ab)c)}|$, ${\vec S}\cdot {\hat
N}_{((ab)c)}= {\check S}^3$ emerges. Then, we can define
$\alpha_{((ab)c)}$ and $\beta_{((ab)c)}$ and identify a final unit
vector ${\hat R}_{((ab)c)}$ with ${\vec S}\cdot {\hat
R}_{((ab)c)}=0$ and $\{ {\hat R}_{((ab)c)}^r,{\hat R}^s_{((ab)c)}
\} =0$.

\bigskip

In conclusion, when $S\not= 0$, we find both a {\it spin frame}
${\hat S}$, ${\hat R}_{((ab)c)}$, ${\hat R}_{((ab)c)}\times {\hat
S}$ and a {\it dynamical body frame} ${\hat \chi}_{((ab)c)}$,
${\hat N}_{((ab)c)}\times {\hat \chi}_{((ab)c)}$, ${\hat
N}_{((ab)c)}$, like in the 3-body case. There is an {\it important
difference}, however: the orthonormal vectors ${\vec N}_{((ab)c)}$
and ${\vec \chi}_{((ab)c)}$ {\it depend on the momenta} of the
relative particles $a$ and $b$ through ${\hat R}_{(ab)}$, so that
our results do not share any relation with the N=4 non-trivial
SO(3) principal bundle of the orientation-shape bundle approach.

\medskip

The final 6 {\it dynamical shape variables} are $q^{\mu} = \{
|{\vec N}_{((ab)c)}|, \gamma_{(ab)}, |{\vec N}_{(ab)}|, \rho_{a},
\rho_{b}, \rho_{c} \}$. While the last four depend only on the
original relative coordinates ${\vec \rho}_{A}$, $A=a,b,c$, the
first two depend also on the original momenta ${\vec \pi}_{A}$:
therefore they are {\it generalized shape variables}. In
Ref.\cite{13} it is shown that, instead of $\rho^r_a=R^{rs}(\tilde
\alpha ,\tilde \beta ,\tilde \gamma ) {\check \rho}^s_a(q^{\mu})$,
it holds

\begin{equation}
  \rho^r_{A}={\cal R}^{rs}(\tilde \alpha ,
 \tilde \beta ,\tilde \gamma )\, {\check \rho}^s_{A}(q^{\mu},
 p_{\mu}, {\check S}^r),\qquad A=a,b,c.
 \label{2.16}
 \end{equation}

\medskip

Clearly, this result stands completely outside the
orientation-shape bundle approach. As a consequence the above
anholonomic bases and the associated {\it evolving dynamical body
frames}, however, have no relations with the N=4 {\it static}
non-trivial SO(3) principal bundle of Ref.\cite{1}, which admits
only local cross sections. Each set of 12 generalized dynamical
canonical shape variables is obviously defined modulo canonical
transformations so that it should even be possible to find local
canonical bases corresponding to the local cross sections of the
N=4 {\it static} non-trivial SO(3) principal bundle of
Ref.\cite{1}.

\bigskip

Finally, it can be shown that, starting from the Hamiltonian
$H_{rel ((ab)c)}$ expressed in the final variables, we can define
a {\it rotational Hamiltonian} $H^{(rot)}_{rel ((ab)c)}$ and a
{\it vibrational Hamiltonian} $H^{(vib)}_{rel ((ab)c)}$ (vanishing
of the physical dynamical angular velocity ${\check
\omega}^r_{((ab)c)}=0$), but $H_{rel ((ab)c)}$ fails to be the sum
of these two Hamiltonians showing once again the non-separability
of rotations and vibrations. Let us stress that in the rotational
Hamiltonian  we find an {\it inertia-like tensor} depending only
on the dynamical shape variables. A similar result, however, does
not hold for the spin-angular velocity relation.

The price to be paid for the existence of 3 global {\it dynamical
body frames} for N=4 is a more complicated form of the Hamiltonian
kinetic energy. On the other hand, {\it dynamical vibrations} and
{\it dynamical angular velocity} are  measurable quantities in
each dynamical body frame.

\bigskip

Our results can be extended to arbitrary N, with $\vec S =
\sum_{a=1}^{N-1}$ ${\vec S}_a$. There are as many independent ways
(say $K$) of spin clustering patterns as in quantum mechanics. For
instance for N=5, $K=15$ : 12 spin clusterings correspond to the
pattern $(((ab)c)d)$ and 3 to the pattern $((ab)(cd))$ [$a, b, c,
d = 1,..,4$]. For N=6, $K=105$: 60 spin clusterings correspond to
the pattern $((((ab)c)d)e)$, 15 to the pattern $(((ab)(cd)e)$ and
30 to the pattern $(((ab)c)(de))$ [$a, b, c, d, e = 1,..,5$]. Each
spin clustering  is associated to: a) a related {\it spin frame};
b) a related {\it dynamical body frame}; c) a related Darboux spin
canonical basis with orientational variables $S^3$, $\beta$, $S$,
$\alpha_{(A)}$, ${\check S}^3_{(A)}$, $\gamma_{(A)}=tg^{-1}\, {{
{\check S}^2_{(A)}}\over {{\check S}_{(A)}^1}}$, $A=1,.., K$
(their anholonomic counterparts are ${\tilde \alpha}_{(A)}$,
${\tilde \beta}_{(A)}$, ${\tilde \gamma}_{(A)}$, ${\check
S}^r_{(A)}$ with uniquely determined orientation angles) and shape
variables $q^{\mu}_{(A)}$, $p_{\mu (A)}$, $\mu =1,.., 3N-6$.
Furthermore, for $N \geq 4$ we find the following relation between
spin and  angular velocity: ${\check S}^r = {\cal
I}^{rs}(q^{\mu}_{(A)})\, {\check \omega}^s_{(A)} +
f^{\mu}(q^{\nu}_{(A)}) p_{(A)\mu}$.

\medskip

Therefore, for $N\geq 4$ our sequence of canonical and
non-canonical transformations leads to the following result, to be
compared with Eq.(\ref{2.13}) of the 3-body case

\begin{equation}
\begin{minipage}[t]{5cm}
\begin{tabular}{|l|} \hline
${\vec \rho}_{A}$ \\ \hline
 ${\vec \pi}_{A}$ \\ \hline
\end{tabular}
\end{minipage}
\ {{\buildrel {non\,\, can.} \over \longrightarrow}
\hspace{.2cm}}\
\begin{minipage}[b]{5cm}
\begin{tabular}{|lll|l|} \hline
$\tilde \alpha$ & $\tilde \beta$ & $\tilde \gamma$& $q^{\mu}({\vec
\rho}_{A},{\vec \pi}_{A})$
\\ \hline
 ${\check S}^1$ & ${\check S}^2$ & ${\check S}^3$ &
 $p_{\mu}({\vec \rho}_{A},{\vec \pi}_{A}$) \\ \hline
 \end{tabular}
 \end{minipage}.
 \label{2.17}
 \end{equation}

Therefore, for $N\geq 4$ and with $S\not= 0$, $S_{A}\not= 0$,
$A=a,b,c$, namely when the standard (3N-3)-dimensional
orientation-shape bundle is not trivial,  the original
(6N-6)-dimensional relative phase space admits as many {\it
dynamical body frames} as spin canonical bases, which are globally
defined (apart isolated coordinate singularities) for the
non-singular N-body configurations with ${\vec S}\not= 0$ (and
with non-zero spin for each spin sub-cluster). These {\it
dynamical body frames} do not correspond to local cross sections
of the static non-trivial orientation-shape SO(3) principal bundle
and the {\it spin canonical bases} do not coincide with the
canonical bases associated to the static theory.

\bigskip

The $\vec S=0$, {\it C-horizontal}, cross section of the {\it
static} SO(3) principal bundle corresponds to N-body
configurations that cannot be included in the previous Hamiltonian
construction based on the canonical realizations of SO(3): these
configurations (which include the singular ones) have to be
analyzed independently since they are related to the exceptional
orbit of SO(3), whose little group is the whole group.
\medskip

While physical observables must be obviously independent of the
gauge-dependent {\it static body frames}, they do depend on the
{\it dynamical body frame}, whose axes are operationally defined
in terms of the relative coordinates and momenta of the particles.
In particular, a {\it dynamical} definition of {\it vibration},
which replaces the $\vec S=0$, {\it C-horizontal}, cross section
of the {\it static} approach \cite{1}, is based on the requirement
that the components of the {\it angular velocity} vanish.
Actually, the angular velocities with respect to the dynamical
body frames become now {\it measurable quantities}, in agreement
with the phenomenology of extended deformable bodies (see, e.g.,
the treatment of spinning stars in astrophysics).
\medskip

In this connection, let us recall that the main efforts done in
developing canonical transformations for the N-body problem have
been done in the context of celestial mechanics. As a consequence,
these transformations are necessarily adapted to the Newtonian
gravitational potential. As well known, the Kepler problem and the
harmonic potential are the only interactions admitting extra
dynamical symmetries besides the rotational one. The particularity
of the final canonical transformations which have been worked out
in that mechanical context is that the Hamiltonian figures as one
of the final momenta. This is in particular true for the N=2 body
problem (Delaunay variables; our $\beta$ is the {\it longitude of
the ascending node} $\Omega$ \cite{10}) and for the N=3 body
problem (Jacobi's method of {\it the elimination of the nodes}).
On the other hand, it is well-known that, for generic
non-integrable interactions, putting the Hamiltonian in the
canonical bases is quite useless, since it does not bring to {\it
isolating integrals} of the motion having the capability of
reducing the dimensionality of phase space. Here we stress again
that our construction rests only on the left and right actions of
the SO(3) group and is therefore completely {\it independent of
form of the interactions}.

\medskip

The expression of the relative Hamiltonian $H_{rel}$ for $N=3$ in
these variables is a complicated function, given in Ref.\cite{13},
which fails to be the sum of a rotational plus a vibrational part.
It is an open problem for N=3 and $N \geq 4$ to check whether
suitable choices of the numerical constants $\gamma_{ai}$,  more
general body frames obtained by exploiting the freedom of making
arbitrary configuration-dependent rotations \cite{34} and/or
suitable canonical transformations of the shape variables among
themselves can simplify the free Hamiltonian and/or some type of
interaction.

\medskip

In conclusion, for each N, we get a finite number of physically
well-defined separations between {\it rotational} and {\it
vibrational} degrees of freedom. The unique {\it body frame} of
rigid bodies is replaced here by a discrete number of {\it
evolving dynamical body frames} and of {\it spin canonical bases}.
Both of them are grounded on patterns of spin couplings which are
the direct analogues of the coupling of quantum-mechanical angular
momenta.

\section{The relativistic center-of-mass problem, relative variables and the rotational
kinematics in special relativity.}

Let us see what happens when we replace Galilean space plus time
with Minkowski space-time, already for the simple model-system of
N free scalar positive-energy particles.
\medskip

First of all we have to describe a relativistic scalar particle.
Among the various possibilities (see Refs.\cite{6,35} for a review
of the various options) we will choose the manifestly Lorentz
covariant approach based on Dirac's first-class constraints

\begin{equation}
 p^2_i-\epsilon m_i^2 \approx 0.
 \label{3.1}
\end{equation}

\noindent  The associated Lagrangian description is based on the
4-vector positions $x^{\mu}_i(\tau )$ and the action $S= \int
d\tau \Big( -\epsilon \sum_im_i \sqrt{\epsilon {\dot x}_i^2(\tau
)} \Big)$, where $\tau$ is a Lorentz scalar {\it mathematical}
time, i.e. an affine parameter for the particle time-like
world-lines. Then, Lorentz covariance implies singular Lagrangians
and the associated Dirac's theory of constraints for the
Hamiltonian description. The individual time variables $x^o_i(\tau
)$ are the {\it gauge variables} associated to the mass-shell
constraints, which have the two topologically disjoint solutions
$p^o_i \approx \pm \sqrt{m^2_i+{\vec p}_i^2}$. As discussed in
Ref.\cite{36} and \cite{6} this implies that:

i) a combination of the time variables can be identified with the
clock of one arbitrary observer labeling the evolution of the
isolated system;

ii) the $N-1$ relative times are related to  observer's freedom of
looking at the N particles either at the same time or with any
prescribed relative delay, or, in other words, to the convention
for synchronization of distant clocks (definition of equal-time
Cauchy surfaces on which particle's clocks are synchronized) used
by the observer to characterize the temporal evolution of the
particles \cite{9}.

\medskip

Introducing interactions in this picture without destroying the
first-class nature of the constraints  is a well-known difficult
problem, reviewed in Refs.\cite{6,35}, where also the models with
second-class constraints are considered and compared.
\medskip

\subsection{Parametrized Minkowski theories.}

If the particle is charged and interacts with a dynamical
electromagnetic field, a {\it problem of covariance} appears. The
standard description is based on the action

\begin{equation}
 S=-\epsilon m \int d\tau \sqrt{\epsilon {\dot x}^2(\tau )} -e
\int d\tau \int d^4z \delta^4(z-x(\tau )) {\dot x}^{\mu}(\tau )
A_{\mu}(z) -{1\over 4}\int d^4z F^{\mu\nu}(z)F_{\mu\nu}(z).
 \label{3.2}
  \end{equation}

\noindent By evaluating the canonical momenta of the isolated
system, {\it charged particle plus electromagnetic field}, we find
two primary constraints:

\begin{equation}
 \chi (\tau ) = \Big( p-eA(x(\tau ))\Big)^2-\epsilon
m^2\approx 0,\qquad
 \pi^o(z^o,\vec z)\approx 0.
 \label{3.3}
 \end{equation}

It is immediately seen that, since there is no concept of {\it
equal time}, it is impossible to evaluate the Poisson bracket of
these constraints. Also, due to the same reason, the Dirac
Hamiltonian, which would be $H_D=H_c+\lambda (\tau ) \chi (\tau)
+\int d^3z \lambda^o(z^o,\vec z)\pi^o(z^o,\vec z)$ with $H_c$ the
canonical Hamiltonian and with $\lambda (\tau )$,
$\lambda^o(z^o,\vec z)$ Dirac's multipliers, does not make sense.
This problem is present even at the level of the Euler-Lagrange
equations, specifically in the formulation of a {\it Cauchy
problem} for a system of coupled equations some of which are
ordinary differential equations in the affine parameter $\tau$
along the particle world-line, while the others are partial
differential equations depending on Minkowski coordinates
$z^{\mu}$. Since the problem is due the lack of a covariant
concept of {\it equal time} between field and particle variables,
a new formulation  is needed.
\medskip

In Ref.\cite{6}, after a discussion of the many-time formalism, a
solution of the problem was found within a context suited to
incorporate the gravitational field. The starting point is an
arbitrary 3+1 splitting of Minkowski space-time with space-like
hyper-surfaces (see Ref.\cite{9} for more details on the
admissible splittings). After choosing the world-line of an
arbitrary observer, a centroid $x^{\mu}_s(\tau)$, as origin, a set
of generalized radar 4-coordinates \cite{9}, adapted to the
splitting, is provided by the $\tau$ affine parameter of the
centroid world-line together with a system of curvilinear
3-coordinates $\vec \sigma \, {\buildrel {def}\over =}\, \{
\sigma^r \}$ vanishing on the world-line. In this way we get an
arbitrary extended non-inertial frame centered on the (in general
accelerated) observer described by the centroid. The coordinates
$(\tau , \vec \sigma )$ are generalized radar coordinates
depending upon the choice of the centroid and the splitting. The
space-like hyper-surfaces are described by their embedding
$z^{\mu}(\tau ,\vec \sigma )$ in Minkowski space-time. The metric
induced by the change of coordinates $x^{\mu} \mapsto \sigma^A =
(\tau ,\vec \sigma )$ is $g_{AB}(\tau ,\vec \sigma ) =
\partial_A\, z^{\mu}(\tau ,\vec \sigma )\, \eta_{\mu\nu}\,
\partial_B\, z^{\nu}(\tau ,\vec \sigma )$ ($\partial_A\, z^{\mu}(\tau
,\vec \sigma ) = \partial\, z^{\mu}(\tau ,\vec \sigma ) / \partial
\sigma^A$). This is essentially Dirac's reformulation \cite{37} of
classical field theory (suitably extended to particles) on
arbitrary space-like hyper-surfaces ({\it equal time} or
simultaneity Cauchy surfaces). Note, incidentally, that it is also
the classical basis of the Tomonaga-Schwinger formulation of
quantum field theory.

Given any isolated system, containing any combination of
particles, strings and fields and described by an action
principle, one is lead to a reformulation of it as a {\it
parametrized Minkowski theory} \cite{6}, with the extra bonus that
the theory is already prepared for the coupling to gravity in its
ADM formulation (Ref.\cite{24}). This is done by coupling its
action to an external gravitational field $g_{\mu\nu}(x)$ and then
by replacing the external 4-metric with $g_{AB}(\tau ,\vec \sigma
)$. In this way we get an action depending on the isolated system
and on the embedding $z^{\mu}(\tau ,\vec \sigma )$ as
configurational variables. Since the action is invariant under
separate $\tau$-reparametrizations and space-diffeomorphisms
(frame-preserving diffeomorphisms \cite{9}), additional
first-class constraints are needed to ensure the independence of
the description from the choice of the 3+1 splitting, namely from
the convention chosen for the synchronization of distant clocks
identifying the instantaneous 3-space to be used as Cauchy
surface. The embedding configuration variables $z^{\mu}(\tau ,\vec
\sigma )$ are the {\it gauge} variables associated with this kind
of special-relativistic general covariance and describe all the
possible {\it inertial effects} compatible with special
relativity.

\medskip

Let us come back to the discussion of free particles within the
parametrized Minkowski theory approach. Since the intersection of
a time-like world-line with a space-like hyper-surface,
corresponding to a value $\tau$ of the time parameter, is
identified by 3 numbers $\vec \sigma =\vec \eta (\tau )$ and {\it
not by four}, each particle must have a well-defined sign of the
energy. We cannot describe, therefore, the two topologically
disjoint branches of the mass hyperboloid simultaneously as in the
standard manifestly Lorentz-covariant approach. Then, there are no
more mass-shell constraints. Each particle with a definite sign of
the energy is described by the canonical coordinates ${\vec
\eta}_i(\tau )$, ${\vec \kappa}_i(\tau )$ (like in
non-relativistic physics), while the 4-position of the particles
is given by $x^{\mu}_i(\tau )=z^{\mu}(\tau ,{\vec \eta}_i(\tau
))$. The 4-momenta $p^{\mu}_i(\tau )$ are ${\vec
\kappa}_i$-dependent solutions of $p^2_i-\epsilon m^2_i =0$ with
the given sign of the energy.

\bigskip

The system of N free scalar and positive energy particles is
described by the action \cite{6}

\begin{eqnarray}
S&=& \int d\tau d^3\sigma \, {\cal L}(\tau ,\vec \sigma )=\int
d\tau L(\tau ),\nonumber \\
 &&{\cal L}(\tau ,\vec \sigma )=-\sum_{i=1}^N\delta^3(\vec \sigma -{\vec \eta}_i
(\tau ))m_i\sqrt{ g_{\tau\tau}(\tau ,\vec \sigma )+2g_{\tau
{\check r}} (\tau ,\vec \sigma ){\dot \eta}^{\check r}_i(\tau
)+g_{{\check r}{\check s}} (\tau ,\vec \sigma ){\dot
\eta}_i^{\check r}(\tau ){\dot \eta}_i^{\check s} (\tau )
},\nonumber \\
 &&L(\tau
)=-\sum_{i=1}^Nm_i\sqrt{ g_{\tau\tau}(\tau ,{\vec \eta}_i (\tau
))+2g_{\tau {\check r}}(\tau ,{\vec \eta}_i(\tau )){\dot
\eta}^{\check r} _i(\tau )+g_{{\check r}{\check s}}(\tau ,{\vec
\eta}_i(\tau )){\dot \eta}_i ^{\check r}(\tau ){\dot
\eta}_i^{\check s}(\tau )  },
 \label{3.4}
\end{eqnarray}

\noindent where the configuration variables are $z^{\mu}(\tau
,\vec \sigma )$ and ${\vec \eta}_i(\tau )$, i=1,..,N. As said the
action is invariant under frame-dependent diffeomorphisms
\cite{9}. In phase space the Dirac Hamiltonian is a linear
combination of the following first-class constraints in strong
involution

\begin{eqnarray}
{\cal H}_{\mu}(\tau ,\vec \sigma )&=& \rho_{\mu}(\tau ,\vec \sigma
)-l_{\mu} (\tau ,\vec \sigma )\sum_{i=1}^N\delta^3(\vec \sigma
-{\vec \eta}_i(\tau )) \sqrt{ m^2_i-\gamma^{{\check r}{\check
s}}(\tau ,\vec \sigma ) \kappa_{i{\check r}}(\tau
)\kappa_{i{\check s}}(\tau ) }-\nonumber \\
 &-&z_{{\check r}\mu}
(\tau ,\vec \sigma )\gamma^{{\check r}{\check s}}(\tau ,\vec
\sigma ) \sum_{i=1}^N\delta^3(\vec \sigma -{\vec \eta}_i(\tau
))\kappa_{i{\check s}} \approx 0.
 \label{3.5}
 \end{eqnarray}

The conserved Poincar\'e generators are

\begin{equation}
 p^{\mu}_s=\int d^3\sigma \rho^{\mu}(\tau ,\vec \sigma
),\qquad J_s^{\mu\nu}=\int d^3\sigma [z^{\mu}(\tau ,\vec \sigma
)\rho^{\nu}(\tau , \vec \sigma )-z^{\nu}(\tau ,\vec \sigma
)\rho^{\mu}(\tau ,\vec \sigma )].
 \label{3.6}
 \end{equation}

\subsection{The rest-frame instant form on the
Wigner hyper-planes.}

Due to the special-relativistic general covariance implying the
gauge equivalence of the descriptions associated to different
admissible 3+1 splitting of Minkowski space-time in parametrized
theories, the foliation can be restricted to space-like
hyper-planes. In particular, for each configuration of the
isolated system with time-like 4-momentum, the leaves are best
chosen as the hyper-planes orthogonal to the conserved total
4-momentum ({\it Wigner hyper-planes}). Note that this special
foliation is {\it intrinsically} determined by the configuration
of the isolated system alone. This leads to the definition of the
{\it Wigner-covariant rest-frame instant form of dynamics}
\cite{6}, for every isolated system whose configurations have
well-defined and finite Poincar\'e generators with time-like total
4-momentum \cite{8}. An inertial rest frame for the system is
obtained by restricting the centroid world-line to a straight line
orthogonal to Wigner hyper-planes.

On a Wigner hyperplane, we obtain the following constraints [the
only remnants of the constraints (\ref{3.5})] and the following
Dirac Hamiltonian \cite{6}

\begin{eqnarray}
 &&\epsilon_s-M_{sys}\approx 0,\qquad {\vec
\kappa}_{+}=\sum_{i=1}^N {\vec \kappa}_i \approx 0,\qquad
M_{sys}=\sum_{i=1}^N\sqrt{m_i^2+{\vec \kappa}_i^2}, \nonumber \\
 &&{}\nonumber \\
 H_D&=& \lambda (\tau ) [\epsilon_s-M_{sys}]-\vec \lambda (\tau )
\sum_{i=1}^N {\vec
\kappa}_i, \nonumber \\
 &&{}\nonumber \\
 {\dot x}_s^{\mu}(\tau ) &\approx& - \lambda (\tau ) u^{\mu}(p_s)+\epsilon^{\mu}_r(u(p_s))
\lambda_r(\tau ),\qquad {\dot {\tilde x}}^{\mu}_s(\tau ) =
-\lambda (\tau ) u^{\mu}(p_s),\nonumber \\
 x^{\mu}_s(\tau ) &=& x^{\mu}_o + u^{\mu}(p_s)\, T_s +
\epsilon^{\mu}_r(u(p_s))\, \int_o^{\tau} d\tau_1\,
\lambda_r(\tau_1).
 \label{3.7}
\end{eqnarray}

The embedding describing Wigner hyper-planes is $z^{\mu}(\tau
,\vec \sigma ) = x^{\mu}_s(\tau ) + \epsilon^{\mu}_r(u(p_s))
\sigma^r$, where $\epsilon^{\mu}_o(u(p)) = u^{\mu}(p) =
p^{\mu}/\sqrt{\epsilon\, p^2}$ and $\epsilon^{\mu}_r(u(p)) = \Big(
- u_r(p); \delta^i_r - {{u^i(p)\, u_r(p)}\over {1 + u^o(p)}}\Big)$
are the columns of the standard Wigner boost, $\epsilon^{\mu
}_{\nu}(u(p_s)) =L^{\mu}{}_{\nu}(p_s,{\buildrel \circ \over
p}_s)$, connecting the time-like 4-vector $p^{\mu}$ to its
rest-frame form, ${\buildrel \circ \over p} =\, \sqrt{\epsilon\,
p^2}\, (1;\vec 0)$. As a consequence the space indices $"r"$ now
transform as Wigner spin-1 3-vectors.
\medskip

The 3 Dirac's multipliers $\vec \lambda (\tau )$ describe the {\it
classical zitterbewegung} of the centroid $x^{\mu}_s(\tau )$: each
gauge-fixing $\vec \chi (\tau )\approx 0$ to the three first-class
constraints ${\vec \kappa}_{+}\approx 0$ gives a different
determination of the multipliers $\vec \lambda (\tau )$ and
therefore  identifies a different world-line $x^{(\vec \chi
)\mu}_s(\tau )$ for the covariant non-canonical centroid.

\medskip

The various spin tensors and vectors evaluated with respect to the
centroid are \cite{6}

\begin{eqnarray*}
J^{\mu\nu}_s&=&x^{\mu}_s p^{\nu}_s- x^{\nu}_s p^{\mu}_s+
S^{\mu\nu}_s,\nonumber \\
 &&{}\nonumber \\
S^{\mu\nu}_s&=&[u^{\mu}(p_s)\epsilon^{\nu}(u(p_s))-u^{\nu}(p_s)\epsilon^{\mu}
(u(p_s))] {\bar S}^{\tau
r}_s+\epsilon^{\mu}(u(p_s))\epsilon^{\nu}(u(p_s)) {\bar S}^{rs}_s,
 \end{eqnarray*}

  \bea
  {\bar S}^{AB}_s&=&\epsilon^A_{\mu}(u(p_s)) \epsilon^B_{\nu}(u(p_s))
S^{\mu\nu}_s = \Big( {\bar S}^{rs}_s\equiv
\sum_{i=1}^N(\eta^r_i\kappa^s_i- \eta^s_i\kappa^r_i);\quad {\bar
S}^{\tau r}_s\equiv - \sum_{i=1}^N\eta^r_i\sqrt{m^2_ic^2+{\vec
\kappa}_i^2}\Big),\nonumber \\
 &&{}\nonumber \\
 {\vec  {\bar S}} &\equiv & {\vec {\bar S}}=\sum_{i+1}^N{\vec
\eta}_i\times {\vec \kappa}_i\approx \sum_{i=1}^N {\vec
\eta}_i\times {\vec \kappa}_i-{\vec \eta}_{+}\times {\vec
\kappa}_{+} = \sum_{a=1}^{N-1} {\vec \rho}_a\times {\vec \pi}_a.
\label{3.8}
\end{eqnarray}

Note that  $L^{\mu\nu}_s = x^{\mu}_s\, p^{\nu}_s - x^{\nu}_s\,
p^{\mu}_s$ and $S^{\mu\nu}_s$ are not constants of the motion due
to the {\it classical zitterbewegung}.
\medskip

Let us now summarize some relevant points of the rest-frame
instant form on Wigner hyper-planes, since this formulation
clarifies the role of the various relativistic centers of mass.
This is a long standing problem which arose just after the
foundation of special relativity in the first decade of the last
century. It became soon clear that the problem of the relativistic
center of mass is highly non-trivial: no definition can enjoy all
the properties of the ordinary non-relativistic center of mass.
See Refs.\cite{38,39,40,41,42,43} for a partial bibliography of
all the existing attempts and Ref.\cite{44} for reviews.

\bigskip

Let us stress that, as said in the Introduction, {\it our approach
leads to a doubling of the usual concepts}: there is an {\it
external} viewpoint (see Subsection C) and an {\it internal}
viewpoint (see Subsection D) with respect to the Wigner
hyper-planes. While in Newton physics for any given absolute time
there exists an absolute instantaneous 3-space and a unique center
of mass with a unique associated 3-momentum $\vec P$ (vanishing in
the rest frame), in special relativity the notion of instantaneous
3-space requires a convention for the synchronization of distant
clocks. In other words we are forced: i) to introduce a 3+1
splitting of Minkowski space-time and an arbitrary accelerated
observer; ii) for the sake of simplicity to restrict ourselves to
inertial frames, namely to an inertial observer with a 4-momentum
$p^{\mu}_s$ and to foliations with space-like hyper-planes
orthogonal to this 4-momentum (Wigner hyper-planes); iii) to
describe the isolated system inside each of these instantaneous
rest-frame 3-spaces, where its total 3-momentum vanishes. As a
consequence, the unique non-relativistic center-of-mass 3-momentum
$\vec P$, vanishing in the rest frame, is replaced by two {\it
independent} notions: i) an {\it external} 4-momentum $p^{\mu}_s =
(p^o_s, {\vec p}_s=\epsilon_s {\vec k}_s)$ describing the
orientation of the instantaneous 3-space with respect to an
arbitrary reference inertial observer; ii) an {\it internal}
3-momentum inside the instantaneous 3-space, which vanishes,
${\vec \kappa}_{+}\approx 0$, as a definition of rest frame. The
rest-frame instant form of dynamics is the natural framework to
visualize this doubling.

\bigskip

In the rest-frame instant form only four first-class constraints
survive so that the original configurational variables
$z^{\mu}(\tau ,\vec \sigma )$, ${\vec \eta}_i(\tau )$ and their
conjugate momenta $\rho_{\mu}(\tau ,\vec \sigma )$, ${\vec
\kappa}_i(\tau )$ are reduced to:

i) a decoupled point ${\tilde x}^{\mu}_s(\tau )$, $p^{\mu}_s$ (the
only remnant of the space-like hyper-surface) with a positive mass
$\epsilon_s=\sqrt{\epsilon p^2_s}$ determined by the first-class
constraint $\epsilon_s-M_{sys} \approx 0$ ($M_{sys}$ is the
invariant mass of the isolated system). Its rest-frame Lorentz
scalar time $T_s={{{\tilde x}_s\cdot p_s}\over {\epsilon_s}}$ is
put equal to the mathematical time as a gauge fixing $T_s-\tau
\approx 0$ to the previous constraint. The unit time-like 4-vector
$u^{\mu}(p_s)=p_s^{\mu}/\epsilon_s$ is orthogonal to the Wigner
hyper-planes and describes their orientation in the chosen
inertial frame.

Here, ${\tilde x}^{\mu}_s(\tau )$ is a {\it non-covariant
canonical} variable describing the decoupled canonical {\it
external 4-center of mass}. It plays the role of a kinematical
external 4-center of mass and of a decoupled observer with his
parametrized clock ({\it point particle clock}). Its velocity
${\dot {\tilde x}}^{\mu}_s(\tau )$ is parallel to $p^{\mu}_s$, so
that it has no {\it classical zitterbewegung}. The connection
between the centroid $x^{\mu}_s(\tau ) = z^{\mu}(\tau ,\vec 0)$
and ${\tilde x}_s^{\mu}(\tau )$ and the associated decomposition
of the angular momentum are

\begin{eqnarray}
{\tilde x}^{\mu}_s(\tau )\, &{\buildrel {def}\over =}&
z^{\mu}(\tau ,{\vec {\tilde \sigma }}) = x^{\mu}_s(\tau )-{1\over
{\epsilon_s(p^o_s+\epsilon_s)}}\Big[
p_{s\nu}S_s^{\nu\mu}+\epsilon_s(S^{o\mu}_s-S^{o\nu}_s{{p_{s\nu}p_s^{\mu}}\over
{\epsilon^2_s}}) \Big],\nonumber \\
 &&{}\nonumber \\
 J^{\mu\nu}_s &=&  {\tilde x}^{\mu}_s p^{\nu}_s - {\tilde x}^{\nu}_s
p^{\mu}_s +{\tilde S} ^{\mu\nu}_s,\nonumber \\
  &&{}\nonumber \\
  {\tilde S}^{\mu\nu}_s&=&S^{\mu\nu}_s+{1\over {\sqrt{\epsilon p^2_s}(p^o_s+
\sqrt{\epsilon p^2_s})}}\Big[ p_{s\beta}(S^{\beta\mu}_s
p^{\nu}_s-S^{\beta\nu}_s p^{\mu}_s)+\sqrt{p^2_s}(S^{o\mu}_s
p^{\nu}_s-S^{o\nu}_s p^{\mu}_s)\Big], \nonumber \\
 &&{\tilde S}^{ij}_s=\delta^{ir}\delta^{js} {\bar S}_s^{rs},\quad\quad
{\tilde S}^{oi}_s=-{{\delta^{ir} {\bar S}^{rs}_s\, p^s_s}\over
{p^o_s+ \sqrt{\epsilon p^2_s}}},
 \label{3.9}
\end{eqnarray}

Now both ${\tilde L}_s^{\mu\nu} = {\tilde x}^{\mu}_s\, p_s^{\nu} -
{\tilde x}^{\nu}_s\, p^{\mu}_s$ and ${\tilde S}^{\mu\nu}_s$ are
conserved.
\medskip

ii) the $6N$ particle canonical variables ${\vec \eta}_i(\tau )$,
${\vec \kappa}_i(\tau )$ {\it inside} the Wigner hyper-planes.
They are restricted by the three first-class constraints (the {\it
rest-frame conditions}) ${\vec \kappa}_{+}=\sum_{i=1}^N {\vec
\kappa}_i \approx 0$. They are Wigner spin-1 3-vectors, like the
coordinates $\vec \sigma$.

\bigskip

The canonical variables ${\tilde x}^{\mu}_s$, $p^{\mu}_s$ for the
{\it external} 4-center of mass, can be replaced by the canonical
pairs \cite{45}

\begin{equation}
T_s = {{p_s\cdot {\tilde x}_s}\over {\epsilon_s}}={{p_s
 \cdot x_s}\over {\epsilon_s}},\qquad
 \epsilon_s = \pm \sqrt{\epsilon p^2_s},\qquad
 {\vec z}_s = \epsilon_s ({\vec {\tilde
x}}_s- {{ {\vec p}_s}\over {p^o_s}} {\tilde x}^o_s), \qquad
 {\vec k}_s = {{ {\vec p}_s}\over {\epsilon_s}}.
 \label{3.10}
\end{equation}

In the rest-frame instant form, this non-point canonical
transformation can be summarized as

\begin{equation}
\begin{minipage}[t]{5cm}
\begin{tabular}{|l|l|} \hline
${\tilde x}_s^{\mu}$& ${\vec \eta}_i$ \\  \hline
 $p^{\mu}_s$& ${\vec \kappa}_i$ \\ \hline
\end{tabular}
\end{minipage} \ {\longrightarrow \hspace{1cm}} \
\begin{minipage}[t]{5 cm}
\begin{tabular}{|l|l|l|} \hline
$\epsilon_s$& ${\vec z}_s$   & ${\vec \eta}_i$   \\ \hline
 $T_s$ & ${\vec k}_s$& ${\vec \kappa}_i$ \\ \hline
\end{tabular}
\end{minipage}
\label{3.11}
\end{equation}

\medskip

The addition of the gauge-fixing $T_s-\tau \approx 0$ for the
first-class constraint $\epsilon_s - M_{sys} \approx 0$, building
a pair of second-class constraints and implying $\lambda (\tau ) =
-1$ (due to the explicit $\tau$-dependence of the gauge fixing),
leads to the elimination of $T_s$ and $\epsilon_s$. We are then
left with a decoupled free point ({\it point particle clock}) of
mass $M_{sys}$ and canonical 3-coordinates ${\vec z}_s$, ${\vec
k}_s$. The position ${\vec q}_s = {\vec z}_s/\epsilon_s$ is {\it
the classical analogue of the Newton-Wigner 3-position operator}
\cite{39} and shares with it the reduced covariance under the
Euclidean subgroup of the Poincar\'e group.

\medskip

The invariant mass $M_{sys}$ of the system, which is also its {\it
internal} energy, replaces the non-relativistic Hamiltonian
$H_{rel}$ for the relative degrees of freedom. For $c\,
\rightarrow \infty$ we have $M_{sys} - m\, c^2 \rightarrow
H_{rel}$. $M_{sys}$ generates the evolution in the rest-frame
Lorentz scalar time $T_s$ in the rest frame: in this way we get
the same equations of motion as before the addition of the gauge
fixing. This reminds of the frozen Hamilton-Jacobi theory, in
which the time evolution can be reintroduced by using the energy
generator of the Poincar\'e group as Hamiltonian. As a
consequence, after the gauge fixing $T_s-\tau \approx 0$, the
final Hamiltonian and the embedding of the Wigner hyperplane into
Minkowski space-time become

\bea
 H_D&=& M_{sys} -\vec \lambda (\tau ) \cdot {\vec \kappa}_{+} ,\nonumber \\
 &&{}\nonumber \\
z^{\mu}(\tau ,\vec \sigma ) &=& x^{\mu}_s(\tau ) + \epsilon^{\mu
}_r(u(p_s)) \sigma^r = x^{\mu}_s(0) + u^{\mu}(p_s) \tau +
\epsilon^{\mu}_r(u(p_s)) \sigma^r,\nonumber \\
 &&{}\nonumber \\
 {\it with} \quad
&& {\dot x}^{\mu}_s(\tau )\, {\buildrel \circ \over =} {{d\,
x^{\mu}_s(\tau )}\over
 {d \tau}}+ \{ x^{\mu}_s(\tau ), H_D \} = u^{\mu}(p_s)+\epsilon^{\mu}_r(u(p_s))
\lambda_r(\tau ),
 \label{3.12}
  \eea

\noindent where $x^{\mu}_s(0)$ is an arbitrary point. This
equation visualizes the role of the Dirac multipliers as sources
of the {\it classical zittebewegung} and consequently the {\it
gauge nature} of this latter. Let us remark that the constant
$x^{\mu}_s(0)$ [and ${\tilde x} ^{\mu}_s(0)$] is arbitrary,
reflecting the arbitrariness in the absolute location of the
origin of the {\it internal} coordinates on each hyperplane in
Minkowski space-time.
\medskip

Inside the Wigner hyperplane three degrees of freedom of the
isolated system, describing an {\it internal} center-of-mass
3-variable ${\vec \sigma}_{com}$ conjugate to ${\vec \kappa}_{+}$
(when the ${\vec \sigma}_{com}$ are canonical variables they are
denoted ${\vec q}_{+}$) are {\it gauge} variables. The natural
gauge fixing in order to eliminate the three first-class
constraints ${\vec \kappa}_{+} \approx 0$ is $\vec \chi ={\vec
q}_{+}\approx 0$ which implies $\lambda_r(\tau )=0$: in this way
the {\it internal} 3-center of mass gets located in the centroid
$x^{\mu}_s(\tau )=z^{\mu}(\tau ,\vec \sigma =0)$ of the Wigner
hyperplane.

\bigskip

\subsection{The external Poincare' group and the external
center-of-mass variables on a Wigner hyper-plane.}

The {\it external} viewpoint is proper to an arbitrary inertial
Lorentz observer, describing the Wigner hyper-planes as leaves of
a foliation of Minkowski space-time and determined by the
time-like configurations of the isolated system. Therefore there
is an {\it external} realization of the Poincar\'e group, whose
Lorentz transformations rotate the Wigner hyper-planes and induce
a Wigner rotation of the 3-vectors inside each Wigner hyperplane.
As a consequence, each such hyperplane inherits an induced {\it
internal Euclidean structure} and an {\it internal} unfaithful
Euclidean action , with an associated unfaithful {\it internal}
realization of the Poincare' group (the internal translations are
generated by the first-class constraints ${\vec \kappa}_{+}
\approx 0$, so that they are eliminable gauge variables), which
will be described in the next Subsection.

\medskip

The {\it external} realization of the Poincar\'e generators with
non-fixed invariants $\epsilon p^2_s = \epsilon_s^2 \approx
M^2_{sys}$ and $W^2 = -\epsilon p^2_s {\vec {\bar S}}_s^2 \approx
-\epsilon M^2_{sys} {\vec {\bar S}}^2$, is obtained from
Eq.(\ref{3.8}) (the four independent Hamiltonians \cite{7} of this
instant form, $p^o_s$ and $J^{oi}_s$, are all functions {\it only}
of the invariant mass $M_{sys}$, which contains the possible
mutual interactions among the particles):

\begin{eqnarray*}
p^{\mu}_s&,&\qquad
 J^{\mu\nu}_s = {\tilde x}^{\mu}_sp^{\nu}_s-{\tilde x}^{\nu}_s p^{\mu}_s
 + {\tilde S}^{\mu\nu}_s,\nonumber \\
 &&{}\nonumber \\
 p^o_s&=& \sqrt{\epsilon_s^2+{\vec p}_s^2}= \epsilon_s \sqrt{1+ {\vec
k}_s^2}\approx  \sqrt{M^2_{sys}+{\vec p}^2_s}=M_{sys}
\sqrt{1+{\vec k}_s^2},\qquad
 {\vec p}_s = \epsilon_s{\vec k}_s\approx M_{sys} {\vec k}_s,
\end{eqnarray*}

\begin{eqnarray}
 J^{ij}_s&=&{\tilde x}^i_sp^j_s-{\tilde x}^j_sp^i_s +
\delta^{ir}\delta^{js}\sum_{i=1}^N(\eta^r_i\kappa^s_i-\eta^s_i\kappa^r_i)=
z^i_sk^j_s-z^j_sk^i_s+\delta^{ir}\delta^{js} \epsilon^{rsu}{\bar
S}^u_s,\nonumber \\
 K^i_s&=&J^{oi}_s= {\tilde x}^o_sp^i_s-{\tilde x}^i_s
\sqrt{\epsilon_s^2+{\vec p}_s^2}-{1\over
{\epsilon_s+\sqrt{\epsilon^2_s+ {\vec p}_s^2}}} \delta^{ir} p^s_s
\sum_{i=1}^N(\eta^r_i\kappa^s_i- \eta^s_i\kappa^r_i)=\nonumber \\
 &=&-\sqrt{1+{\vec k}_s^2} z^i_s-{{\delta^{ir} k^s_s\epsilon
^{rsu}{\bar S}^u_s}\over {1+\sqrt{1+{\vec k}_s^2} }}\approx
{\tilde x}^o_sp^i_s -{\tilde x}^i_s\sqrt{M^2_{sys}+{\vec
p}^2_s}-{{\delta^{ir}p^s_s\epsilon^{rsu}{\bar S}^u_s}\over
{M_{sys}+\sqrt{M_{sys}^2+{\vec p}^2_s}}}.\nonumber \\
 &&{}
 \label{3.13}
 \end{eqnarray}

\bigskip

Given a canonical realization of the ten Poincar\'e generators,
one can build \cite{18} three {\it external} 3-variables, the
canonical 3-center of mass ${\vec q}_s$, the Moller 3-center of
energy ${\vec R}_s$ and the Fokker-Pryce 3-center of inertia
${\vec Y} _s$ by using only these generators. For the rest-frame
realization of the Poincar\'e algebra given in Eqs.(\ref{3.13}) we
get

\begin{eqnarray*}
{\vec R}_s&=& -{{1}\over {p^o_s}}{\vec K}_s=({\vec {\tilde
x}}_s-{{{\vec p}_s}\over {p^o_s}} {\tilde x}^o_s)-{{{\vec {\bar
 S}}_s\times {\vec p}_s} \over {p^o_s(p^o_s+\epsilon_s)}},\nonumber \\
 {\vec q}_s&=&{\vec {\tilde x}}_s-{{{\vec p}_s}\over
{p^o_s}}{\tilde x}^o_s= {{{\vec z}_s}\over {\epsilon_s}}= {\vec
R}_s+{{ {\vec {\bar S}}_s\times {\vec p}_s}\over {p^o_s(p^o_s+
\epsilon_s)}}={{p^o_s {\vec R}_s+\epsilon_s {\vec Y}_s}\over
{p^o_s+\epsilon_s}} ,\end{eqnarray*}

\bea
 {\vec Y}_s&=&{\vec q}_s+{{ {\vec
{\bar S}}_s\times {\vec p}_s}\over {\epsilon_s
(p^o_s+\epsilon_s)}}={\vec R}_s+{{ {\vec {\bar S}}_s\times {\vec
p}_s}\over {p^o_s\epsilon_s}},\nonumber \\
 &&{}\nonumber \\
 &&\{ R^r_s,R^s_s \} =-{{1}\over
{(p^o_s)^2}}\epsilon^{rsu}\Omega^u_s, \quad\quad {\vec
\Omega}_s={\vec J}_s-{\vec R}_s\times {\vec p}_s,\nonumber \\
 &&{}\nonumber \\
 &&\{ q^r_s, q^s_s \} =0,\qquad
 \{ Y^r_s,Y^s_s \} ={1\over {\epsilon_sp^o_s}}\epsilon^{rsu}\Big[
{\bar S}^u_s +{{ {\vec {\bar S}}_s\cdot {\vec p}_s\, p^u_s}\over
{\epsilon_s(p^o_s+ \epsilon_s)}}\Big] ,\nonumber \\
 &&{}\nonumber \\
 {\vec p}_s\cdot {\vec q}_s&=&{\vec p}_s\cdot {\vec
 R}_s={\vec p}_s\cdot {\vec Y}_s={\vec k}_s\cdot {\vec z}_s,\qquad
{\vec p}_s=0 \Rightarrow {\vec q}_s={\vec Y}_s={\vec R}_s.
\label{3.14}
\end{eqnarray}

\noindent All of these have the same velocity and coincide in the
Lorentz rest frame where ${\buildrel \circ \over
p}^{\mu}_s=\epsilon_s (1;\vec 0)$

\bigskip

Then, three {\it external} concepts of 4-center of mass can be
defined (each having an {\it internal} 3-location inside the
Wigner hyper-planes) starting from the kinematics of the Wigner
hyper-planes and from the above concepts of 3-centers of mass
\cite{38} :\hfill\break \hfill\break
 a) The {\it external} non-covariant canonical {\it 4-center of mass}
${\tilde x}^{\mu}_s$ (with 3-location ${\vec {\tilde \sigma}}$),
extension of the {\it canonical 3-position vector} ${\vec q}_s$
(also named {\it center of spin} \cite{42}). ${\vec q}_s$ is the
classical analogue of the Newton-Wigner position operator
\cite{39}. ${\tilde x}^{\mu}_s$ is a frame-dependent pseudovector
(${\vec q}_s$ does not satisfy the {\it world line condition}
\cite{38}), but it is canonical: $\{ {\tilde x}^{\mu}_s, {\tilde
x}^{\nu}_s \} =0$. \hfill\break \hfill\break
 b) The {\it external} non-covariant and non-canonical M\o ller {\it
4-center of energy} $R^{\mu}_s$ (with 3-location ${\vec
\sigma}_R$), extension of the M\o ller 3-center of mass ${\vec
R}_s$ \cite{40}, which corresponds to the standard
non-relativistic definition of center of mass of a system of
particles with masses replaced by energies. ${\vec R}_s$ does not
satisfy the world line condition, so that $R^{\mu}_s$ is  a
frame-dependent pseudovector and moreover $\{ R^{\mu}_s, R^{\nu}_s
\} \not= 0$. \hfill\break \hfill\break
 c) The {\it external} non-canonical but covariant Fokker-Pryce {\it
 4-center of inertia} $Y^{\mu}_s$ (with 3-location ${\vec
\sigma}_Y$), extension of the Fokker-Pryce 3-center of inertia
\cite{41,42}. $Y^{\mu}_s$ is a 4-vector by construction: it is the
Lorentz transform  of the rest-frame pseudo-world-line
$R^{(rest)\, \mu}_s = {\tilde x}^{(rest)\mu}_s$ of the M\o ller
center of energy to an arbitrary frame. It holds $\{ Y^{\mu}_s,
Y^{\nu}_s \} \not= 0$.
\medskip

Note that while the Fokker-Pryce $Y^{\mu}_s$ is the only 4-vector
by construction, only ${\tilde x}^{\mu}_s(\tau )$ can be an {\it
adapted coordinate in a Hamiltonian treatment with Dirac
constraints}.

\medskip

To find the 3-locations on the Wigner hyper-planes, with respect
to the centroid world-line $x^{\mu}_s(\tau )=z^{\mu}(\tau ,\vec
0)$, of these quantities, we note \cite{11} that, since we have
$T_s=u(p_s)\cdot x_s=u(p_s)\cdot {\tilde x}_s\equiv \tau$ on the
Wigner hyperplane labeled by $\tau$, we can require $Y^{\mu}_s$
and $R^{\mu}_s$ to have time components such that $u(p_s)\cdot
Y_s=u(p_s)\cdot R_s=T_s\equiv \tau$. As a consequence, the
following 3-locations are determined in Ref.\cite{11}: a) for the
world-line $Y^{\mu}_s(\tau )=z^{\mu}(\tau ,{\vec \sigma}_Y) =
\Big( {\tilde x}^o(\tau ); {\vec Y}_s(\tau )\Big)$ we get $
\sigma^r_Y = R^r_+$, with ${\vec R}_+$ defined in Eq.(\ref{3.16})
of next Subsection; b) for the pseudo-world-line ${\tilde
x}^{\mu}_s(\tau )=z^{\mu}(\tau ,{\tilde {\vec \sigma}}) =
\Big({\tilde x}^o_s(\tau); {\vec {\tilde x}}_s(\tau )\Big)$ we get
${\tilde \sigma}^r = \sigma^r_Y + {{{\bar S}^{rv}_s\,
u^v(p_s)}\over {1 + u^o(p_s)}}$; c) for the pseudo-world-line
$R^{\mu}_s=z^{\mu}(\tau ,{\vec \sigma}_R) = \Big({\tilde x}^o(\tau
); {\vec R}_s(\tau )\Big)$ we get $\sigma^r_R = \sigma^r_Y + {{[1
- u^o(p_s)]\, {\bar S}^{rv}_s\, u^v(p_s)}\over {u^o(p_s)\, [1 +
u^o(p_s)]}}$.

\medskip

It is seen that {\it the external Fokker-Pryce non-canonical
4-center of inertia coincides with the centroid $x^{({\vec
q}_{+})\mu}_s(\tau )$ carrying the internal 3-center of mass}.

\medskip

In each Lorentz frame one has different pseudo-world-lines
describing $R^{\mu}_s$ and ${\tilde x}^{\mu}_s$: the canonical
4-center of mass ${\tilde x}^{\mu}_s$ {\it lies in between}
$Y^{\mu}_s$ and $R^{\mu}_s$ in every (non rest)-frame. In an
arbitrary Lorentz frame, the pseudo-world-lines associated with
${\tilde x}^{\mu}_s$ and $R^{\mu}_s$ fill a world-tube (see the
book in Ref.\cite{40}) around the world-line $Y^{\mu}_s$ of the
covariant non-canonical Fokker-Pryce 4-center of inertia
$Y_s^{\mu}$. The {\it invariant radius} of the tube is $\rho
=\sqrt{-\epsilon W^2}/p^2=|\vec S|/\sqrt{\epsilon p^2}$ where
($W^2=-\epsilon p^2{\vec S}^2$ is the Pauli-Lubanski invariant
when $\epsilon p^2 > 0$). This classical intrinsic radius
delimitates the non-covariance effects (the pseudo-world-lines) of
the canonical 4-center of mass ${\tilde x}_s^{\mu}$. See
Ref.\cite{7} for a discussion of the properties of the {\it
M$\o$ller radius}. At the quantum level $\rho$ becomes the Compton
wavelength of the isolated system times its spin eigenvalue
$\sqrt{s(s+1)}$ , $\rho \mapsto \hat \rho = \sqrt{s(s+1)} \hbar
/M=\sqrt{s(s+1)} \lambda_M$ with $M=\sqrt{\epsilon p^2}$ the
invariant mass and $\lambda_M=\hbar /M$ its Compton wavelength.
The critique of classical relativistic physics argued from quantum
pair production concerns testing of distances where, due to the
Lorentz signature of space-time, intrinsic classical covariance
problems emerge: the canonical 4-center of mass ${\tilde
x}_s^{\mu}$ adapted to the first-class constraints of the system
cannot be localized in a frame-independent way. Remember \cite{6},
finally, that $\rho$ is also a remnant of the energy conditions of
general relativity in flat Minkowski space-time: since the
M$\o$ller non-canonical, non-covariant 4-center of energy $R^{\mu
}$ has non-covariance properties localized inside the world-tube
with radius $\rho$ (see the book in \cite{40}), it turns out that
for an extended relativistic system with the material radius
smaller of its intrinsic radius $\rho$ one has: i) its peripheral
rotation velocity can exceed the velocity of light; ii) its
classical energy density cannot be positive definite everywhere in
every frame.

\subsection{The internal Poincare' group and the internal
center-of-mass variables on a Wigner hyper-plane.}

Let us consider now the notions defined according to the {\it
internal} viewpoint. They correspond to an unfaithful {\it
internal} realization of the Poincar\'e algebra: the {\it
internal} 3-momentum ${\vec \kappa}_{+}$ vanishes due to the
rest-frame conditions; the {\it internal} energy and angular
momentum are given by the invariant mass $M_{sys}$ and by the
external spin (angular momentum with respect to ${\tilde
x}^{\mu}_s(\tau )$) of the isolated system, respectively.

This {\it internal} realization of the Poincar\'e algebra is built
inside the Wigner hyperplane by using the expression of ${\bar
S}_s^{AB}$ given by Eq.(\ref{3.8}) (the invariants are
$M^2_{sys}-{\vec \kappa}_{+}^2 \approx M^2_{sys} >
 0$ and $W^2=-\epsilon (M^2_{sys}-{\vec \kappa}^2_{+}) {\vec {\bar
S}}^2_s \approx -\epsilon M^2_{sys} {\vec {\bar S}}^2_s$; in the
interacting case $M_{sys}$ and $\vec K$ are modified by the mutual
interactions among the particles)

\begin{eqnarray}
&&M_{sys}=H_M=\sum_{i=1}^N  \sqrt{m^2_i+{\vec \kappa}_i^2},\qquad
 {\vec \kappa}_{+}=\sum_{i=1}^N {\vec
\kappa}_i\, (\approx 0),\nonumber \\
 &&\vec J=\sum_{i=1}^N {\vec \eta}_i\times {\vec
\kappa}_i,\quad\quad J^r={\bar S}^r={1\over 2}\epsilon^{ruv}{\bar
S}^{uv} \equiv {\bar S}^r_s,\nonumber \\
 &&\vec K=- \sum_{i=1}^N
\sqrt{m^2_i+{\vec \kappa}_i^2}\,\, {\vec \eta}_i,\quad\quad
K^r=J^{or}={\bar S}_s^{\tau r}.
 \label{3.15}
\end{eqnarray}
\medskip

As we shall see, $\vec K\approx 0$ are the natural gauge fixings
for the first-class constraints ${\vec \kappa}_{+}\approx 0$: this
makes the internal realization even more unfaithful.
\medskip

In analogy with the external viewpoint, the determination of the
three {\it internal} 3-center of mass can be achieved using again
the group theoretical methods of Ref.\cite{18}: \hfill\break

i) a canonical {\it internal} 3-center of mass  (or {\it 3-center
of spin}) ${\vec q}_{+}$; \hfill\break
 ii) a non-canonical {\it internal} M\o ller {\it
3-center of energy} ${\vec R}_{+}$ ;\hfill\break
 iii) a non-canonical {\it internal} Fokker-Pryce {\it 3-center of inertia}
${\vec y}_{+}$. \hfill\break

Starting from the {\it internal} realization (\ref{3.15}) of the
Poincar\'e algebra, we get the following Wigner spin 1  3-vectors:
\medskip
i) The {\it internal} Moller 3-center of energy ${\vec R}_+$ and
the associated spin vector ${\vec S}_R$

\begin{eqnarray}
{\vec R}_{+}&=& - {1\over {M_{sys}}} \vec K =
{{\sum_{i=1}^N\sqrt{m^2_i+ {\vec \kappa}^2_i}\,\, {\vec
\eta}_i}\over {\sum_{k=1}^N\sqrt{m_k^2+{\vec
\kappa}_k^2}}},\nonumber \\ {\vec S}_R &=& \vec J -{\vec
R}_{+}\times {\vec \kappa}_{+},\nonumber \\
 &&\{ R^r_{+},\kappa^s_{+} \} =\delta^{rs},\quad\quad \{ R^r_{+},M_{sys}
\} = {{\kappa^r_{+}}\over {M_{sys}}},\qquad
 \{ R^r_{+},R^s_{+} \} =-{1\over {M_{sys}^2}} \epsilon^{rsu}S_R^u,
\nonumber \\
 &&\{ S_R^r,S_R^s \} =\epsilon^{rsu}(S_R^u-{1\over {M_{sys}^2}}
{\vec S_R} \cdot {\vec \kappa}_{+}\,\, \kappa_{+}^u),\quad\quad \{
S_R^r ,M_{sys} \} =0.
 \label{3.16}
 \end{eqnarray}

Note that the gauge fixing ${\vec R}_{+} \approx  0$ gives

\begin{eqnarray}
{\vec R}_{+}\approx 0 & \Rightarrow &
 {\dot {\vec R}}_{+}\, {\buildrel \circ \over =} \,
 \{ {\vec R}_{+},H_D \} = \nonumber \\
&=& { {{\vec \kappa}_{+}} \over
      {\sum_{k=1}^N\sqrt{m^2_k+{\vec \kappa}_k^2}} }
   -\vec \lambda (\tau )
      { {\sum_{i=1}^N\sqrt{m^2_i+{\vec \kappa}_i^2}} \over
        {\sum_{k=1}^N\sqrt{m^2_k+{\vec \kappa}_k^2} }}
  \approx -\vec \lambda (\tau ) \approx 0.
\label{3.17}
\end{eqnarray}

Furthermore, the {\it internal} boost generator of Eq.(\ref{3.15})
may be rewritten as $\vec K=-M_{sys} {\vec R}_{+}$, so that ${\vec
R}_{+}\approx 0$ implies $\vec K \approx 0$.

\medskip

ii) The canonical {\it internal} 3-center of mass  ${\vec q}_+$
and the associated spin vector ${\vec S}_q$ [$\{ q^r_+, q^s_+\} =
0$, $\{ q^r_+, \kappa^s_+\} = \delta^{rs}$, $\{J^r, q_+^s\} =
\epsilon^{rsu}\, q^u_+$, $\{ S_q^r,S_q^s \} =\epsilon^{rsu}S_q^u$]

\begin{eqnarray*}
{\vec q}_{+}&=& {\vec R}_{+}- {{ \vec J\times \vec \Omega}\over
{\sqrt{M^2_{sys}-{\vec
\kappa}^2_{+}}(M_{sys}+\sqrt{M^2_{sys}-{\vec
\kappa}^2_{+}})}}=\nonumber \\
 &=&-{{\vec K}\over
{\sqrt{M^2_{sys}-{\vec \kappa}^2_{+}}}}+{{ \vec J\times {\vec
\kappa}_{+}}\over {\sqrt{M_{sys}^2-{\vec
\kappa}^2_{+}}(M_{sys}+\sqrt{M^2_{sys}-{\vec \kappa}^2_{+}})}}+
 \end{eqnarray*}

\begin{eqnarray*}
 &+&{{\vec K\cdot {\vec \kappa}_{+}\,\, {\vec
\kappa}_{+}}\over {M_{sys}\sqrt{M^2_{sys}-{\vec \kappa}^2_{+}}
\Big( M_{sys}+\sqrt{M^2_{sys} -{\vec \kappa}^2_{+}}\Big)
}},\nonumber \\
 &&\approx {\vec
R}_{+}\quad for\quad {\vec \kappa}_{+}\approx 0;\quad\quad \{
{\vec q}_{+},M_{sys} \} ={{{\vec \kappa}_{+}}\over {M_{sys}}},
 \end{eqnarray*}

\bea
 {\vec S}_q &=&\vec J-{\vec q}_{+}\times {\vec
\kappa}_{+}= {{M_{sys}\vec J}\over {\sqrt{M^2_{sys}-{\vec
\kappa}^2_{+}}}}+\nonumber \\
 &+&{{ \vec K\times {\vec \kappa}_{+}}\over
{\sqrt{M^2_{sys}-{\vec \kappa}^2_{+}}}}-{{\vec J\cdot {\vec
\kappa}_{+}\,\, {\vec \kappa}_{+}}\over {\sqrt{M^2_{sys}-{\vec
\kappa}^2_{+}}\Big( M_{sys}+\sqrt{M^2_{sys}-{\vec
 \kappa}^2_{+}}\Big) }} \approx {\vec {\bar S}} =\vec J.
 \label{3.18}
\end{eqnarray}

\medskip

iii) The {\it internal} non-canonical Fokker-Pryce 3-center of
inertia ${\vec y}_{+}$

\begin{eqnarray}
{\vec y}_{+}&=& {\vec q}_{+}+{{ {\vec S}_q\times {\vec
\kappa}_{+}}\over {\sqrt{M^2_{sys}-{\vec \kappa}_{+}^2}
(M_{sys}+\sqrt{M^2_{sys}-{\vec \kappa}_{+}^2})}} ={\vec R}_{+}+{{
{\vec S}_q\times {\vec \kappa}_{+}}\over {M_{sys}\sqrt{M^2_{sys}
-{\vec \kappa}_{+}^2}}},\nonumber \\
 {\vec q}_{+}&=&{\vec
R}_{+}+{{ {\vec S}_q\times {\vec \kappa}_{+}}\over
{M_{sys}(M_{sys}+\sqrt{M^2_{sys}-{\vec \kappa}_{+}^2})}} =
{{M_{sys}{\vec R}_{+}+ \sqrt{M^2_{sys}-{\vec \kappa}_{+}^2} {\vec
y}_{+}}\over {M_{sys}+\sqrt{M^2_{sys} -{\vec
\kappa}_{+}^2}}},\nonumber \\
 &&\{ y^r_{+},y^s_{+} \}
={1\over {M_{sys}\sqrt{M^2_{sys}-{\vec \kappa}_{+}^2} }}
\epsilon^{rsu}\Big[ S^u_q+{{ {\vec S}_q\cdot {\vec \kappa}_{+}\,
\kappa^u_{+}} \over {\sqrt{M^2_{sys}-{\vec
\kappa}_{+}^2}(M_{sys}+\sqrt{M^2_{sys}-{\vec \kappa}
_{+}^2})}}\Big].\nonumber \\
 &&{}
 \label{3.19}
 \end{eqnarray}

\bigskip

Note that on the Wigner hyper-planes, due to the rest-frame
conditions ${\vec \kappa}_{+}\approx 0$, all the internal
3-centers of mass coincide, ${\vec q}_{+} \approx {\vec R}_{+}
\approx {\vec y}_{+}$, and become essentially a unique {\it gauge}
variable conjugate to ${\vec \kappa}_{+}$. As a natural gauge
fixing for the rest-frame conditions ${\vec \kappa}_{+}\approx 0$,
we can add the vanishing of the {\it internal} Lorentz boosts
$\vec K = - M_{sys}\, {\vec R}_+ \approx 0$: this implies $\vec
\lambda (\tau )\approx 0$ and is equivalent to locate the internal
canonical 3-center of mass ${\vec q}_{+}$ in $\vec \sigma =0$,
i.e. in the external centroid $x^{\mu}_s(\tau )=z^{\mu}(\tau ,\vec
0)$ origin of the internal 3-coordinates in each Wigner
hyperplane. Remember that the centroid $x^{\mu}_s(\tau )$
corresponds to the {\it unique} special-relativistic
center-of-mass-like world-line of Refs.\cite{46}. With these gauge
fixings and with $T_s-\tau \approx 0$, the world-line
$x^{\mu}_s(\tau )$ of the centroid becomes uniquely determined
except for the arbitrariness in the choice of $x^{\mu}_s(0)$

\begin{equation}
 x^{({\vec q}_+)\mu}_s(\tau = T_s )=x^{\mu}_s(0) + u^{\mu}(p_s) T_s,
 \label{3.20}
 \end{equation}

\noindent   and coincides with the {\it external} covariant
non-canonical Fokker-Pryce 4-center of inertia, $x^{\mu}_s(\tau )
= x^{\mu}_s(0) + Y^{\mu}_s$. It can also be shown that the {\it
centroid} $x_s^{(\mu )}(\tau )$ coincides with the {\it Dixon
center of mass} of an extended object \cite{29} as well as with
the {\it Pirani} \cite{47} and the {\it Tulczyjew} \cite{48} {\it
centroids}.

Note that in the non-relativistic limit all the quantities ${\vec
q}_{+}$, ${\vec R}_+$, ${\vec y}_+$ tends the the non-relativistic
center of mass ${\vec q}_{nr}= {{\sum_{i=1}^Nm_i{\vec
\eta}_i}\over {\sum_{i=1}^Nm_i}}$.

\bigskip

We are left with the problem of the construction of a canonical
transformation bringing from the basis ${\vec \eta}_i$, ${\vec
\kappa}_i$, to a new canonical basis ${\vec q}_{+}$, ${\vec
\kappa}_{+} (\approx 0)$, ${\vec \rho}_{q,a}$, ${\vec \pi}_{q,a}$,
in which ${\vec S}_q= \sum_{a=1}^{N-1} {\vec \rho}_{q,a}\times
{\vec \pi}_{q,a}$:

\begin{equation}
\begin{minipage}[t]{1cm}
\begin{tabular}{|l|} \hline
${\vec \eta}_i$ \\  \hline
 ${\vec \kappa}_i$ \\ \hline
\end{tabular}
\end{minipage} \ {\longrightarrow \hspace{.2cm}} \
\begin{minipage}[t]{2 cm}
\begin{tabular}{|l|l|} \hline
${\vec q}_{+}$   & ${\vec \rho}_{qa}$   \\ \hline
 ${\vec \kappa}_{+}$&${\vec \pi}_{qa}$ \\ \hline
\end{tabular}
\end{minipage}
\label{3.21}
\end{equation}

Let us stress that this cannot be a point transformation, because
of the momentum dependence of the relativistic internal 3-center
of mass ${\vec q}_{+}$. The canonical transformation (\ref{3.21})
will be constructed in the next Subsection  by using the method of
Gartenhaus-Schwartz \cite{49} as delineated in Ref.\cite{50}.
\medskip

In conclusion, at the relativistic level the non-relativistic
Abelian translation symmetry generating the non-relativistic
Noether constants ${\vec P}=const.$ gets split into the two
following symmetries: i) the {\it external} Abelian translation
symmetry whose Noether constants of motion are ${\vec
p}_s=\epsilon_s {\vec k}_s\approx M_{sys}{\vec k}_s = const.$ (its
conjugate variable being the {\it external} 3-center of mass
${\vec z}_s$); ii) the {\it internal} Abelian gauge symmetry
generating the three first-class constraints ${\vec
\kappa}_{+}\approx 0$ (rest-frame conditions) inside the Wigner
hyperplane (the conjugate {\it gauge} variable being the {\it
internal} 3-center of mass ${\vec q}_{+}\approx {\vec
R}_{+}\approx {\vec y}_{+}$). Of course, its non-relativistic
counterpart is the non-relativistic rest-frame condition ${\vec
P}\approx 0$.

\subsection{The canonical transformation to the internal
center-of-mass and relative variables for N free particles.}

Since ${\vec q}_+$ and ${\vec \kappa}_+$ are known, we have only
to find the internal conjugate variables appearing in the
canonical transformation (\ref{3.21}). They have been determined
in Ref.\cite{11} by using the technique of Ref.\cite{49} and
starting from a set of canonical variables defined in
Ref.\cite{6}. Precisely, starting from the naive {\it internal}
center-of-mass variable ${\vec \eta}_{+} ={1\over N}\sum_{i=1}^N
{\vec \eta}_i$, we applied with definition of relative variables
${\vec \rho}_a$, ${\vec \pi}_a$ based on the following family of
point canonical transformations [the numerical parameters
$\gamma_{ai}$ satisfy the relations in Eqs.(\ref{2.2})]

\begin{eqnarray*}
&&\begin{minipage}[t]{3cm}
\begin{tabular}{|l|} \hline
${\vec \eta}_i$ \\  \hline
 ${\vec \kappa}_i$ \\ \hline
\end{tabular}
\end{minipage} \ {\longrightarrow \hspace{.2cm}} \
\begin{minipage}[t]{2 cm}
\begin{tabular}{|l|l|} \hline
${\vec \eta}_{+}$   & ${\vec \rho}_a$   \\ \hline
 ${\vec \kappa}_{+}$&${\vec \pi}_a$ \\ \hline
\end{tabular}
\end{minipage}  , \quad\quad a=1,..,N-1,
 \end{eqnarray*}

\bea
 {\vec \eta}_i&=&{\vec \eta}_{+}+{1\over {\sqrt{N}}}
\sum_{a=1}^{N-1} \gamma_{ai}{\vec \rho}_a,\qquad
 {\vec \kappa}_i = {1\over N} {\vec \kappa}_{+}+\sqrt{N} \sum_{a=1}^{N-1}
\gamma_{ai} {\vec \pi}_a,\nonumber \\
 &&{}\nonumber \\
  {\vec \eta}_{+}&=&{1\over N} \sum_{i=1}^N {\vec \eta}_i,\qquad
{\vec \kappa}_{+} = \sum_{i=1}^N {\vec \kappa}_i\approx
 0,\nonumber \\
 {\vec \rho}_a&=&\sqrt{N} \sum_{i=1}^N \gamma_{ai}
{\vec \eta}_i,\qquad
 {\vec \pi}_a = {1\over {\sqrt{N}}}
\sum_{i=1}^N \gamma_{ai} {\vec \kappa}_i,\nonumber \\
 &&{}\nonumber \\
 &&\{ \eta_i^r,\kappa^s_j\}
=\delta_{ij}\delta^{rs},\quad\quad \{ \eta^r_{+},\kappa^s_{+} \}
=\delta^{rs},\quad\quad \{ \rho^r_a,\pi^s_b \} =
\delta_{ab}\delta^{rs}.
 \label{3.22}
 \end{eqnarray}

\medskip

Then, (in Appendix B of Ref.\cite{12}), we gave the closed form of
the canonical transformation for arbitrary N, which turned out to
be {\it point in the momenta} but, unlike the non-relativistic
case, {\it non-point} in the configurational variables.
\medskip

Explicitly, for $N=2$ we have

\bea
 M_{sys} &=& \sqrt{m_1^2 + {\vec \kappa}_1^2} + \sqrt{m_1^2 + {\vec
 \kappa}_1^2},\qquad {\vec S}_q = {\vec \rho}_q \times {\vec \pi}_q,\nonumber \\
 &&{}\nonumber \\
 {\vec q}_+ &=& {{\sqrt{m_1^2 + {\vec \kappa}_1^2}\, {\vec \eta}_1 +
 \sqrt{m_2^2 + {\vec \kappa}_2^2}\, {\vec \eta}_2}\over {\sqrt{M^2 -
 {\vec \kappa}_+^2}}} + {{({\vec \eta}_1 \times {\vec \kappa}_1 +
 {\vec \eta}_2 \times {\vec \kappa}_2) \times {\vec \kappa}_+}\over
 {\sqrt{M_{sys}^2 - {\vec \kappa}_+^2}\, (M_{sys} + \sqrt{M_{sys}^2 - {\vec \kappa}_+^2})}}
 -\nonumber \\
 &-& {{(\sqrt{m_1^2 + {\vec \kappa}_1^2}\, {\vec \eta}_1 + \sqrt{m_2^2 + {\vec \kappa}_2^2}\,
 {\vec \eta}_2) \cdot {\vec \kappa}_+\,\, {\vec \kappa}_+}\over {M_{sys}\,
 \sqrt{M_{sys}^2 - {\vec \kappa}_+^2}\, (M_{sys} +
 \sqrt{M_{sys}^2 - {\vec \kappa}_+^2})}},\nonumber \\
 {\vec \kappa}_+ &=& {\vec \kappa}_1 + {\vec \kappa}_2 \approx
 0,\nonumber \\
 {\vec \pi}_q &=& \vec \pi - {{{\vec \kappa}_+}\over {\sqrt{M_{sys}^2 - {\vec \kappa}_+^2}}}\,
 \Big[{1\over 2}\, (\sqrt{m_1^2 + {\vec \kappa}_1^2} - \sqrt{m_2^2 + {\vec
 \kappa}_2^2}) -\nonumber \\
 &-& {{{\vec \kappa}_+ \cdot \vec \pi}\over {{\vec \kappa}_+^2}}\, (M_{sys} -
 \sqrt{M_{sys}^2 - {\vec \kappa}_+^2})\Big] \approx \vec \pi,\nonumber \\
 {\vec \rho}_q &=& \vec \rho + \Big({{\sqrt{m_1^2 + {\vec \kappa}_1^2}}\over
 {\sqrt{m_2^2 + {\vec \pi}_q^2}}} + {{\sqrt{m_2^2 + {\vec \kappa}_2^2}}\over
 {\sqrt{m_1^2 + {\vec \pi}_q^2}}}\Big)\, {{{\vec \kappa}_+ \cdot \vec \rho\,\,
 {\vec \pi}_q}\over {M_{sys}\, \sqrt{M_{sys}^2 -
 {\vec \kappa}_+^2}}} \approx \vec \rho,\nonumber  \\
 &&{}\nonumber \\
 \Rightarrow&& M_{sys} = \sqrt{{\cal M}^2 + {\vec \kappa}_+^2} \approx
 {\cal M} = \sqrt{m_1^2 + {\vec \pi}_q^2} + \sqrt{m_2^2 + {\vec \pi}_q^2}.
 \label{3.23}
 \eea

The inverse canonical transformation is

\begin{eqnarray*}
 {\vec \eta}_i &=& {\vec q}_+ - {{{\vec S}_q \times {\vec \kappa}_+}\over
 {\sqrt{{\cal M}^2 + {\vec \kappa}_+^2}\, ({\cal M} + \sqrt{{\cal M}^2 +
 {\vec \kappa}_+^2}}} + {1\over 2}\, \Big[ (-)^{i+1} -\nonumber \\
 &-& {{2\, {\cal M}\,
 {\vec \pi}_q \cdot {\vec \kappa}_+ + (m_1^2 - m_2^2)\,
 \sqrt{{\cal M}^2  + {\vec \kappa}_+^2}}\over
 {{\cal M}^2\, \sqrt{{\cal M}^2 + {\vec \kappa}_+^2}}}\Big] \cdot
 \end{eqnarray*}

\bea
 &&\Big[{\vec \rho}_q - {{{\vec \rho}_q \cdot {\vec \kappa}_+\,\,
 {\vec \pi}_q}\over {{\cal M}\, \sqrt{{\cal M}^2 + {\vec \kappa}_+^2}\,
 \Big({{\sqrt{m_1^2 + {\vec \kappa}_1^2}}\over {\sqrt{m_2^2 + {\vec \pi}_q^2}}} +
 {{\sqrt{m_2^2 + {\vec \kappa}_2^2}}\over {\sqrt{m_1^2 + {\vec \pi}_q^2}}}\Big)^{-1}
 + {\vec \pi}_q \cdot {\vec \kappa}_+}}\Big] \approx \nonumber \\
 &\approx& {\vec q}_+ + {1\over 2}\, \Big[ (-)^{i+1} - {{m_1^2 - m_2^2}\over
 {{\cal M}^2}}\Big]\, {\vec \rho}_q,\nonumber \\
 {\vec \kappa}_i &=& \Big[{1\over 2} + {{(-)^{i+1}}\over {{\cal M}\, \sqrt{{\cal M}^2
 + {\vec \kappa}_+^2}}}\, \Big({\vec \pi}_q \cdot {\vec
 \kappa}_+\, [1 - {{{\cal M}}\over {{\vec \kappa}_+^2}}\, (\sqrt{{\cal M}^2 +
 {\vec \kappa}_+^2} - {\cal M})] +\nonumber \\
 &+& (m_1^2 - m_2^2)\, \sqrt{{\cal M}^2 + {\vec
 \kappa}_+^2}\Big)\Big]\, {\vec \kappa}_+ + (-)^{i+1}\, {\vec
 \pi}_q \approx (-)^{i+1}\, {\vec \pi}_q.
 \label{3.24}
 \eea

\medskip

For $N \geq 2$, the Hamiltonian $M_{sys} = \sum_{i=1}^N\,
\sqrt{m_i^2 + N\, \sum_{ab}^{1..N-1}\, \gamma_{ai}\, \gamma_{bi}\,
{\vec \pi}_{qa}\cdot {\vec \pi}_{qb}} $ for the relative motions
in the rest frame instant form, is a sum of square roots each
containing a $(N-1)\times (N-1)$ matrix
$K^{-1}_{(i)ab}=N\gamma_{ai}\gamma_{bi}=K^{-1}_{(i)ba}$ [note that
in the non-relativistic limit only one such matrix appears, namely
$k^{-1}_{ab} = \sum_{i=1}^N\, {1\over {m_i}}\, K^{-1}_{(i)ab}$ of
Eqs.(\ref{2.3})]. The existence of relativistic normal Jacobi
coordinates would require the simultaneous diagonalization of
these N matrices. This is impossible, however, because

\begin{equation}
 [ K^{-1}_{(i)}, K^{-1}_{(j)}
]_{ab}=G_{(ij)ab}=-G_{(ij)ba}=-G_{(ji)ab}=
-N[\gamma_{ai}\gamma_{bj}-\gamma_{aj}\gamma_{bi}].
 \label{3.25}
  \end{equation}

\noindent There are ${1\over 2}N(N-1)$ matrices $G_{ij}$, each one
with ${1\over 2}(N-1)(N-2)$ independent elements. While the
conditions $G_{(ij)ab}=0$ are ${1\over 4}N(N-1)^2(N-2)$, the free
parameters at our disposal in the $\gamma_{ai}$ are only ${1\over
2}(N-1)(N-2)$. For N=3, there are 3 conditions and only 1
parameter; for N=4, 18 conditions and 3 parameters.
\medskip

In conclusion: it is impossible to diagonalize the N quadratic
forms under the square roots simultaneously, {\it no relativistic
normal Jacobi coordinates exist, and reduced masses and inertia
tensor cannot be defined}.

\bigskip

The relative Lagrangian can be worked out in the special case of
N=2 with equal masses ($m_1=m_2=m$). It results

\begin{equation}
L_{rel}(\vec \rho ,{\dot {\vec \rho}})=-m \sqrt{4-{\dot {\vec
\rho}}^2}.
 \label{3.26}
 \end{equation}

\noindent so that the relative velocity is bounded by $|{\dot
{\vec \rho}}|\leq 2$.

Let us write $\vec \rho = \rho \hat \rho$ with $\rho = |\vec \rho
|$ and $\hat \rho ={{\vec \rho}\over {|\vec \rho |}}$. With a
single relative variable the three Euler angles $\theta^{\alpha}$
are redundant: there are only two independent angles, identifying
the position of the unit 3-vector $\hat \rho$ on $S^2$. We shall
use the parametrization (Euler angles $\theta^1=\phi$,
$\theta^2=\theta$, $\theta^3=0$)

\begin{equation}
 {\hat \rho}^r = R^{rs}(\theta ,\phi )\, {\hat \rho}_o^s =
 \Big(R_z(\theta )R_y(\phi )\Big)^{rs}\, {\hat \rho}_o^s ,\qquad
 {\hat \rho}_o=(0,0,1).
 \label{3.27}
 \end{equation}

In analogy with Subsection IA of the non-relativistic case, we get
the following body frame velocity and angular velocity ($\rho$ is
the only shape variable for N=2)

\begin{eqnarray}
{\check v}^r&=&R^{T\, rs}{\dot \rho}^s=\rho (R^T\dot R)^{rs}{\hat
\rho}^s_o+\dot \rho {\hat \rho}_o^r=\rho \epsilon^{ru3}{\check
\omega}^u +\dot \rho {\hat \rho}_o^r=
 \rho ({\vec \omega}\times {\hat \rho}_o)^r+\dot \rho {\hat
\rho}^r_o,\nonumber \\
  &&{}\nonumber \\
{\vec \omega}&=&({1\over 2}\, \epsilon^{urs}\, (R^T\dot R)^{rs}) =
({\check \omega}^1=-sin\, \theta \dot \phi ,{\check \omega}^2=\dot
\theta ,0),\nonumber \\
 &&{}\nonumber \\
{\vec v}^2&=&\check I(\rho ){\vec \omega}^2+{\dot \rho}^2,\qquad
\check I(\rho ) = \rho^2.
 \label{3.28}
 \end{eqnarray}

The non-relativistic inertia tensor of the dipole ${\check
I}_{nr}=\mu \rho^2$ ($\quad \mu ={{m_1m_2}\over {m_1+m_2}}$ is the
reduced mass) is replaced by $\check I = {\check I}_{nr}/ \mu
=\rho^2$. The Lagrangian in anholonomic variables become

\begin{equation}
{\tilde L}({\vec \omega},\rho ,\dot \rho )=-m\sqrt{4-\check I(\rho
){\vec \omega}^2-{\dot \rho}^2}.
 \label{3.29}
 \end{equation}

\noindent It is clear that the bound $|{\dot {\vec \rho}}|\leq 2$
puts upper limitations upon the kinetic energy of both the
rotational and vibrational motions.

The canonical momenta are

\begin{equation}
{\vec S} = {{\partial \tilde L}\over {\partial {\vec \omega} }}
={{m \check I(\rho ) {\vec \omega}}\over {\sqrt{4-\check I(\rho
){\vec \omega}^2-{\dot \rho}^2}}},\qquad
 \pi = {{\partial
\tilde L}\over {\partial \dot \rho}}={{m\dot \rho}\over
{\sqrt{4-\check I(\rho ){\vec \omega}^2-{\dot \rho}^2}}}.
 \label{3.30}
\end{equation}

Note that {\it there is no more a linear relation between spin and
angular velocity}.

When $|{\dot {\vec \rho}}|$ varies between $0$ and 2 the momenta
vary between $0$ and $\infty$, so that {\it in phase space there
is no bound from the limiting light velocity}. This shows once
more that in special relativity it is convenient to work  in the
Hamiltonian framework avoiding relative and angular velocities.

Since we have $\sqrt{4-\check I(\rho ){\vec \omega}^2-{\dot
\rho}^2}={{2m}\over {\sqrt{m^2+{\check I}^{-1}(\rho ){\vec
S}^2+\pi^2} }}$, the inversion formulas become

\begin{eqnarray}
{\vec \omega}&=&{{ {\vec S}}\over {m\check I(\rho )}}\sqrt{4-
\check I(\rho ){\vec \omega}^2-{\dot \rho}^2}={{2{\check
I}^{-1}(\rho ) {\vec S}}\over {\sqrt{m^2+ {\check I}^{-1}(\rho
){\vec S}^2+\pi^2}}},\nonumber \\
 \dot \rho &=&{{\pi}\over
m}\sqrt{4-\check I(\rho ){\vec \omega}^2-{\dot
\rho}^2}={{2\pi}\over {\sqrt{m^2+{\check I}^{-1}(\rho ){\vec
S}^2+\pi^2}}}.
 \label{3.31}
 \end{eqnarray}

Then, the Hamiltonian results

\begin{equation}
M_{sys} = \pi \dot \rho +{\vec S}\cdot {\vec \omega}-\tilde L=
2\sqrt{m^2+{\check I}^{-1}(\rho ){\vec S}^2+\pi^2}.
 \label{3.32}
 \end{equation}

 \subsection{Dynamical body frames and canonical spin bases for $N
$ relativistic particles.}

For isolated systems the constraint manifold \cite{7} is a
stratified manifold with each stratum corresponding to a type of
Poincar\'e orbit. The main stratum (dense in the constraint
manifold) corresponds to all configurations of the isolated system
belonging to time-like Poincar\'e orbits with $\epsilon
p^2_s\approx \epsilon M^2_{sys} > 0$. As said in Ref.\cite{20},
this implies that the internal 3-center-of-mass coordinates have
been adapted to the co-adjoint orbits of the internal realization
Poincar\'e group. But, since the second Poincar\'e invariant (the
Pauli-Lubanski invariant ${\vec W}_s^2=-p_s^2{\vec S}_s^2$) does
not appear among the canonical variables, this canonical basis is
not adapted, as yet, to a {\it typical form} of canonical action
of the Poincar\'e group \cite{18} on the phase space of the
isolated system. However, remember that the {\it scheme A} for the
{\it internal} realization of the Poincar\'e group \cite{18}
contains the canonical pairs ${\vec \kappa}_{+}$, ${\vec q}_{+}$,
$S^3_q$, $arctg\, {{S^2_q}\over {S_q^1}}$, and the two Casimirs
invariants $|{\vec S}_q|=\sqrt{-W^2/M_{sys}^2}$,
$M_{sys}=\sum_{i=1}^N \sqrt{m^2_i+{\vec \kappa}_i^2}$.
\medskip

As a consequence, as shown in Ref.\cite{20}, it is possible to
construct a canonical basis including both Poincar\'e invariants
in such a way that all of the internal coordinates are adapted to
the co-adjoint action of the group and the new internal relative
variables are thereby adapted to the SO(3) group.

\bigskip

In conclusion, in the rest-frame instant form of dynamics {\it the
construction of the internal dynamical body frames and of the
internal canonical spin bases is identical to that of the
non-relativistic case}. Only the form of the relative Hamiltonian,
the invariant mass $M_{sys}$, is modified.

\subsection{Action-at-a-distance interacting particles and relativistic orbit
theory.}

As shown in Ref.\cite{6} and its bibliography (see also
Ref.\cite{51}), the action-at-a-distance interactions inside the
Wigner hyperplane may be introduced under either the square roots
(scalar and vector potentials) or outside (scalar potential like
the Coulomb one) appearing in the free Hamiltonian. Since a
Lagrangian density in presence of action-at-a-distance mutual
interactions is not known and since we are working in an instant
form of dynamics,  the potentials  in the constraints restricted
to hyper-planes must be introduced {\it by hand}. The only
restriction is that the Poisson brackets of the modified
constraints must generate the same algebra of the free ones.

\medskip

In the rest-frame instant form the most general Hamiltonian with
action-at-a-distance interactions is

\begin{equation}
 M_{sys, int} = \sum_{i=1}^N \sqrt{ m_i^2+U_i+[{\vec k}_i-{\vec V}_i]^2}
+ V,
 \label{3.33}
  \end{equation}

\noindent where $U=U({\vec \kappa}_k, {\vec \eta}_h-{\vec
\eta}_k)$, ${\vec V}_i={\vec V}_i({\vec k}_{j\not= i}, {\vec
\eta}_i-{\vec \eta}_{j\not= i})$, $V=V_o(|{\vec \eta}_i-{\vec
\eta}_j|)+V^{'}({\vec k}_i, {\vec \eta}_i-{\vec \eta}_j)$.
\medskip

If we use the canonical transformation (\ref{3.21}) defining the
relativistic canonical internal 3-center of mass (now it is
interaction-dependent, ${\vec q}_+^{(int)}$) and relative
variables on the Wigner hyperplane, with the rest-frame conditions
${\vec \kappa}_+ \approx 0$, the rest frame Hamiltonian for the
relative motion becomes

\begin{equation}
 M_{sys, int} \approx \sum_{i=1}^N\sqrt{m_i^2+{\tilde
U}_i+[\sqrt{N}\sum_{a=1}^{N-1}\gamma_{ai}{\vec \pi}_{qa}- {\tilde
{\vec V}}_i]^2} +\tilde V, \label{3.34}
\end{equation}

\noindent where

\bea
 {\tilde U}_i&=&U([\sqrt{N}\sum_{a=1}^{N-1}\gamma_{ak}{\vec
\pi}_{qa},
{1\over{\sqrt{N}}}\sum_{a=1}^{N-1}(\gamma_{ah}-\gamma_{ak}){\vec
\rho}_{qa}),\nonumber \\
 {\tilde {\vec V}}_i&=&{\vec V}_i([\sqrt{N}\sum_{a=1}^{N-1}
\gamma_{aj\not= i}{\vec
\pi}_{qa},{1\over{\sqrt{N}}}\sum_{a=1}^{N-1}
(\gamma_{ai}-\gamma_{aj\not= i}){\vec \rho}_{qa}),\nonumber \\
 \tilde V &=& V_o(|{1\over
{\sqrt{N}}}\sum_{a=1}^{N-1}(\gamma_{ai}-\gamma_{aj}){\vec
\rho}_{qa}|) + V^{'}([\sqrt{N}\sum_{a=1}^{N-1}\gamma_{ai}{\vec
\pi}_{qa},
{1\over{\sqrt{N}}}\sum_{a=1}^{N-1}(\gamma_{ai}-\gamma_{aj}){\vec
\rho}_{qa}).
 \label{3.35}
  \eea

\medskip

In order to build a realization of the internal Poincare' group,
besides $M_{sys, int}$ we need to know the potentials appearing in
the boosts ${\vec K}_{int}$ (being an instant form, ${\vec
\kappa}_+ \approx 0$ and $\vec J$ are the free ones). We need
therefore the rest-frame energy-momentum tensor of the isolated
system (see later).

\bigskip

Since the 3-centers ${\vec R}_{+}$ and ${\vec q}_{+}$ become
interaction dependent, the final canonical basis ${\vec q}_{+}$,
${\vec \kappa}_{+}$, ${\vec \rho}_{qa}$, ${\vec \pi}_{qa}$ is not
explicitly known the interacting case. For an isolated system,
however, we have $M_{sys} = \sqrt{{\cal M}^2 + {\vec
\kappa}^2_{+}} \approx {\cal M}$ with ${\cal M}$ independent of
${\vec q}_{+}$ ($\{ M_{sys}, {\vec \kappa}_{+} \} = 0$ in the
internal Poincare' algebra). This suggests that the same result
should hold true even in the interacting case. Indeed, by its
definition, the Gartenhaus-Schwartz transformation gives ${\vec
\rho}_{qa} \approx {\vec \rho}_a$, ${\vec \pi}_{qa} \approx {\vec
\pi}_a$ also in presence of interactions, so that we get

\bea
 M_{sys, int}{|}_{{\vec \kappa}_{+} =0} &=& \Big( \sum_i\, \sqrt{m_i^2 + U_i +
  ({\vec \kappa}_i - {\vec V}_i)^2} + V \Big){|}_{{\vec \kappa}_{+} =0}
= \sqrt{{\cal M}_{int}^2 + {\vec \kappa}^2_{+}} {|}_{{\vec
\kappa}_{+}} =\nonumber \\
 &=& {\cal M}_{int}{|}_{{\vec \kappa}_{+} =0} = \sum_i\, \sqrt{m_i^2 +
  {\tilde U}_i + ({\vec \kappa}_i - {\vec {\tilde V}}_i)^2} + \tilde  V,
  \label{3.36}
  \eea

 \noindent where the potentials ${\tilde U}_i$, ${\vec {\tilde
 V}}_i$, $\tilde V$ are now functions of ${\vec \pi}_{qa} \cdot {\vec
 \pi}_{qb}$, ${\vec \pi}_{qa} \cdot {\vec \rho}_{qb}$, ${\vec
 \rho}_{qa} \cdot {\vec \rho}_{qb}$.

\bigskip

Unlike in the non-relativistic case, the canonical transformation
(\ref{3.23}) is now {\it interaction dependent} (and no more point
in the momenta), since ${\vec q}_+$ is determined by a set of
Poincare' generators depending on the interactions. The only thing
to do in the generic situation is therefore to use the free
relative variables (\ref{3.23}) even in the interacting case. We
cannot impose anymore, however, the natural gauge fixings ${\vec
q}_+ \approx 0$ ($\vec K \approx 0$) of the free case, since it is
replaced by ${\vec q}_+^{(int)} \approx 0$ (namely by ${\vec
K}_{int} \approx 0$), the only gauge fixing identifying the
centroid with the external Fokker-Pryce 4-center of inertia also
in the interacting case. Once written in terms of the canonical
variables (\ref{3.23}) of the free case, these equations can be
solved for ${\vec q}_+$, which takes a form ${\vec q}_+ \approx
\vec f({\vec \rho}_{aq}, {\vec \pi}_{aq})$ as a consequence of the
potentials appearing in the boosts. Therefore, for $N=2$, the
reconstruction of the relativistic orbit by means of
Eqs.(\ref{3.24}) in terms of the relative motion is given by
(similar equations hold for arbitrary N)

\bea
 {\vec \eta}_i(\tau ) &\approx&
{\vec q}_+({\vec \rho}_q, {\vec \pi}_q) + {1\over 2}\, \Big[
(-)^{i+1} - {{m_1^2 - m_2^2}\over
 {{\cal M}^2}}\Big]\, {\vec \rho}_q {\rightarrow}_{c \rightarrow \infty}\,
 {1\over 2}\Big[(-)^{i+1} - {{m_1-m_2}\over {m}}\Big]\, {\vec \rho}_q,\nonumber \\
 &&{}\nonumber \\
 \Rightarrow && x^{\mu}_i(\tau ) = z^{\mu}_{wigner}(\tau , {\vec
 \eta}_i(\tau )) = x^{\mu}_s(0) + u^{\mu}(p_s)\, \tau +
 \epsilon^{\mu}_r(u(p_s))\, \eta^r_i(\tau ).
 \label{3.37}
 \eea

\medskip

While the potentials in $M_{sys, int}$ determine ${\vec
\rho}_q(\tau )$ and ${\vec \pi}_q(\tau )$ through Hamilton
equations, the potentials in ${\vec K}_{int}$ determine ${\vec
q}_+({\vec \rho}_q, {\vec \pi}_q)$. It is seen, therefore - as it
should be expected - that the relativistic theory of orbits is
much more complicated than in the non-relativistic case, where the
absolute orbits ${\vec \eta}_i(t)$ are proportional to the
relative orbit ${\vec \rho}_q(t)$.

\bigskip

A relevant example of this type of isolated system has been
studied in the second paper of Ref.\cite{6} starting from the
isolated system of $N$ charged positive-energy particles (with
Grassmann-valued electric charges $Q_i = \theta^*\, \theta$,
$Q^2_i =0$, $Q_i\, Q_j = Q_j\, Q_i \not= 0$ for $i \not= j$) plus
the electro-magnetic field. After a Shanmugadhasan canonical
transformation, this system can be expressed only in terms of
transverse Dirac observables corresponding to a radiation gauge
for the electro-magnetic field.  The expression of the
energy-momentum tensor in this gauge will be shown in the next
Subsection IVC (where ${\vec K}_{int}$ can be calculated). In the
semi-classical approximation of the second paper of Ref.\cite{6},
the electro-magnetic degrees of freedom are re-expressed in terms
of the particle variables by means of the Lienard-Wiechert
solution, in the framework of the rest-frame instant form. In this
way the exact semi-classical relativistic form of the
action-at-a-distance Darwin potential in the reduced phase space
of the particles has been obtained. Note that this form is
independent of the choice of the Green function in the
Lienard-Wiechert solution. In the second paper  of Ref.\cite{6}
the associated energy-momentum tensor for the case $N = 2$
[Eqs.(6.48)] is also given. The internal energy is $M_{sys} =
\sqrt{{\cal M}^2 + {\vec \kappa}_{+}^2} \approx {\cal M} =
\sum_{i=1}^2\, \sqrt{m^2_i + {\vec \pi}^2} + {{Q_1\, Q_2}\over
{4\pi\, \rho}}\, [1 + \tilde V({\vec \pi}^2, {\vec \pi} \cdot
{{\vec \rho}\over {\rho}})]$ where $\tilde V$ is given in
Eqs.(6.34), (6.35) [in Eqs. (6.36), (6.37) for $m_1 = m_2$]. The
internal boost ${\vec K}_{int}$ [Eq.(6.46)] allows the
determination of the 3-center of energy ${\vec R}_{+} = - {{{\vec
K}_{int}}\over M_{sys}} \approx {\vec q}_{+} \approx {\vec y}_{+}$
in the present interacting case.

\subsection{An Example of  Deformable Continuous System:
the Classical Klein-Gordon Field.}

Consider now the problem of separating the relativistic center of
mass for isolated special-relativistic {\it extended systems}. In
Ref.\cite{46}, mainly devoted to the same problem in general
relativity, it is shown that in special relativity there is a
unique world-line describing this notion, corresponding to the
centroid $x^{\mu}_s(\tau )$ of the rest-frame instant form. The
first attempt to define a {\it collective variable} for a
relativistic extended body in the Hamiltonian formulation was done
in Ref.\cite{25} in the case of a configuration of the
Klein-Gordon field (which is used, e.g., in the description of
bosonic stars). Let us show how the previous kinematical formalism
is working in this case.

\medskip

The rest-frame instant form for a classical real free Klein-Gordon
field \cite{26} $\phi (\tau ,\vec \sigma ) = \tilde \phi
(z^{\mu}(\tau ,\vec \sigma ))$ ($\tilde \phi (x)$ is the standard
field) on the Wigner hyper-planes is built starting from the
reformulation of Klein-Gordon action as a parametrized Minkowski
theory. The relevant quantities, as well as the internal and
external Poincare' generators, can be worked out as in Section
III, Subsection A, B, C and D, for N free relativistic particles
[$q^A=(q^{\tau}=\omega ( q);\, q^r)$, $\omega = \sqrt{m^2+ {\vec
q}^2}$, $\Omega (q) = (2\pi )^3\, 2\, \omega (q)$, $d\tilde q =
d^3q/\Omega (q)$]

 \bea
 a\left( \tau ,\vec q\right) &=& \int d^3\sigma \, \left[ \omega
\left( q\right) \phi \left( \tau ,\vec \sigma \right) +i\pi \left(
\tau ,\vec \sigma \right) \right] e^{i\left( \omega \left(
q\right) \tau -\vec
q\cdot \vec \sigma \right) }, \nonumber \\
 a^{*}\left( \tau ,\vec q\right) &=& \int d^3\sigma  \, \left[ \omega
\left( q\right) \phi \left( \tau ,\vec \sigma \right) -i\pi \left(
\tau ,\vec \sigma \right) \right] e^{-i\left( \omega \left(
q\right) \tau -\vec
q\cdot \vec \sigma \right) }, \nonumber \\
 &&{}\nonumber \\
 \phi \left( \tau ,\vec \sigma \right) &=& \int d\tilde q \, \left[
a\left( \tau ,\vec q\right) e^{-i\left( \omega \left(  q\right)
\tau -\vec q\cdot \vec \sigma \right) }+a^{*}\left( \tau ,\vec
q\right) e^{+i\left( \omega
\left(  q\right) \tau -\vec q\cdot \vec \sigma \right) }\right] , \nonumber \\
 \pi \left( \tau ,\vec \sigma \right) &=& -i\int d\tilde q \, \omega
\left( q\right) \left[ a\left( \tau ,\vec q\right) e^{-i\left(
\omega \left( q\right) \tau -\vec q\cdot \vec \sigma \right)
}-a^{*}\left( \tau ,\vec q\right) e^{+i\left( \omega \left( \vec
q\right) \tau -\vec q\cdot \vec \sigma \right) }\right],
 \nonumber \\
 &&{}\nonumber \\
&& \{\phi (\tau \vec \sigma ), \pi (\tau ,{\vec \sigma}_1\} =
\delta^3(\vec \sigma - {\vec \sigma}_1),\qquad \left\{ a\left(
\tau ,\vec q\right) ,a^{*}\left( \tau ,\vec k \right) \right\}
=-i\Omega \left( q\right) \delta ^3\left( \vec q-\vec
k\right).\nonumber \\
 &&{}
 \label{3.38}
 \eea

\begin{eqnarray*}
 M_{\phi} &=& P^{\tau}_{\phi} = {1\over 2} \int d^3\sigma \Big[ \pi^2+(\vec \partial
\phi )^2+m^2\phi^2\Big] (\tau ,\vec \sigma ) = \int d\tilde q \,
\omega \left(  q\right) a^{*}\left( \tau ,\vec q\right) a\left(
\tau ,\vec q\right),\nonumber \\
 &&{}\nonumber \\
 {\vec P}_{\phi} &=&
\int d^3\sigma [\pi \vec \partial \phi ](\tau ,\vec \sigma ) =
\int d\tilde q \, \vec q\, a^{*}\left( \tau ,\vec q\right) a\left(
\tau ,\vec q\right) \approx 0,
\end{eqnarray*}

 \begin{eqnarray*}
  J_{\phi}^{rs}&=& \int d^3\sigma [\pi (\sigma^r\partial^s-\sigma^s\partial^r)\phi
](\tau ,\vec \sigma )=\nonumber \\
&=&-i\int d\tilde q \, a^{*}\left( \tau ,\vec q\right) \left(
q^r\frac
\partial {\partial q^s}-q^s\frac \partial {\partial q^r}\right) a\left(
\tau ,\vec q\right) , \nonumber \\
J^{\tau r}_{\phi}&=&-\tau P^r_{\phi}+{1\over 2} \int d^3\sigma \,
\sigma^r\,\,
[\pi^2+(\vec \partial \phi )^2+m^2\phi^2](\tau ,\vec \sigma )=\nonumber \\
&=&- \tau P_{\phi}^r+i\int d\tilde q \, \omega \left( q\right)
a^{*}\left( \tau ,\vec q\right) \frac \partial {\partial
q^r}a\left( \tau ,\vec q\right) ,\end{eqnarray*}

\bea
 p^{\mu}_s&&\nonumber \\
 J^{ij}_s&=&{\tilde x}_s^ip_s^j-{\tilde x}_s^jp_s^i+\delta^{ir}\delta^{js}
{\bar S}_s^{rs},\qquad J^{oi}_s = {\tilde x}_s^op_s^i-{\tilde
x}^i_sp^o_s-{{\delta^{ir}{\bar S}_s^{rs}
p_s^s}\over {p^o_s+\epsilon_s}},\nonumber \\
&&{}\nonumber \\
{\bar S}_s^{rs}&\equiv& S^{rs}_{\phi}=J^{rs}_{\phi}{|}_{{\vec
P}_{\phi}=0}= \int d^3\sigma \{ \sigma^r[\pi \partial^s \phi ]
(\tau ,\vec \sigma )-(r \leftrightarrow s) \} {|}_{{\vec
P}_{\phi}=0}.
 \label{3.39}
 \eea

Here ${\tilde x}^{\mu}_s$ is the usual canonical non-covariant
4-center of mass of Eq.(\ref{3.9}). We are working on Wigner
hyper-planes with $\tau \equiv T_s$, so that the Hamiltonian
(\ref{3.12}) is $H_D = M_{\phi} - \vec \lambda (\tau ) \cdot {\vec
P}_{\phi}$.
\medskip

We want to construct four variables $X_{\phi}^A[\phi ,\pi
]=(X_{\phi}^\tau ;{\vec X}_{\phi})$ canonically conjugated to
$P_{\phi}^A[\phi ,\pi ]=(P_{\phi}^\tau ;{\vec P}_{\phi})$. First
of all we make a canonical transformation to {\it modulus-phase}
canonical variables

\bea
 a\left( \tau ,\vec q\right) &=& \sqrt{I\left( \tau ,\vec
q\right) }\, e^{\left[ i\varphi \left( \tau ,\vec q\right)
\right]},\qquad
 I\left( \tau ,\vec q\right) = a^{*}\left( \tau ,\vec q\right)
a\left( \tau ,\vec q\right), \nonumber \\
 a^{*}\left( \tau ,\vec q\right) &=& \sqrt{I\left( \tau ,\vec q\right)
}\, e^{\left[
 -i\varphi \left( \tau ,\vec q\right) \right]},\qquad
 \varphi \left( \tau ,\vec q\right) = \frac 1{2i}\ln \left[
\frac{a\left( \tau , \vec q\right) }{a^{*}\left( \tau ,\vec
q\right) }\right], \nonumber \\
 &&{}\nonumber \\
 &&\left\{ I\left( \tau ,\vec q\right) ,\varphi \left( \tau ,\vec
q^{\prime }\right) \right\} =\Omega \left(  q\right) \delta
^3\left( \vec q-\vec q^{\prime }\right) .
 \label{3.40}
 \eea

In terms of the original canonical variables $\phi$, $\pi$, we
have

\begin{eqnarray*}
I( \tau ,\vec q) &=&\int d^3\sigma \int d^3\sigma ^{\prime }\,
e^{i\vec q\cdot ( \vec \sigma -\vec \sigma ^{\prime }) }[ \omega (
q) \phi ( \tau ,\vec \sigma ) -i\pi ( \tau ,\vec \sigma ) ] [
\omega (  q) \phi ( \tau ,\vec \sigma ^{\prime }) +i\pi ( \tau
,\vec
\sigma ^{\prime }) ] , \nonumber \\
\varphi ( \tau ,\vec q) &=&\frac 1{2i}\ln\, \Big[ \frac{\int
d^3\sigma \, [ \omega (  q) \phi ( \tau ,\vec \sigma ) +i\pi (
\tau ,\vec \sigma ) ] e^{i( \omega (  q) \tau -\vec q\cdot \vec
\sigma ) }}{\int d^3\sigma ^{\prime }\, [ \omega (  q) \phi ( \tau
,\vec \sigma ^{\prime }) -i\pi ( \tau ,\vec \sigma ^{\prime }) ]
e^{-i( \omega (  q) \tau -\vec q\cdot
\vec \sigma ^{\prime }) }}\Big] =\nonumber \\
&=&\omega (q)\, \tau + \frac 1{2i}\ln\, \Big[ \frac{\int d^3\sigma
\, [ \omega (  q) \phi ( \tau ,\vec \sigma ) +i\pi ( \tau ,\vec
\sigma ) ] e^{-i\vec q\cdot \vec \sigma  }}{\int d^3\sigma
^{\prime }\, [ \omega (  q) \phi ( \tau ,\vec \sigma ^{\prime })
-i\pi ( \tau ,\vec \sigma ^{\prime }) ] e^{ i\vec q\cdot \vec
\sigma ^{\prime } }}\Big],\end{eqnarray*}

\bea
 \phi \left( \tau ,\vec \sigma \right) &=&\int d\tilde q\,
\sqrt{I\left( \tau ,\vec q\right) }\left[ e^{i\varphi \left( \tau
,\vec q\right) -i\left( \omega \left(  q\right) \tau -\vec q\cdot
\vec \sigma \right) }+e^{-i\varphi \left( \tau ,\vec q\right)
+i\left( \omega \left(  q\right) \tau -\vec
q\cdot \vec \sigma \right) }\right],\nonumber \\
\pi \left( \tau ,\vec \sigma \right) &=& -i\int d\tilde q\, \omega
\left( q\right) \sqrt{I\left( \tau ,\vec q\right) }\left[
e^{i\varphi \left( \tau ,\vec q\right) -i\left( \omega \left(
q\right) \tau -\vec q\cdot \vec \sigma \right) }-e^{-i\varphi
\left( \tau ,\vec q\right) +i\left( \omega \left( q\right) \tau
-\vec q\cdot \vec \sigma \right) }\right].
 \label{3.41}
  \eea

Then, the internal Poincar\'e charges of the field configuration
take the form

\bea
 P_{\phi}^\tau &=& \int d\tilde q\, \omega \left( q\right)
I\left( \tau ,\vec q\right), \qquad {\vec P}_{\phi} = \int d\tilde
q\,  \vec q\, I\left( \tau ,\vec q\right) \approx 0,\nonumber \\
 &&{}\nonumber \\
J^{rs}_{\phi} &=& \int d\tilde q\, I\left( \tau ,\vec q\right)
\left( q^r\frac \partial {\partial q^s}-q^s\frac \partial
{\partial q^r}\right) \varphi \left( \tau
,\vec q\right),\nonumber \\
J_{\phi}^{\tau r} &=& -\tau P_{\phi}^r-\int d\tilde q\, \omega
\left( q\right) I\left( \tau ,\vec q\right) \frac \partial
{\partial q^r}\varphi \left( \tau ,\vec q\right),
 \label{3.42}
\eea

while the classical analogue of the occupation number is
[$\triangle =-{\vec \partial}^2$]

\begin{eqnarray}
N_{\phi}&=&\int d\tilde q\, a^{*}(\tau ,\vec q) a(\tau ,\vec q) =
\int d\tilde q\, I(\tau ,\vec q)=\nonumber \\
&=&{1\over 2} \int d^3\sigma [\pi {1\over
{\sqrt{m^2+\triangle}}}+\phi \sqrt{m^2+\triangle}] \phi (\tau
,\vec \sigma ).
 \label{3.43}
\end{eqnarray}

\subsubsection{Definition of the collective variables.}

Define the four functionals of the phases

\begin{eqnarray}
&&X_{\phi}^\tau =\int d\tilde q \omega (q) F^\tau \left( q\right)
\varphi \left( \tau ,\vec q\right) , \qquad {\vec X}_{\phi}=\int
d\tilde q  \vec q\, F\left( q\right) \varphi
\left( \tau ,\vec q\right) ,\nonumber \\
&&\Rightarrow \lbrace X^r_{\phi},X^s_{\phi} \rbrace =0,\quad\quad
\lbrace X_{\phi}^\tau ,X_{\phi}^r \rbrace =0.
 \label{3.44}
\end{eqnarray}

\noindent depending on two Lorentz scalar functions $F^{\tau}(q)$,
$F(q)$. Their form will be restricted by the following
requirements implying that $X^A _{\phi}$ and $P^A_{\phi}$ are
canonical variables

\begin{equation}
\begin{array}{cccc}
\left\{ P_{\phi}^\tau ,X_{\phi}^\tau \right\} =1, & \left\{
P^r_{\phi},X_{\phi} ^s \right\} =-\delta ^{rs}, & \left\{
P_{\phi}^r,X_{\phi}^\tau \right\} =0, & \left\{ P_{\phi}^\tau
,X_{\phi}^r\right\} =0,
\end{array}
\label{3.45}
\end{equation}

Since $\left\{ P_{\phi}^\tau ,X_{\phi}^\tau \right\} =\int d\tilde
q \omega^2 \left( q\right) F^\tau \left( q\right)$ and $\left\{
P_{\phi}^r,X_{\phi}^s\right\} =\int d\tilde q \, q^rq^s\, F\left(
q \right) $, we must impose the following normalizations for
$F^{\tau }(q)$, $F(q)$

\begin{equation}
 \int d\tilde q \omega^2(q) F^\tau \left( q\right) =1,\qquad
 \int d\tilde q\, q^rq^s F\left( q\right) =-\delta ^{rs}.
 \label{3.46}
\end{equation}

Moreover, $\left\{ P_{\phi}^r,X_{\phi}^\tau \right\} =\int d\tilde
q \omega (q) q^r F^\tau \left( q\right)$ and $\left\{
P_{\phi}^\tau ,X_{\phi} ^r\right\} =\int d\tilde q \omega (q)\,
q^r F\left( q \right)$ , imply the conditions

\begin{equation}
 \int d\tilde q \omega (q)\, q^r F^\tau \left( q\right) =0,
\qquad \int d\tilde q \omega (q)\, q^r F\left( q\right) =0.
 \label{3.47}
 \end{equation}

\noindent that are automatically satisfied since $F^{\tau}(q)$,
$F(q)$, $q=|\vec q|$, are even under $q^r\, \rightarrow -q^r$.

\medskip

A solution of Eqs.(\ref{3.47})  is

\begin{equation}
 F^{\tau}(q) = {{16 \pi^2}\over {m\, q^2\, \sqrt{m^2+q^2} }}
e^{-{{4\pi}\over {m^2}}\, q^2},\qquad  F(q) = - {{48 \pi^2}\over
{m\, q^4}} \sqrt{m^2+q^2}\, e^{-{{4\pi}\over {m^2}} q^2}.
 \label{3.48}
\end{equation}

The singularity in $\vec q =0$ requires $\varphi (\tau ,\vec q)\,
{\rightarrow}_{q\rightarrow 0}\, q^{\eta}$, $\eta > 0$ for the
existence of $X^{\tau}_{\phi}$, ${\vec X}_{\phi}$.

\medskip

Note that for field configurations $\phi (\tau ,\vec \sigma )$
such that the Fourier transform $\hat \phi (\tau ,\vec q)$ has
compact support in a sphere centered at $\vec q=0$ of volume V, we
get $X^{\tau}_{\phi}=-{1\over V} \int {{d^3q}\over {\omega (q)}}
\varphi (\tau ,\vec q)$, ${\vec X}_{\phi}= {1\over V}\int d^3q
{{3\vec q}\over {{\vec q}^2}} \varphi (\tau ,\vec q)$.

\subsubsection{Auxiliary relative variables.}

As in Ref.\cite{25}, let us define an auxiliary relative action
variable and an auxiliary relative phase variable

\bea
 &&\hat I\left( \tau ,\vec q\right) =I\left( \tau ,\vec q\right)
-F^\tau \left( q\right) P_{\phi}^\tau \omega \left( q\right)
+F\left( q\right) \vec q\cdot {\vec P}_{\phi},\nonumber \\
 &&\hat \varphi \left( \tau ,\vec q\right) =\varphi \left( \tau ,\vec
q\right) -\omega \left( q\right) X_{\phi}^\tau +\vec q\cdot {\vec
X}_{\phi}.
 \label{3.49}
\eea

The previous canonicity conditions on $F^{\tau}(q)$, $F(q)$, imply

\begin{eqnarray}
&&\int d\tilde q\, \omega \left( q\right) \hat I\left( \tau ,\vec
q\right) =0, \qquad \int d\tilde q\, q^i\, \hat I\left( \tau ,\vec
q\right) =0,\nonumber \\
 &&\int d\tilde q F^\tau \left( \mathrm{q}\right) \omega \left(
q\right) \hat \varphi \left( \tau ,\vec q\right) d\tilde
q=0,\qquad \int d\tilde q F\left( q\right) q^i\, \hat \varphi
\left( \tau ,\vec q\right)=0.
 \label{3.50}
\end{eqnarray}

Such auxiliary variables have the following non-zero Poisson
brackets

\bea
 \left\{ \hat I\left( \tau ,\vec k\right) ,\hat \varphi \left(
\tau ,\vec q \right) \right\} &=& \Delta \left( \vec k,\vec
q\right) =\nonumber \\
 &=&\Omega \left( k\right) \delta ^3\left( \vec k-\vec q\right)
-F^\tau \left( k\right) \omega \left( k\right) \omega \left(
q\right) +F\left( k\right) \vec k\cdot \vec q.
 \label{3.51}
 \eea

The distribution $\Delta \left( \vec k,\vec q\right) $ has the
semigroup property and satisfies the four constraints

\bea
 &&\int d\tilde q\Delta \left( \vec k,\vec q\right) \Delta
\left( \vec q,\vec k^{\prime }\right) =\Delta \left( \vec k,\vec
k^{\prime }\right),\nonumber \\
 &&{}\nonumber \\
 &&\int d\tilde q\, \omega \left( q\right) \Delta \left( \vec q,\vec
k\right) =0, \qquad \int d\tilde q\, q^r\, \Delta \left( \vec
q,\vec k\right)=0,\nonumber \\
 &&{}\nonumber \\
 &&\int d\tilde q\, F^\tau \left( q\right) \omega \left( q\right)
\Delta \left( \vec k,\vec q\right) =0, \qquad \int d\tilde q\,
q^r\, F\left( q\right) \Delta \left( \vec k,\vec q\right) =0.
 \label{3.52}
 \eea

At this stage the canonical variables $I(\tau ,\vec q)$, $\varphi
(\tau ,\vec q)$ for the Klein-Gordon field are replaced by the
non-canonical set $X_{\phi}^\tau$ ,  $P_{\phi}^\tau$ , ${\vec
X}_{\phi}$, ${\vec P}_{\phi}$, $\hat I(\tau ,\vec q)$, $\hat
\varphi (\tau ,\vec q)$ with Poisson brackets

\bea
 &&\left\{ P_{\phi}^\tau ,X_{\phi}^\tau \right\} =1, \quad \left\{
P_{\phi}^r, X_{\phi}^s\right\} =-\delta ^{rs}, \quad \left\{
P_{\phi}^r,X_{\phi}^\tau \right\} =0, \quad \left\{ P_{\phi}^\tau
,X_{\phi}^r\right\} =0,\nonumber \\
 &&\left\{ X_{\phi}^r,X_{\phi}^s\right\} =0, \quad \left\{ X_{\phi}^\tau
,X_{\phi}^r \right\} =0, \quad \left\{ P^A_{\phi}, P^B_{\phi}
\right\} =0,\quad A,B=(\tau , r),\nonumber \\
 &&\left\{ P_{\phi}^\tau ,\hat I\left( \tau ,\vec q\right) \right\}
=0, \quad \left\{ P_{\phi}^r,\hat I\left( \tau ,\vec q\right)
\right\} =0, \quad \left\{ \hat I\left( \tau ,\vec q\right)
,X_{\phi}^\tau \right\} =0, \quad \left\{ \hat I\left( \tau ,\vec
q\right), X_{\phi}^r\right\} =0,\nonumber \\
 &&\left\{ X_{\phi}^r,\hat \varphi \left( \tau ,\vec q\right)
\right\} =0, \quad \left\{ X_{\phi}^\tau ,\hat \varphi \left( \tau
,\vec q\right) \right\} =0, \quad \left\{ P^r_{\phi},\hat \varphi
\left( \tau ,\vec q\right) \right\} =0, \quad \left\{
P_{\phi}^\tau ,\hat \varphi \left( \tau ,\vec q\right) \right\} =0,\nonumber \\
 &&\left\{ \hat I\left( \tau ,\vec k\right) ,\hat \varphi \left( \tau
,\vec q \right) \right\} =\Omega \left( k\right) \delta ^3\left(
\vec k-\vec q\right) -F^\tau \left( k\right) \omega \left(
k\right) \omega \left( q\right) +F\left( k\right) \vec k\cdot \vec
q .
 \label{3.53}
\eea

Note finally that the generators of the internal Lorentz group are
already decomposed into the collective and the relative parts,
each satisfying the Lorentz algebra and having vanishing mutual
Poisson brackets

\bea
J^{rs}_{\phi} &=& L_{\phi}^{rs}+{\hat S}_{\phi}^{rs}, \nonumber \\
 &&{}\nonumber \\
 &&L_{\phi}^{rs}=X_{\phi}^rP_{\phi}^s-X_{\phi}^sP_{\phi}^r,\qquad
{\hat S}_{\phi}^{rs}=\int d\tilde q\, \hat I\left( \tau ,\vec
q\right) \left( q^r \frac \partial {\partial q^s}-q^s\frac
\partial {\partial q^r}\right) \hat \varphi \left( \tau ,\vec q\right) ,
\nonumber \\
 &&{}\nonumber \\
 J^{\tau r}_{\phi} &=&L^{\tau r}_{\phi}+{\hat S}^{\tau r}_{\phi}, \nonumber \\
 &&{}\nonumber \\
 &&L_{\phi}^{\tau r}=[X_{\phi}^\tau -\tau ] P_{\phi}^r-X_{\phi}^r
P_{\phi}^\tau , \qquad {\hat S}_{\phi}^{\tau r}=-\int d\tilde q\,
\omega \left( q\right) \hat I\left( \tau ,\vec q\right) \frac
\partial {\partial q^r}\hat \varphi \left( \tau ,\vec q\right) .
 \label{3.54}
\eea

\subsubsection{Canonical relative variables.}

Now we must find the relative canonical variables hidden inside
the auxiliary ones. They are not free but satisfy
Eqs.(\ref{3.50}). As in Ref.\cite{25}, let us introduce the
following differential operator [$\triangle_{LB}$ is the
Laplace-Beltrami operator of the mass shell sub-manifold $H^1_3$
(see Ref.\cite{25,26})]

\begin{eqnarray}
{\cal D}_{\vec q}&=&3-m^2 \triangle_{LB}=\nonumber \\
&=&3-m^2\left[ \sum\limits_{i=1}^3\left( \frac \partial {\partial
q^i}\right) ^2+\frac 2{m^2}\sum\limits_{i=1}^3q^i\, \frac \partial
{\partial q^i}+\frac 1{m^2}\left( \sum\limits_{i=1}^3q^i\, \frac
\partial {\partial q^i}\right) ^2\right].
 \label{3.55}
\end{eqnarray}

\noindent Note that, being invariant under Wigner's rotations, is
a scalar on the Wigner hyperplane.

Since $\omega \left( \mathrm{q}\right) $ and $\vec q$ are null
modes of this operator \cite{25},  we can put

\begin{equation}
 \hat I\left( \tau ,\vec q\right) = {\cal D}_{\vec q}{\bf H}(
\tau ,\vec q) ,\qquad {\bf H}\left( \tau ,\vec q\right) = \int
d\tilde k\, {\cal G}\left( \vec q,\vec k\right) \hat I\left( \tau
,\vec k\right),
 \label{3.56}
 \end{equation}

\noindent with  ${\cal G}\left( \vec q,\vec k\right) $ the Green
function of  ${\cal D}_{\vec q}$ (see Refs.\cite{25} for its
expression)

\begin{equation}
{\cal D}_{\vec q}{\cal G}\left( \vec q,\vec k\right) =\Omega
\left( k \right) \delta ^3\left( \vec k-\vec q\right).
 \label{3.57}
\end{equation}

Like in Ref.\cite{25}, for each zero mode $f_o(\vec q)$ of ${\cal
D}_{\vec q}$ [${\cal D}_{\vec q}\, f_o(\vec q)=0$] for which $|
\int d\tilde q\, f_o(\vec q) \hat I(\tau ,\vec q) | < \infty$,
integrating by parts, we get

\begin{eqnarray}
\int d\tilde q && f_o(\vec q) \hat I(\tau ,\vec q) = \int d\tilde
q\, f_o(\vec
q) {\cal D}_{\vec q} {\bf H}(\tau ,\vec q)=\nonumber \\
&=&-{1\over {2(2\pi )^3}} \int d^3q\, {{\partial}\over {\partial
q^r}} \Big( {{m^2\delta^{rs}+q^rq^s}\over {\omega (q)}} \Big[
f_o(\vec q) {{\partial}\over {\partial q^s}} {\bf H}(\tau ,\vec
q)-{\bf H}(\tau ,\vec q) {{\partial}\over {\partial q^s}} f_o(\vec
q)\Big] \Big).\nonumber \\
&&{}
 \label{3.58}
\end{eqnarray}

The boundary conditions (ensuring finite Poincar\'e generators)

\begin{eqnarray}
{\bf H}(\tau ,\vec q)\, &{\rightarrow}_{q \rightarrow 0}\,&
q^{-1+\epsilon},
\quad\quad \epsilon > 0,\nonumber \\
{\bf H}(\tau ,\vec q)\, &{\rightarrow}_{q \rightarrow \infty}\,&
q^{-3-\sigma}, \quad\quad \sigma > 0,
 \label{3.59}
\end{eqnarray}

\noindent imply $\int d\tilde q\, f_o(\vec q) \hat I(\tau ,\vec
q)=0$ (so that the first two conditions (\ref{3.50}) are also
satisfied), or

\begin{equation} \int d\tilde q\, f_o(\vec q) I(\tau ,\vec q) =
P^{\tau}_{\phi}\, \int d\tilde q\, \omega (q) f_o(\vec q)
F^{\tau}(q)- {\vec P}_{\phi}\, \cdot \, \int d\tilde q\, \vec q\,
f_o(\vec q) F(q).
 \label{3.60}
\end{equation}

It is shown in Ref.\cite{25} that, by restricting ourselves to
field configurations for which $I(\tau ,\vec q)\,
{\rightarrow}_{q\rightarrow 0}\, q^{-3+\eta}$ with $\eta \in
(0,1]$ and by imposing the following restriction on $\phi (\tau
,\vec \sigma )$ and $\pi (\tau ,\vec \sigma ) =\partial_{\tau}
\phi (\tau ,\vec \sigma )$

\begin{eqnarray}
P_{l0}&=&const. \int d\tilde q\, q^l\, {}_2F_1 ({{l-1}\over 2},
{{l+3}\over 2}; l+{3\over 2}; -q^2)\, Y_{l0}(\theta ,\varphi )
\nonumber \\
&&\int d^3\sigma \int d^3\sigma^{'} e^{i\vec q\cdot (\vec \sigma
-{\vec \sigma} ^{'})} \Big[ (m^2+q^2)\phi (\tau ,\vec \sigma )\phi
(\tau ,{\vec \sigma}^{'})+ \pi (\tau ,\vec \sigma )\pi (\tau
,{\vec \sigma}^{'})\Big] =0.
 \label{3.61}
\end{eqnarray}

\noindent we can identify the class of the Klein-Gordon field
configurations that is compatible with the previous canonical
transformation and lead to a unique realization of the Poincar\'e
group without any ambiguity

We can satisfy the constraints (\ref{3.50}) on $\hat \varphi (\tau
,\vec q)$ with the definition [${\cal D}_{\vec q} \omega (q)={\cal
D}_{\vec q} \vec q =0$]

\begin{eqnarray}
\hat \varphi (\tau ,\vec q) &=&\int d\tilde k\int d\tilde
k^{\prime } {\bf K}(\tau ,\vec k) {\cal G}(\vec k,\vec k^{\prime})
\Delta (\vec k^{\prime},
\vec q) ,\nonumber \\
{\bf K}(\tau ,\vec q) &=&{\cal D}_{\vec q}\hat \varphi (\tau ,\vec
q) =
{\cal D}_{\vec q}\varphi (\tau ,\vec q),\nonumber \\
&&{}\nonumber \\
&&{\rightarrow}_{q\rightarrow \infty}\, q^{1-\epsilon},\quad
\epsilon > 0, \quad\quad {\rightarrow}_{q\rightarrow 0}\, q^{\eta
-2},\quad \eta > 0,
 \label{3.62}
\end{eqnarray}

\noindent which also imply

\bea
 &&\left\{ {\bf H}\left( \tau ,\vec q\right) ,X_{\phi}^\tau
\right\} =0, \qquad \left\{
{\bf H}\left( \tau ,\vec q\right) ,P_{\phi}^\tau \right\} =0,\nonumber  \\
 &&\left\{
{\bf H}\left( \tau ,\vec q\right) ,X_{\phi}^r\right\} =0,
\qquad\left\{ {\bf H} \left( \tau ,\vec q\right)
,P_{\phi}^r\right\} =0, \nonumber \\
 &&\left\{ {\bf K}\left( \tau ,\vec q\right) ,X_{\phi}^\tau \right\}
=0, \qquad \left\{ {\bf K}\left( \tau ,\vec q\right)
,P_{\phi}^\tau \right\} =0, \nonumber \\
 &&\left\{ {\bf K}\left(
\tau ,\vec q\right) ,X_{\phi}^r\right\} =0, \qquad \left\{ {\bf K}
\left( \tau ,\vec q\right) ,P_{\phi}^r\right\} =0, \nonumber \\
 &&\left\{ {\bf H}\left( \tau ,\vec q\right) ,\mathbf{K}\left( \tau
,\vec q^{\prime }\right) \right\} =\Omega \left(  q\right) \delta
^3\left( \vec q-\vec q^{\prime }\right).
 \label{3.63}
\eea

The final decomposition of the internal Lorentz generators is

\bea
 J_{\phi}^{rs} &=& L_{\phi}^{rs}+S_{\phi}^{rs}, \nonumber \\
 &&{}\nonumber \\
 &&L_{\phi}^{rs}=X_{\phi}^rP_{\phi}^s-X_{\phi}^sP_{\phi}^r,
\qquad S_{\phi}^{rs}=\int d\tilde k{\bf H}\left( \tau ,\vec
k\right) \left( k^r \frac \partial {\partial k^s}-k^s\frac
\partial {\partial k^r}\right) {\bf K}\left( \tau ,\vec k\right) ,
\nonumber \\
 &&{}\nonumber \\
 J_{\phi}^{\tau r} &=& L_{\phi}^{\tau r}+S_{\phi}^{\tau r},\nonumber \\
 &&{}\nonumber \\
 &&L_{\phi}^{\tau r}=(X_{\phi}^\tau -\tau )P_{\phi}^r-X_{\phi}^rP_{\phi}^\tau ,
\qquad S_{\phi}^{\tau r}=-\int d\tilde q\omega \left( q\right)
{\bf H} \left( \tau ,\vec q\right) \frac \partial {\partial
q^r}{\bf K} \left( \tau ,\vec q\right) .
 \label{3.64}
\eea

\subsubsection{Field variables in terms of collective-relative
variables.}

We have found the canonical transformation

\begin{eqnarray*}
I\left( \tau ,\vec q\right) &=& F^{\tau}(q)\omega (q)
P^{\tau}_{\phi}-F(q)\vec q\cdot {\vec P}_{\phi}+
{\cal D}_{\vec q} {\bf H}(\tau ,\vec q),\nonumber \\
 \varphi (\tau ,\vec q)&=&\int d\tilde k\int d\tilde k^{\prime }
{\bf K} ( \tau ,\vec k) {\cal G}\left( \vec k,\vec k^{\prime
}\right) \Delta \left( \vec k^{\prime },\vec q\right) +\omega
\left( q\right) X_{\phi}^\tau -\vec q\cdot {\vec X}_{\phi},
 \end{eqnarray*}

\bea
 N_{\phi}&=&P^{\tau}_{\phi} \int d\tilde q\, \omega (q)
F^{\tau}(q)-{\vec P} _{\phi}\cdot \,\, \int d\tilde q\, \vec q
F(q) +\int d\tilde q\,
{\cal D}_{\vec q} {\bf H}(\tau ,\vec q)=\nonumber \\
&=& \tilde c {{ P^{\tau}_{\phi}}\over m}+\int d\tilde q\,
{\cal D}_{\vec q} {\bf H}(\tau ,\vec q),\nonumber \\
&&\tilde c = \tilde c(m) = m \int d\tilde q \omega (q)
F^{\tau}(q)=2\int_0^{\infty}{{dq}\over
{\sqrt{m^2+q^2}}}e^{-{{4\pi}\over {m^2}}q^2}=2 e^{4\pi}
\int^{\infty}_m
{{dx}\over {\sqrt{x^2-m^2}}}e^{-{{4\pi}\over {m^2}}x^2},\nonumber \\
&&{}
 \label{3.65}
\end{eqnarray}

\noindent with the two functions $F^{\tau}(q)$, $F(q)$ given in
Eqs.(\ref{3.48}). Its inverse is

\begin{eqnarray*}
P^{\tau}_{\phi}&=&\int d\tilde q \omega (q) I(\tau ,\vec
q)={1\over 2}\int d^3\sigma \Big[ \pi^2 +(\vec \partial \phi
)^2+m^2\phi^2\Big] (\tau ,\vec
\sigma ),\nonumber \\
{\vec P}_{\phi}&=&\int d\tilde q \vec q\, I(\tau ,\vec q)=\int
d^3\sigma
\Big[ \pi \vec \partial \phi \Big] (\tau ,\vec \sigma ),\nonumber \\
X^{\tau}_{\phi}&=&\int d\tilde q \omega (q) F^{\tau}(q) \varphi
(\tau ,\vec q)=
\tau +\nonumber \\
&+&{1\over {2\pi i m}} \int d^3q {{e^{-{{4\pi}\over {m^2}} q^2}
}\over {q^2\, \omega (q)}} ln\, \Big[ {{\omega (q) \int d^3\sigma
e^{i\vec q\cdot \vec \sigma} \phi (\tau ,\vec \sigma ) +i\int
d^3\sigma e^{i\vec q\cdot \vec \sigma} \pi (\tau ,\vec \sigma
)}\over {\omega (q) \int d^3\sigma e^{-i\vec q\cdot \vec \sigma}
\phi (\tau ,\vec \sigma ) -i \int d^3\sigma e^{-i\vec q\cdot \vec
\sigma} \pi (\tau ,\vec \sigma )}}\Big] =\nonumber \\
&{\buildrel {def} \over =}\,& \tau + {\tilde
X}^{\tau}_{\phi},\quad\quad \Rightarrow \quad L^{\tau
r}_{\phi}={\tilde X}^{\tau}_{\phi}P^r_{\phi}-
X^r_{\phi}P^{\tau}_{\phi},\nonumber \\
&&{}\nonumber \\
{\vec X}_{\phi}&=&\int d\tilde q \vec q\, F(q) \varphi (\tau ,\vec
q)=
\nonumber \\
&=&{{2i}\over {\pi m}} \int d^3q\, {{q^i}\over {q^4}}\,
e^{-{{4\pi}\over {m^2}} q^2}  ln\, \Big[ {{\sqrt{m^2+q^2} \int
d^3\sigma e^{i\vec q\cdot \vec \sigma} \phi (\tau ,\vec \sigma )
+i \int d^3\sigma e^{i\vec q\cdot \vec \sigma} \pi (\tau ,\vec
\sigma )}\over {\sqrt{m^2+q^2} \int d^3\sigma e^{-i\vec q\cdot
\vec \sigma} \phi (\tau ,\vec \sigma ) -i \int d^3\sigma e^{-i\vec
q\cdot \vec \sigma} \pi (\tau ,\vec \sigma )}} \Big] ,
 \end{eqnarray*}

\bea
 {\bf H}(\tau ,\vec q)&=&\int d\tilde k {\cal G}(\vec q,\vec
k) [I(\tau ,\vec k) -F^{\tau}(k)\omega (k) \int d{\tilde q}_1
\omega (q_1) I(\tau ,{\vec q}_1)+
\nonumber \\
&+&F(k) \vec k \cdot \, \int d{\tilde q}_1 {\vec q}_1\, I(\tau
,{\vec q}_1) ]=
\nonumber \\
&=&\int d^3\sigma_1d^3\sigma_2 \Big[ \pi (\tau ,{\vec \sigma}_1)
\pi (\tau ,{\vec \sigma}_2) \int d\tilde k {\cal G}(\vec q,\vec k)
\int d{\tilde k}_1 \triangle (\vec k,{\vec k}_1)e^{i{\vec
k}_1\cdot ({\vec \sigma}_1-{\vec
\sigma}_2)}+\nonumber \\
&+&\phi (\tau ,{\vec \sigma}_1) \phi (\tau ,{\vec \sigma}_2) \int
d\tilde k {\cal G}(\vec q,\vec k) \int d{\tilde k}_1
\omega^2(k_1)\triangle (\vec k,{\vec
k}_1) e^{i{\vec k}_1\cdot ({\vec \sigma}_1-{\vec \sigma}_2)}-\nonumber \\
&-&i\Big( \pi (\tau ,{\vec \sigma}_1) \phi (\tau ,{\vec
\sigma}_2)+\pi (\tau
,{\vec \sigma}_2) \phi (\tau ,{\vec \sigma}_1)\Big) \nonumber \\
&& \int d\tilde k {\cal G}(\vec q,\vec k) \int d{\tilde k}_1
\omega (k_1) \triangle (\vec k.{\vec k}_1) e^{i{\vec k}_1\cdot
({\vec \sigma}_1-{\vec
\sigma}_2)} \Big] ,\nonumber \\
&&{}\nonumber \\
{\bf K}(\tau ,\vec q)&=&{\cal D}_{\vec q} \hat \varphi (\tau ,\vec
q)={\cal D}
_{\vec q}\varphi (\tau ,\vec q)=\nonumber \\
&=&{1\over {2i}} {\cal D}_{\vec q} \ln \left[ \frac{\int d^3\sigma
\, \left[ \omega \left(  q\right) \phi \left( \tau ,\vec \sigma
\right) +i\pi \left( \tau ,\vec \sigma \right) \right] e^{-i\vec
q\cdot \vec \sigma  }}{\int d^3\sigma ^{\prime }\, \left[ \omega
\left(  q\right) \phi \left( \tau ,\vec \sigma ^{\prime }\right)
-i\pi \left( \tau ,\vec \sigma ^{\prime }\right) \right] e^{i\vec
q\cdot \vec \sigma ^{\prime } }}\right].
 \label{3.66}
\end{eqnarray}

\noindent The remaining canonical variables $a(\tau ,\vec q)$,
$\phi (\tau ,\vec \sigma )$, $\pi (\tau ,\vec \sigma )$, can be
worked out in terms of the final ones

\begin{eqnarray}
a(\tau ,\vec q)&=&\sqrt{F^{\tau}(q)\omega (q)P^{\tau}
_{\phi}-F(q)\vec q\cdot {\vec P}_{\phi}+{\cal D}_{\vec q}
{\bf H}(\tau , \vec q) }\nonumber \\
&&e^{i[\omega (q) X_{\phi}^\tau -\vec q\cdot {\vec X}_{\phi}]
+i\int d\tilde k\int d\tilde k^{\prime }\, {\bf K}( \tau ,\vec k)
{\cal G}\left( \vec k,\vec k^{\prime }\right) \Delta \left( \vec
k^{\prime }
,\vec q\right)  },\nonumber \\
&&{}\nonumber \\
N_{\phi}&=& \tilde c {{P^{\tau}_{\phi}}\over m}- +\int d\tilde k\,
{\cal D}_{\vec k} {\bf H}(\tau ,\vec k),
 \label{3.67}
\end{eqnarray}

\begin{eqnarray*}
\phi (\tau ,\vec \sigma ) &=&  \int d\tilde
q\sqrt{F^{\tau}(q)\omega (q)P^{\tau} _{\phi}-F(q)\vec q\cdot {\vec
P}_{\phi}+{\cal D}_{\vec q}
{\bf H}(\tau , \vec q) }\nonumber \\
&&\left[ e^{-i[\omega (q) \left( \tau -X_{\phi}^\tau \right) -
\vec q\cdot \left(\vec \sigma -{\vec X}_{\phi}\right)] +i\int
d\tilde k\int d\tilde k^{\prime }\, {\bf K}( \tau ,\vec k) {\cal
G}\left( \vec k,\vec k^{\prime }\right) \Delta \left( \vec
k^{\prime }
,\vec q\right)  }\right. + \nonumber \\
&&+\left. e^{i[\omega \left( q\right) \left( \tau -X_{\phi}^\tau
\right) -\vec q\cdot \left( \vec \sigma -{\vec X}_{\phi}\right) ]
-i\int d\tilde k\int d\tilde k^{\prime }\, {\bf K}\left( \tau
,\vec k\right) {\cal G}\left( \vec k,\vec k^{\prime }\right)
\Delta \left( \vec
k^{\prime },\vec q\right) }\right] =\nonumber \\
&=&2 \int d\tilde q {\bf A}_{\vec q}(\tau ;P^A_{\phi},{\bf H}]\,
cos\, \Big[ \vec q\cdot \vec \sigma +{\bf B}_{\vec q}(\tau
;X^A_{\phi},{\bf K}] \Big] ,\end{eqnarray*}

\begin{eqnarray*}
 \pi (\tau ,\vec \sigma )&=&-i \int d\tilde q \omega
(q)\sqrt{F^{\tau}(q) \omega (q)P^{\tau}_{\phi}-F(q)\vec q\cdot
{\vec P}_{\phi}+{\cal D}_{\vec q}
{\bf H}(\tau , \vec q) }\nonumber \\
&&\left[ e^{-i[\omega (q) \left( \tau -X_{\phi}^\tau \right)- \vec
q\cdot \left(\vec \sigma -{\vec X}_{\phi}\right)] +i\int d\tilde
k\int d\tilde k^{\prime }\, {\bf K}( \tau ,\vec k) {\cal G}\left(
\vec k,\vec k^{\prime }\right) \Delta \left( \vec k^{\prime }
,\vec q\right)  }\right. - \nonumber \\
&&-\left. e^{+i[\omega \left( q\right) \left( \tau -X_{\phi}^\tau
\right) -\vec q\cdot \left( \vec \sigma -{\vec X}_{\phi}\right)
]-i\int d\tilde k\int d\tilde k^{\prime }\, {\bf K}\left( \tau
,\vec k\right) {\cal G}\left( \vec k,\vec k^{\prime }\right)
\Delta \left( \vec
k^{\prime },\vec q\right) }\right] =\nonumber \\
&=&-2\int d\tilde q \omega (q) {\bf A}_{\vec q}(\tau
;P^A_{\phi},{\bf H}]\, sin\, \Big[ \vec q\cdot \vec \sigma +{\bf
B}_{\vec q}(\tau ;X^A_{\phi},{\bf K}] \Big],
 \end{eqnarray*}

\bea
 {\bf A}_{\vec q}(\tau ;P^A_{\phi},{\bf
H}]&=&\sqrt{F^{\tau}(q)\omega (q) P^{\tau}_{\phi}-F(q)\vec q\cdot
{\vec P}_{\phi}+{\cal D}_{\vec q}{\bf H}(\tau
,\vec q)}=\sqrt{I(\tau ,\vec q)},\nonumber \\
{\bf B}_{\vec q}(\tau ;X^A_{\phi},{\bf K}]&=&-\vec q\cdot {\vec
X}_{\phi}- \omega (q)(\tau -X^{\tau}_{\phi})+\int d\tilde
kd{\tilde k}^{'} {\bf K}(\tau ,\vec k){\cal G}(\vec k,{\vec
k}^{'})\triangle ({\vec k}^{'},\vec q)=
\nonumber \\
&=&\varphi (\tau ,\vec q)-\omega (q) \tau .
 \label{3.68}
\end{eqnarray}

Summarizing, the Klein-Gordon field configuration is described by:

\hfill\break i) its energy $P^{\tau} _{\phi} = M_{\phi}$, i.e. its
invariant mass, and the conjugate field time-variable
$X^{\tau}_{\phi}\, {\buildrel {def}\over  =}\, \tau + {\tilde
X}^{\tau}_{\phi}$ ($\tau \equiv T_s$), which is equal to $\tau$
plus some kind of {\it internal time} ${\tilde X}^{\tau}_{\phi}$
(note that in the N-body case this variable does not exist:
$M_{sys}$ is a given function of the other canonical variables and
not an independent canonical variable like here); \hfill\break ii)
the conjugate reduced canonical variables of a free point ${\vec
X}_{\phi}$, ${\vec P}_{\phi} \approx 0$ (see ${\vec \eta}_+$ and
${\vec \kappa}_+ \approx 0$ of Eqs.(\ref{3.22}) in the N-body
case);\hfill\break iii) an infinite set of canonically conjugate
relative variables ${\bf H}(\tau ,\vec q)$, ${\bf K}(\tau ,\vec
q)$ (${\vec \rho}_a$ and ${\vec \pi}_a$ of Eq.(\ref{3.22}) in the
N-body case). \hfill\break

While the set iii) describes an infinite set of {\it canonical
relative variables} with respect to the relativistic collective
variables of the sets i) and ii), the sets i) and ii) describe a
{\it monopole} field configuration, which depends only on 8
degrees of freedom like a scalar particle at rest [${\vec
P}_{\phi}\approx 0$] with mass $M_s =
\sqrt{(P^{\tau}_{\phi})^2-{\vec P}_{\phi}^2}\approx
P^{\tau}_{\phi}$ but without the mass-shell condition $M_s \approx
const.$, corresponding to the decoupled collective variables of
the field configuration. The conditions ${\bf H}(\tau ,\vec
q)={\bf K}(\tau ,\vec q)=0$ select the class of field
configurations, solutions of the Klein-Gordon equation, which are
of the {\it monopole} type on the Wigner hyperplanes

\begin{eqnarray}
\phi_{mon}(\tau ,\vec \sigma )&=&2 \int d\tilde q
\sqrt{F^{\tau}(q)\omega (q) P^{\tau}_{\phi}-F(q)\vec q\cdot {\vec
P}_{\phi}} cos\, \Big[ \vec q\cdot (\vec \sigma -{\vec
X}_{\phi})-\omega (q)(\tau -X^{\tau}_{\phi})\Big] \approx
\nonumber \\
&\approx& 2 \sqrt{P^{\tau}_{\phi}} \int d\tilde q
\sqrt{F^{\tau}(q)\omega (q)} cos\, \Big[ \vec q\cdot (\vec
\sigma -{\vec X}_{\phi})-\omega (q)(\tau -X^{\tau}_{\phi})\Big] ,\nonumber \\
\pi_{mon}(\tau ,\vec \sigma )&=& -2 \int d\tilde q
\sqrt{F^{\tau}(q)\omega (q)P^{\tau}_{\phi}-F(q)\vec q\cdot {\vec
P}_{\phi}} sin\, \Big[ \vec q\cdot (\vec \sigma -{\vec
X}_{\phi})-\omega (q)(\tau -X^{\tau}_{\phi})\Big] \approx
\nonumber \\
&\approx& -2 \sqrt{P^{\tau}_{\phi}} \int d\tilde q
\sqrt{F^{\tau}(q)\omega (q)} sin\, \Big[ \vec q\cdot (\vec \sigma
-{\vec X}_{\phi})-\omega (q)(\tau -X^{\tau}_{\phi})\Big] .
\label{3.69}
\end{eqnarray}

\medskip

If we add the gauge-fixings ${\vec X}_{\phi}\approx 0$ to ${\vec
P}_{\phi} \approx 0$ (this implies $\vec \lambda (\tau )=0$ and
$H_D = M_{\phi}$) and go to Dirac brackets, the rest-frame
instant-form canonical variables of the Klein-Gordon field in the
gauge $\tau \equiv T_s$ are (in the following formulas it holds
$T_s-X^{\tau}_{\phi}=-{\tilde X}^{\tau}_{\phi}$)

\begin{eqnarray*}
a(T_s ,\vec q)&=&\sqrt{F^{\tau}(q)\omega (q)P^{\tau}_{\phi}
+{\cal D}_{\vec q} {\bf H}(T_s , \vec q) }\nonumber \\
&&e^{i[\omega (q) {\tilde X}_{\phi}^\tau  + \vec q\cdot \vec
\sigma ]+ i\int d\tilde k\int d\tilde k^{\prime }\, {\bf K}(T_s
,\vec k) {\cal G}\left( \vec k,\vec k^{\prime }\right) \Delta
\left( \vec k^{\prime }
,\vec q\right)   },\nonumber \\
&&{}\nonumber \\
N_{\phi}&=&\tilde c {{P^{\tau}_{\phi}}\over m} +\int d\tilde q
{\cal D}_{\vec q} {\bf H}(T_s,\vec q),
 \end{eqnarray*}

\begin{eqnarray*}
 \phi (T_s ,\vec \sigma ) &=&  \int d\tilde
q\sqrt{F^{\tau}(q)\omega (q)P^{\tau} _{\phi}+{\cal D}_{\vec q}
{\bf H}(T_s , \vec q) }
\nonumber \\
&&\left[ e^{i[\omega (q) {\tilde X}_{\phi}^\tau + \vec q\cdot \vec
\sigma ]+ i\int d\tilde k\int d\tilde k^{\prime }\, {\bf K}( T_s
,\vec k) {\cal G}\left( \vec k,\vec k^{\prime }\right) \Delta
\left( \vec k^{\prime }
,\vec q\right)  }\right. + \nonumber \\
&&+\left. e^{-i[\omega \left( q\right) {\tilde X}_{\phi}^\tau
+\vec q\cdot \vec \sigma ] -i\int d\tilde k\int d\tilde k^{\prime
}\, {\bf K}\left( T_s ,\vec k\right) {\cal G}\left( \vec k,\vec
k^{\prime }\right) \Delta \left( \vec
k^{\prime },\vec q\right) }\right] =\nonumber \\
&=&2 \int d\tilde q\, {\bf A}_{\vec q}(T_s;P^{\tau} _{\phi},{\bf
H}] \, cos\, \Big[ \vec q\cdot \vec \sigma +{\bf B}_{\vec q}(T_s;
{\tilde X}^{\tau}_{\phi},{\bf K}]\, \Big] ,\nonumber \\
&&{}\nonumber \\
\pi (T_s ,\vec \sigma )&=&-i \int d\tilde q \omega
(q)\sqrt{F^{\tau}(q)
\omega (q)P^{\tau}_{\phi}+{\cal D}_{\vec q} {\bf H}(T_s , \vec q) }\nonumber \\
&&\left[ e^{i[\omega (q) {\tilde X}_{\phi}^\tau + \vec q\cdot\vec
\sigma ] +i\int d\tilde k\int d\tilde k^{\prime }\, {\bf K}( T_s
,\vec k) {\cal G}\left( \vec k,\vec k^{\prime }\right) \Delta
\left( \vec k^{\prime }
,\vec q\right)  }\right. - \nonumber \\
&&-\left. e^{-i[\omega \left( q\right) {\tilde X}_{\phi}^\tau
+\vec q\cdot \vec \sigma ] -i\int d\tilde k\int d\tilde k^{\prime
}\, {\bf K}\left( T_s ,\vec k\right) {\cal G}\left( \vec k,\vec
k^{\prime }\right) \Delta \left( \vec
k^{\prime },\vec q\right) }\right] =\nonumber \\
&=&-2 \int d\tilde q\, \omega (q) {\bf A}_{\vec q}(T_s;P^{\tau}
_{\phi},{\bf H}] \, sin\, \Big[ \vec q\cdot \vec \sigma +{\bf
B}_{\vec q}(T_s; {\tilde X}^{\tau}_{\phi},{\bf K}]\, \Big] ,
 \end{eqnarray*}

\begin{eqnarray}
 {\bf A}_{\vec q}(T_s;P^{\tau}_{\phi},{\bf H}]&=& \sqrt{
F^{\tau}(q) \omega (q) P^{\tau}_{\phi}+ {\cal D}_{\vec q} {\bf
H}(T_s,\vec q)}=
\sqrt{I(T_s,\vec q)},\nonumber \\
{\bf B}_{\vec q}(T_s;X^{\tau}_{\phi},{\bf K}]&=& \int d\tilde k
d\tilde k^{'}\, {\bf K}(T_s,\vec k){\cal G}(\vec k,{\vec k}^{'})
\triangle
({\vec k}^{'},\vec q) +\omega (q) {\tilde X}^{\tau}_{\phi}=\nonumber \\
&=&\varphi (T_s,\vec q)-\omega (q) T_s.
 \label{3.70}
\end{eqnarray}

The Hamiltonian  $H_D = M_{\phi}=P^{\tau}_{\phi}$ generates the
following evolution in $T_s$ ($\stackrel{\circ }{=}$ means
evaluated on the equations of motion)

\begin{eqnarray*}
&&{{\partial}\over {\partial T_s}}\, X^{\tau}_{\phi}\, {\buildrel
\circ \over =}\, \{ X^{\tau}_{\phi},P^{\tau}_{\phi} \}
=-1,\quad\quad \Rightarrow \quad X^{\tau}_{\phi}\, {\buildrel
\circ \over =}\, -T_s,
\nonumber \\
&&{{\partial}\over {\partial T_s}}\, P^{\tau}_{\phi}\, {\buildrel
\circ \over
=}\, 0,\nonumber \\
&&{{\partial}\over {\partial T_s}}\, {\bf H}(T_s,\vec q)\,
{\buildrel \circ \over =}\, 0,\qquad {{\partial}\over {\partial
T_s}}\, {\bf K}(T_s,\vec q)\, {\buildrel \circ
\over =}\, 0,\nonumber \\
&&\Rightarrow \,\, {{\partial}\over {\partial T_s}}{\bf A}_{\vec
q}(T_s;P ^{\tau}_{\phi},{\bf H}]={{\partial}\over {\partial
T_s}}{\bf B}_{\vec q}(T_s;{\tilde X}^{\tau}_{\phi},{\bf K}]\,
{\buildrel \circ \over =}\, 0, \end{eqnarray*}

\bea
 &&{{\partial}\over {\partial T_s}}\, \phi (T_s,\vec \sigma )\,
{\buildrel \circ \over =}\, -{{\partial}\over {\partial
X^{\tau}_{\phi}}}\, \phi (T_s,
\vec \sigma )=\pi (T_s,\vec \sigma ),\nonumber \\
&&{{\partial}\over {\partial T_s}}\, \pi (T_s,\vec \sigma )\,
{\buildrel \circ \over =}\, -{{\partial}\over {\partial
X^{\tau}_{\phi}}}\, \pi (T_s,\vec
\sigma )=-[\triangle +m^2]\phi (T_s,\vec \sigma ),\nonumber \\
\Rightarrow&& ({{\partial^2}\over {\partial
T_s^2}}-{{\partial^2}\over {\partial {\vec \sigma}^2}}+m^2) \phi
(T_s,\vec \sigma )\, {\buildrel \circ \over =}\, 0.
 \label{3.71}
\end{eqnarray}

Therefore, in the free case ${\bf H}(T_s,\vec q)$, ${\bf
K}(T_s,\vec q)$ are constants of the motion (complete
integrability and Liouville theorem for the free Klein-Gordon
field). Since the canonical variable $P^{\tau}_{\phi}$ is the
Hamiltonian for the evolution in $T_s\equiv \tau$, we need the
{\it internal} variable $X^{\tau}_{\phi}=\tau +{\tilde
X}^{\tau}_{\phi}$ (i.e. the {\it internal time variable} ${\tilde
X}^{\tau}_{\phi}$) to write Hamilton's equations ${{\partial}\over
{\partial T_s}} F\, {\buildrel \circ \over =}\, \{ F,P^{\tau}
_{\phi} \} =-{{\partial F}\over {\partial X^{\tau}_{\phi}}}=
-{{\partial F}\over {\partial {\tilde X}^{\tau}_{\phi}}}$; in the
free case we have ${{\partial}\over {\partial T_s}}\, {\buildrel
\circ \over =}\, -{{\partial}\over {\partial X^{\tau}_{\phi}}}$ on
$\phi (T_s,\vec \sigma )[X ^{\tau}_{\phi},P^{\tau}_{\phi},{\bf
H},{\bf K}]$ and $\pi (T_s,\vec \sigma )[X
^{\tau}_{\phi},P^{\tau}_{\phi},{\bf H},{\bf K}]$, so that the
evolution in the time $X^{\tau}_{\phi}=T_s+{\tilde
X}^{\tau}_{\phi}$, which takes place inside the Wigner hyperplane
and which can be interpreted as an evolution in the internal time
${\tilde X}^{\tau}_{\phi}$, is equal and opposite to the evolution
in the rest-frame time $T_s$ from a Wigner hyperplane to the next
one in the free case.

\medskip

By adding the two second-class constraints $X^{\tau}_{\phi}-T_s
={\tilde X} ^{\tau}_{\phi} \approx 0$, $P^{\tau}_{\phi}-const.
\approx 0$, and by going to Dirac brackets, we get the rest-frame
Hamilton-Jacobi formulation corresponding to the given constant
value of the total energy: the field $\phi (T_s,\vec \sigma )$,
which is $T_s$-independent depending only upon the internal time
${\tilde X}^{\tau}_{\phi}$, becomes now even ${\tilde
X}^{\tau}_{\phi}$- independent. We find in this way a symplectic
subspace (spanned by the canonical variables ${\bf H}$, ${\bf K}$)
of each constant energy ($P^{\tau} _{\phi}=const.$) surface of the
Klein-Gordon field. {\it Each constant energy surface is not a
symplectic manifold. However, it turns out to be the disjoint
union (over ${\tilde X}^{\tau}_{\phi}$) of the symplectic
manifolds determined by ${\tilde X}^{\tau}_{\phi}=const.$,
$P^{\tau}_{\phi}=const.$}
\bigskip

While the definitions of Subsection IIIC of the three external 4-
and 3-center of mass is not changed, the three internal 3-centers
of mass ${\vec q}_{\phi}$, ${\vec R}_{\phi}$ and ${\vec Y}_{\phi}$
of Subsection IIID have to be built by using the internal
Poincare' generators (\ref{3.64}) in Eqs.
(\ref{3.16})-(\ref{3.19}). In particular, the boosts are ${\vec
K}_{\phi} = {\tilde X}_{\phi}\, {\vec P}_{\phi} - M_{\phi}\, {\vec
X}_{\phi} + {\vec K}_{S, \phi} {\buildrel {def}\over =}\, -
M_{\phi}\, {\vec R}_{\phi}$ with $K^r_{S, \phi} = - {\hat S}^{\tau
r}_{\phi}$. Again, ${\vec P}_{\phi} \approx 0$ implies ${\vec
q}_{\phi} \approx {\vec R}_{\phi} \approx {\vec Y}_{\phi}$, the
gauge fixing ${\vec q}_{\phi} \approx 0$ implies ${\vec K}_{\phi}
\approx 0$, $\vec \lambda (\tau ) = 0$, and the selection of the
centroid (\ref{3.20}) that coincides with the external
Fokker-Pryce 4-center of inertia. The 4-momentum of the field
configuration is peaked on this world-line while the canonical
variables ${\bf H}^{'}(\tau ,\vec q)$, ${\bf K}^{'}(\tau ,\vec q)$
characterize the relative motions with respect to the {\it
monopole} configuration of Eq.(\ref{3.69}) describing the center
of mass of the field configuration.
\medskip

As in the N-body case (${\vec \eta}_+ \not= {\vec q}_+$), the
canonical 3-center of mass ${\vec q}_{\phi}$ does not coincide
with ${\vec X}_{\phi}$, which in turn could be better defined as
the {\it center of phase} of the field configuration. While the
gauge fixing ${\vec X}_{\phi} \approx 0$, used to get
Eqs.(\ref{3.70}), implies ${\vec K}_{\phi} \approx {\vec K}_{S,
\phi} \not= 0$, the gauge fixing ${\vec q}_{\phi} \approx 0$
implies ${\vec X}_{\phi} \approx - {\vec K}_{S,\phi}/ M_{\phi}$
(in fact it is M$\o$ller's definition of the 3-center of energy
but only in terms of the spin part of the boost).
 \medskip

Note that, like in Subsection IIIE for the N-body case, there
should exist a canonical transformation from the canonical basis
${\tilde X}^{\tau}_{\phi}$, $M_{\phi} = P^{\tau}_{\phi}$, ${\vec
X}_{\phi}$, ${\vec P} _{\phi} \approx 0$, ${\bf H}(\tau ,\vec k)$,
${\bf K}(\tau ,\vec k)$, to a new basis $q^{\tau}_{\phi}$,
$M^q_{\phi} =\sqrt{ (P^{\tau}_{\phi})^2-{\vec P}^2 _{\phi}}
\approx M_{\phi}$ (since $\{ {\vec q}_{\phi},P^{\tau}_{\phi} \} =
{\vec P}_{\phi}/P^{\tau}_{\phi} \approx 0$ is only weakly zero),
${\vec q}_{\phi}$, ${\vec P}_{\phi} \approx 0$, ${\bf H}_q(\tau
,\vec k)$, ${\bf K}_q(\tau ,\vec k)$ containing relative variables
with respect to the true center of mass of the field configuration

\bea
 &&\begin{minipage}[t]{4cm}
\begin{tabular}{|l|l|l|} \hline
${\tilde X}_{\phi}$ & ${\vec X}_{\phi}$ & ${\bf H}(\tau  ,\vec k)$\\
\hline
 $M_{\phi}$ & ${\vec P}_{\phi}\approx 0$ & ${\bf K}(\tau ,\vec k)$\\ \hline
\end{tabular}
\end{minipage}
\ {\longrightarrow \hspace{.2cm}}\
\begin{minipage}[b]{4cm}
\begin{tabular}{|l|l|l|} \hline
$q^{\tau}_{\phi}$ & ${\vec q}_{\phi}$ & ${\bf H}_q(\tau ,\vec k)$ \\
\hline
 $M^q_{\phi}\approx M_{\phi}$ & ${\vec P}_{\phi}\approx 0$ &
 ${\bf K}_q(\tau ,\vec k)$ \\ \hline
\end{tabular}
\end{minipage}.
 \label{3.72}
 \eea

It does not seem easy, however, to characterize this final
canonical basis. Besides the extension of the Gartenhaus-Schwartz
methods from particles to fields (so that ${\bf H}_q \approx {\bf
H}$, ${\bf K}_q \approx {\bf K}$), the real problem is finding the
new {\it internal time} variable $q^{\tau}_{\phi}$.
\medskip

In final canonical basis we would have still Eqs.(\ref{3.70}) but
with ${\bf H}_q(\tau ,\vec k)$ and  ${\bf K}_q(\tau ,\vec k)$
replacing  ${\bf H}(\tau ,\vec k)$  and  ${\bf K}(\tau ,\vec k)$
and with $q^{\tau}_{\phi}$ replacing ${\tilde X}_{\phi}$ as
internal time. Since both the gauge fixings ${\vec q}_{\phi}
\approx 0$ and ${\vec X}_{\phi} \approx 0$ give $\vec \lambda
(\tau ) = 0$, both of them identify the same centroid (\ref{3.20})
but lead to different internal times and relative variables
connected by the canonical transformation (\ref{3.72}).

\bigskip

It is under investigation \cite{52} the problem of characterizing
the configurations of the real Klein-Gordon field in terms of
dynamical body frames and canonical spin bases, to the effect of
finding its orientation-shape variables. Ref.\cite{26} contains
also the treatment of the coupling of the real Klein-Gordon field
to scalar particles and that of the charged (complex) Klein-Gordon
field to electro-magnetic field. The collective and relative
variables of the electro-magnetic field are now under
investigation: they could be relevant for the problem of phases in
optics and laser physics \cite{53}. Finally, Ref.\cite{27}
contains the analysis of relativistic perfect fluids along these
lines.

\section{The multipolar expansion.}

In practice one is neither able to follow the motion of the
particles of an open cluster in interaction with the environment
nor to describe extended continuous bodies (for instance a
satellite or a star), unless either the cluster or the body is
approximated with a multi-polar expansion (often a pole-dipole
approximation is enough).

In this Section, after the treatment of the non-relativistic case
in Subsection A, we show that the rest-frame instant form provides
the natural framework for studying relativistic multipolar
expansions of N free particles (Subsection B), of open clusters of
particles (Subsection C) and of the classical Klein-Gordon field
(Subsection D) as a prototype of extended continuous systems
(perfect fluids, electro-magnetic field,...).

\subsection{The non-relativistic case.}

The review paper of Ref.\cite{29} contains the Newtonian
multipolar expansions for a continuum isentropic distribution of
matter characterized by a mass density $\rho (t,\vec \sigma )$,  a
velocity field $U^r(t,\vec \sigma )$, and a stress tensor
$\sigma^{rs}(t,\vec \sigma )$, with $\rho (t,\vec \sigma ) \vec
U(t,\vec \sigma )$ the momentum density. In case the system is
isolated, the only dynamical equations are the mass conservation
and the continuity equations of motion
\medskip

\bea
 &&{{\partial \rho (t,\vec \sigma )}\over {\partial
t}}-{{\partial \rho (t,\vec \sigma ) U^r(t,\vec \sigma )}\over
{\partial \sigma^r}}=0,\nonumber  \\
 &&{{\partial \rho (t,\vec \sigma )U^r(t,\vec \sigma )}\over {\partial
t}}-{{\partial [\rho U^r U^s -\sigma^{rs}](t,\vec \sigma )}\over
{\partial \sigma^s}}\, {\buildrel \circ \over =}\, 0,
 \label{4.1}
\eea

\noindent respectively.

This description can be adapted to an isolated system of N
particles in the following way. The mass density

\begin{equation}
\rho (t,\vec \sigma )=\sum_{i=1}^N m_i \delta^3(\vec \sigma -{\vec
\eta}_i(t)),
 \label{4.2}
\end{equation}

\noindent satisfies

\begin{equation}
{{\partial \rho (t,\vec \sigma )}\over {\partial
t}}=-\sum_{i=1}^Nm_i {\dot {\vec \eta}}_i(t)\cdot {\vec
\partial}_{{\vec \eta}_i} \delta^3(\vec \sigma -{\vec
\eta}_i(t))\, {\buildrel {def} \over =}\, {{\partial}\over
{\partial \sigma^r}}[\rho U^r](t,\vec \sigma ),
 \label{4.3}
\end{equation}

\noindent while the momentum density (this can be taken as the
definitory equation for the velocity field) is

\begin{equation}
\rho (t,\vec \sigma ) U^r(t,\vec \sigma )= \sum_{i=1}^N m_i {\dot
{\vec \eta}}_i(t) \delta^3(\vec \sigma -{\vec \eta}_i(t )).
 \label{4.4}
\end{equation}

\noindent The associated constant of motion is the total mass
$m=\sum_{i=1}^N$.
\medskip

Introducing a function $\zeta (\vec \sigma ,{\vec \eta}_i)$
concentrated in the N points ${\vec \eta}_i$, i=1,..,N, such that
$\zeta (\vec \sigma ,{\vec \eta}_i)=0$ for $\vec \sigma \not=
{\vec \eta}_i$ and $\zeta ({\vec \eta}_i,{\vec
\eta}_j)=\delta_{ij}$ (it is a limit concept deriving from the
characteristic function of a manifold), the velocity field
associated to N particles becomes

\begin{equation}
\vec U(t,\vec \sigma )= \sum_{i=1}^N {\dot {\vec \eta}}_i(t) \zeta
(\vec \sigma ,{\vec \eta}_i(t)).
 \label{4.5}
\end{equation}

The continuity equations of motion are replaced by

\begin{eqnarray}
&&{{\partial}\over {\partial t}} [\rho (t,\vec \sigma ) U^r(t,\vec
\sigma )]\, {\buildrel \circ \over =}\, {{\partial}\over {\partial
\sigma^s}} \sum_{i=1}^N m_i {\dot \eta}^r_i(t) {\dot \eta}^s_i(t)
\delta^3(\vec \sigma -{\vec \eta}_i(t)) +\sum_{i=1}^N m_i {\ddot
\eta}^r_i(t)=\nonumber \\
 &&{\buildrel {def} \over =}\,
 {{\partial [\rho U^r U^s -\sigma^{rs}](t,\vec \sigma )}\over
{\partial \sigma^s}}.
 \label{4.6}
\end{eqnarray}

For a system of free particles we have ${\ddot {\vec \eta}}_i(t)\,
{\buildrel \circ \over =}\, 0$ so that $\sigma^{rs}(t,\vec \sigma
)=0$. If there are inter-particle interactions, they will
determine the effective stress tensor.

\bigskip
Consider now an arbitrary point $\vec \eta (t)$. The {\it
multipole moments} of the mass density $\rho$, the momentum
density $\rho \vec U$ and the stress-like density $\rho U^rU^s$,
with respect to the point $\vec \eta (t)$, are defined by setting
($N \geq 0$)

\begin{eqnarray}
 m^{r_1...r_n}[\vec \eta (t) ]&=& \int d^3\sigma
[\sigma^{r_1}-\eta^{r_1}(t)]...[\sigma^{r_n}-\eta^{r_n}(t)]\rho
(\tau ,\vec \sigma )=\nonumber \\
 &=&\sum_{i=1}^N m_i
[\eta^{r_1}_i(\tau
)-\eta^{r_1}(t)]...[\eta^{r_n}_i(t)-\eta^{r_n}(t)],
 \nonumber \\
 &&n=0\quad\quad m[\vec \eta (t)]=m=\sum_{i=1}^N m_i,\nonumber \\
 &&{}\nonumber \\
 p^{r_1...r_nr}[\vec \eta (t) ]&=& \int d^3\sigma
[\sigma^{r_1}-\eta^{r_1}(t)]...[\sigma^{r_n}-\eta^{r_n}(t)]\rho
(t,\vec \sigma )U^r(t,\vec \sigma )=\nonumber \\
 &=&\sum_{i=1}^N m_i {\dot \eta}^r_i(t)
[\eta^{r_1}_i(t)-\eta^{r_1}(t)]...[\eta^{r_n}_i(t)-\eta^{r_n}(t)],
\nonumber \\
 && n=0\quad\quad p^r[\vec \eta (t)]=\sum_{i=1}^Nm_i{\dot \eta}_i(t)
 =\sum_{i=1}^N\kappa_i^r=\kappa^r_{+}\approx 0,\nonumber \\
  &&{}\nonumber \\
  p^{r_1...r_nrs}[\vec \eta (t) ]&=& \int d^3\sigma
[\sigma^{r_1}-\eta^{r_1}(t)]...[\sigma^{r_n}-\eta^{r_n}(t)]\rho
(t,\vec \sigma )U^r(t,\vec \sigma )U^s(t,\vec \sigma )=\nonumber
\\
 &=&\sum_{i=1}^N m_i {\dot \eta}^r_i(t){\dot \eta}^s_i(t)
[\eta^{r_1}_i(t)-\eta^{r_1}(t)]...[\eta^{r_n}_i(t)-\eta^{r_n}(t)].
 \label{4.7}
 \end{eqnarray}

The {\it mass monopole} is the conserved mass, while the {\it
momentum monopole} is the total 3-momentum, vanishing in the rest
frame.
\medskip

If the {\it mass dipole} vanishes, the point $\vec \eta (t)$ is
the {\it center of mass}:

\begin{equation}
m^r[\vec \eta (t)]=\sum_{i=1}^Nm_i[\eta^r_i(t)-\eta^r(t)]=0
\Rightarrow \vec \eta (t)={\vec q}_{nr}.
 \label{4.8}
\end{equation}

The time derivative of the mass dipole is

\begin{equation}
{{d m^r[\vec \eta (t)]}\over {dt}} = p^r[\vec \eta (t)]-m {\dot
\eta}^r(t)=\kappa_{+}^r-m{\dot \eta}^r(t).
 \label{4.9}
\end{equation}

When $\vec \eta (t)={\vec q}_{nr}$, from the vanishing of this
time derivative we get the {\it momentum-velocity relation for the
center of mass}

\begin{equation}
p^r[{\vec q}_{nr}]=\kappa^r_{+} = m {\dot q}_{+}^r \quad [\approx
0\, \, \, in\, the\, rest\, frame].
 \label{4.10}
\end{equation}

The {\it mass quadrupole} is

\begin{equation}
m^{rs}[\vec \eta
(t)]=\sum_{i=1}^Nm_i\eta^r_i(t)\eta^s_i(t)-m\eta^r(t)\eta^s(t)-
\Big( \eta^r(t)m^s[\vec \eta (t)]+\eta^s(t)m^r[\vec \eta
(t)]\Big),
 \label{4.11}
\end{equation}

\noindent so that the {\it barycentric mass quadrupole and  tensor
of inertia} are, respectively

\begin{eqnarray}
m^{rs}[{\vec q}_{nr}]&=&\sum_{i=1}^N m_i\eta^r_i(t)\eta^s_i(t)-m
q^r_{nr}q^s_{nr},\nonumber \\
 &&{}\nonumber \\
 I^{rs}[{\vec q}_{nr}]&=&\delta^{rs} \sum_um^{uu}[{\vec q}_{nr}]-m^{rs}[{\vec
 q}_{nr}]=\nonumber \\
 &=&\sum_{i=1}m_i[\delta^{rs} {\vec \eta}_i^2(t)-\eta^r_i(t)\eta^s_i(t)]-
 m[\delta^{rs}{\vec q}^2_{nr}-q^r_{nr}q^s_{nr}]=\nonumber \\
 &=&\sum_{a,b}^{1...N-1}k_{ab}({\vec \rho}_a\cdot {\vec
\rho}_b\delta^{rs}-\rho^r_a\rho^s_b) ,\nonumber \\ \Rightarrow&&
m^{rs}[{\vec q}_{nr}]=\delta^{rs}\sum_{a,b=1}^{N-1}k_{ab}\, {\vec
\rho}_a\cdot {\vec \rho}_b-I^{rs}[{\vec q}_{nr}].
 \label{4.12}
\end{eqnarray}

The antisymmetric part of the barycentric momentum dipole gives
rise to the {\it spin vector} in the following way

\begin{eqnarray}
p^{rs}[{\vec q}_{nr}]&=&\sum_{i=1}^Nm_i\eta^r_i(t){\dot
\eta}^s_i(t)-q^r_{nr}p^s[{\vec
q}_{nr}]=\sum_{i=1}^N\eta^r_i(t)\kappa_i^s(t)-
q^r_{+}\kappa^s_{+},\nonumber \\
 &&{}\nonumber \\
 S^u&=&{1\over 2}\epsilon^{urs} p^{rs}[{\vec q}_{nr}]=\sum_{a=1}^{N-1}
 ({\vec \rho}_a\times {\vec \pi}_{qa})^u.
 \label{4.13}
\end{eqnarray}

The {\it multipolar expansions} of the mass and momentum densities
around the point  $\vec \eta (t)$ are

\begin{eqnarray}
\rho (t,\vec \sigma )&=& \sum_{n=0}^{\infty} {{m^{r_1....r_n}[\vec
\eta ]}\over {n!}} {{\partial^n}\over {\partial
\sigma^{r_1}...\partial \sigma^{r_n}}} \delta^3(\vec \sigma -{\vec
\eta }(t)),\nonumber \\ &&{}\nonumber \\ \rho (t,\vec \sigma )
U^r(t,\vec \sigma )
 &=&\sum_{n=0}^{\infty}{{p^{r_1....r_nr}[\vec \eta ]}\over {n!}}
{{\partial^n}\over {\partial \sigma^{r_1}...\partial
\sigma^{r_n}}} \delta^3(\vec \sigma -{\vec \eta }(t)).
 \label{4.14}
\end{eqnarray}

Finally, we get the {\it barycentric multipolar expansions} as

\begin{eqnarray}
\rho (t,\vec \sigma )&=&m \delta^3(\vec \sigma -{\vec q}_{nr})-
{1\over 2}(I^{rs}[{\vec q}_{nr}]-{1\over
2}\delta^{rs}\sum_uI^{uu}[{\vec q}_{nr}]){{\partial^2}\over
{\partial \sigma^r\partial \sigma^s}} \delta^3(\vec \sigma -{\vec
q}_{nr})+ \nonumber \\ &+&\sum_{n=3}^{\infty}
{{m^{r_1....r_n}[{\vec q}_{nr}]}\over {n!}} {{\partial^n}\over
{\partial \sigma^{r_1}...\partial \sigma^{r_n}}} \delta^3(\vec
\sigma -{\vec q}_{nr}),\nonumber \\ &&{}\nonumber \\ \rho (t,\vec
\sigma ) U^r(t,\vec \sigma )&=&\kappa^r_{+} \delta^3(\vec \sigma
-{\vec q}_{nr})+\Big[ {1\over 2}\epsilon^{rsu}S^u+p^{(sr)}[{\vec
q}_{nr}] \Big] {{\partial}\over {\partial \sigma^s}} \delta^3(\vec
\sigma - {\vec q}_{nr})+\nonumber \\
 &+&\sum_{n=2}^{\infty}{{p^{r_1....r_nr}[{\vec q}_{nr}]}\over {n!}}
{{\partial^n}\over {\partial \sigma^{r_1}...\partial
\sigma^{r_n}}} \delta^3(\vec \sigma -{\vec q}_{nr}).
 \label{4.15}
\end{eqnarray}

\subsection{The relativistic case on a Wigner hyper-plane.}

It is shown in Ref.\cite{12} that, on a Wigner hyperplane with
$\tau \equiv T_s$, the energy-momentum of N free scalar particles
has the form

\begin{eqnarray}
T^{\mu\nu}[x^{\beta}_s(T_s)+\epsilon^{\beta}_u(u(p_s))\sigma^u]&=&
\epsilon^{\mu}_A(u(p_s)) \epsilon^{\nu}_B(u(p_s)) T^{AB}(T_s,\vec
\sigma ) =\nonumber \\
 &=&\sum_{i=1}^N \delta^3(\vec \sigma -{\vec \eta}_i(T_s))
\Big[\sqrt{m_i^2+{\vec \kappa}^2_i(T_s)}
u^{\mu}(p_s)u^{\nu}(p_s)+\nonumber \\
&+&k^r_i(T_s)\Big(u^{\mu}(p_s)\epsilon^{\nu}_r(u(p_s))
+u^{\nu}(p_s)\epsilon^{\mu}_r(u(p_s))\Big) +\nonumber \\
&+&{{\kappa^r_i(T_s)\kappa^s_i(T_s)}\over {\sqrt{m_i^2+{\vec
\kappa}^2 _i(T_s)} }}
\epsilon^{\mu}_r(u(p_s))\epsilon^{\nu}_s(u(p_s))\Big] , \nonumber
\\ T^{\tau\tau}(T_s,\vec \sigma )&=&\sum_{i=1}^N\delta^3(\vec
\sigma -{\vec \eta}_i(T_s)) \sqrt{m_i^2+{\vec
\kappa}_i^2(T_s))},\nonumber \\ T^{\tau r}(T_s,\vec \sigma
)&=&\sum_{i=1}^N\delta^3(\vec \sigma -{\vec \eta}_i(T_s))
\kappa_i^r(T_s),\nonumber \\ T^{rs}(T_s,\vec \sigma
)&=&\sum_{i=1}^N\delta^3(\vec \sigma -{\vec \eta}_i(T_s)) {{
\kappa_i^r(T_s) \kappa_i^s(T_s)}\over { \sqrt{m_i^2+{\vec
\kappa}_i^2(T_s)}}},\nonumber \\
 &&{}\nonumber \\
P^{\mu}_T&=&p^{\mu}_s=M_{sys}\,
u^{\mu}(p_s)+\epsilon^{\mu}_r(u(p_s))\kappa^r_{+} \approx
M_{sys}\, u^{\mu}(p_s),\nonumber \\
 M_{sys}&=&\sum_{i=1}^N
\sqrt{m^2_ic^2+{\vec \kappa}_i^2(T_s)}.
 \label{4.16}
 \end{eqnarray}
\medskip

We will now define the special relativistic Dixon multi-poles on
the Wigner hyperplane with $T_s - \tau \equiv 0$ for the N-body
problem (see the bibliography of Ref.\cite{11} for older
attempts). Note that, since we have not yet added the gauge
fixings ${\vec q}_+ \approx 0$, the centroid origin of the
3-coordinates has the form $x^{\mu}_s(\tau ) = x_s^{({\vec
q}_{+})\mu}(\tau ) + \int_o^{\tau} d\tau_1\, \lambda_r(\tau_1)$
according to Eqs.(\ref{3.7}) and (\ref{3.20}).
\medskip

Consider an arbitrary time-like world-line $w^{\mu}(\tau
)=z^{\mu}(\tau ,\vec \eta (\tau ))= x^{\mu}_s(\tau ) +
\epsilon^{\mu}_r(u(p_s))\, \eta^r(\tau ) = x_s^{({\vec
q}_{+})\mu}(\tau ) + \epsilon^{\mu}_r(u(p_s))\, {\tilde
\eta}^r(\tau )$ [${\tilde \eta}^r(\tau ) = \eta^r(\tau ) +
\int_o^{\tau} d\tau_1\, \lambda_r(\tau_1)$] and evaluate the Dixon
multi-poles \cite{28} on the Wigner hyper-planes in the natural
gauge with respect to the given world-line. A generic point will
be parametrized by

\begin{eqnarray}
z^{\mu}(\tau ,\vec \sigma ) &=& x^{\mu}_s(\tau ) +
\epsilon^{\mu}_r(u(p_s))\, \sigma^r =\nonumber \\
 &=& w^{\mu}(\tau ) + \epsilon^{\mu}_r(u(p_s)) [\sigma^r-\eta^r(\tau
)]\, {\buildrel {def} \over =}\, w^{\mu}(\tau )+\delta
z^{\mu}(\tau ,\vec \sigma ),
 \label{4.17}
\end{eqnarray}

\noindent so that $\delta z_{\mu}(\tau ,\vec \sigma
)u^{\mu}(p_s)=0$.\medskip

While for ${\vec {\tilde \eta}} (\tau )=0$  [$\vec \eta (\tau ) =
\int_o^{\tau} d\tau_1\, \lambda_r(\tau_1)$] we get the multi-poles
relative to the centroid $x^{\mu}_s(\tau )$, for $\vec \eta (\tau
) = 0$ we get those relative to the centroid $x_s^{({\vec
q}_{+})\mu}(\tau )$. In the gauge ${\vec R}_{+} \approx {\vec
q}_{+} \approx {\vec y}_{+} \approx 0$, where $\vec \lambda (\tau
) = 0$, it follows that $\vec \eta (\tau ) = {\vec {\tilde
\eta}}(\tau ) = 0$ identifies the {\it barycentric} multi-poles
with respect to the centroid $x_s^{({\vec q}_{+})\mu}(\tau )$,
that now carries the internal 3-center of mass.
\medskip

Lorentz covariant {\it Dixon's multi-poles} and their Wigner
covariant counterparts on the Wigner hyper-planes are then defined
as
\medskip

\begin{eqnarray*}
t_T^{\mu_1...\mu_n\mu\nu}(T_s,\vec \eta
)&=&t_T^{(\mu_1...\mu_n)(\mu\nu)}(T_s,\vec \eta )= \nonumber \\
 &&{}\nonumber \\
 &=&\epsilon^{\mu_1}_{r_1}(u(p_s))...\epsilon^{\mu_n}_{r_n}(u(p_s))\,
 \epsilon^{\mu}_A(u(p_s))\epsilon^{\nu}_B(u(p_s)) q_T^{r_1..r_nAB}(T_s,\vec \eta )
 =\nonumber \\
 &&{}\nonumber \\
&=&\int d^3\sigma \delta z^{\mu_1}(T_s,\vec \sigma )...\delta
z^{\mu_n}(T_s,\vec \sigma ) T^{\mu\nu}[x^{({\vec
q}_{+})\beta}_s(T_s)+\epsilon^{\beta}_u(u(p_s))
\sigma^u]=\nonumber \\
&=&\epsilon^{\mu}_A(u(p_s))\epsilon^{\nu}_B(u(p_s)) \int d^3\sigma
\delta z^{\mu_1}(T_s,\vec \sigma )....\delta z^{\mu_n}(T_s,\vec
\sigma ) T^{AB}(T_s,\vec \sigma )=\nonumber \\
 &&{}\nonumber \\
&=&\epsilon^{\mu_1}_{r_1}(u(p_s))...\epsilon^{\mu_n}_{r_n}(u(p_s))
\nonumber \\
 &&\Big[ u^{\mu}(p_s) u^{\nu}(p_s)
\sum_{i=1}^N[\eta^{r_1}_i(T_s)-\eta^{r_1}(T_s)]...[\eta^{r_n}
_i(T_s)-\eta^{r_n}(T_s)]\sqrt{m^2_i+{\vec
\kappa}^2_i(T_s)}+\nonumber \\
 &+&\epsilon^{\mu}_r(u(p_s))\epsilon^{\nu}_s(u(p_s))\nonumber \\
 &&\sum_{i=1}^N[\eta^{r_1}
_i(T_s)-\eta^{r_1}(T_s)]...[\eta^{r_n}_i(T_s)-\eta^{r_n}(T_s)]
{{\kappa^r_i(T_s)\kappa^s_i(T_s)}\over {\sqrt{m_i^2+{\vec
\kappa}^2_i(T_s)}}}+\nonumber \\
&+&[u^{\mu}(p_s)\epsilon^{\nu}_r(u(p_s))+u^{\nu}(p_s)\epsilon^{\mu}_r(u(p_s))]
  \nonumber \\
&&\sum_{i=1}^N[\eta^{r_1}_i(T_s)-\eta^{r_1}(T_s)]...[\eta^{r_n}_i(T_s)-
\eta^{r_n}(T_s)]\kappa^r_i(T_s)\Big],
 \end{eqnarray*}

\begin{eqnarray*}
 q_T^{r_1...r_nAB}(T_s,\vec \eta ) &=& \int d^3\sigma\,
 [\sigma^{r_1} - \eta^{r_1}(T_s)] ... [\sigma^{r_n} -
 \eta^{r_n}]\, T^{AB}(T_s, \vec \sigma ) =
 \end{eqnarray*}

\bea
 &=&\delta^A_{\tau}\delta^B_{\tau}
 \sum_{i=1}^N[\eta^{r_1}_i(T_s)-\eta^{r_1}(T_s)]...[\eta^{r_n}
_i(T_s)-\eta^{r_n}(T_s)]\sqrt{m^2_i+{\vec
\kappa}^2_i(T_s)}+\nonumber \\ &+&\delta^A_u\delta^B_v
\sum_{i=1}^N[\eta^{r_1}
_i(T_s)-\eta^{r_1}(T_s)]...[\eta^{r_n}_i(T_s)-\eta^{r_n}(T_s)]
{{\kappa^u_i(T_s)\kappa^v_i(T_s)}\over {\sqrt{m_i^2+{\vec
\kappa}^2_i(T_s)}}}+\nonumber \\
&+&(\delta^A_{\tau}\delta^B_u+\delta^A_u\delta^B_{\tau})
\sum_{i=1}^N[\eta^{r_1}_i(T_s)-\eta^{r_1}(T_s)]...[\eta^{r_n}_i(T_s)-
\eta^{r_n}(T_s)]\kappa^r_i(T_s),\nonumber \\
 &&{}\nonumber \\
 u_{\mu_1}(p_s)&& t_T^{\mu_1...\mu_n\mu\nu}(T_s,\vec \eta )=0.
\label{4.18}
\end{eqnarray}

\medskip

\noindent Related multi-poles are $p_T^{\mu_1..\mu_n\mu}(T_s,\vec
\eta ) = t_T^{\mu_1...\mu_n\mu\nu}(T_s,\vec \eta )\, u_{\nu}(p_s)
=$ $\epsilon^{\mu_1}_{r_1}(u(p_s)) ...
\epsilon^{\mu_n}_{r_n}(u(p_s)) \epsilon^{\mu}_A(u(p_s))\,
q_T^{r_1...r_nA\tau}(T_s,\vec \eta )$. They satisfy
$u_{\mu_1}(p_s) p_T^{\mu_1...\mu_n\mu}(T_s,\vec \eta )=0$ and for
$n = 0$ they imply $p^{\mu}_T(T_s,\vec \eta
)=\epsilon^{\mu}_A(u(p_s)) q_T^{A\tau}(T_s)=P^{\mu}_T\approx
p^{\mu}_s$.

\medskip

The inverse formulas, giving the {\it multipolar expansion}, are
\medskip

\begin{eqnarray}
T^{\mu\nu}[w^{\beta}(T_s)+\delta z^{\beta}(T_s,\vec \sigma )]&=&
T^{\mu\nu}[x^{({\vec q}_{+})\beta}_s(T_s) +
\epsilon^{\beta}_r(u(p_s))\, \sigma^r]=\nonumber \\
 &&{}\nonumber \\
 &=&\epsilon^{\mu}_A(u(p_s))\epsilon^{\nu}_B(u(p_s)) T^{AB}(T_s,\vec \sigma )
 =\nonumber \\
  &&{}\nonumber \\
  =\epsilon^{\mu}_A(u(p_s))\epsilon^{\nu}_B(u(p_s)) \sum_{n=0}^{\infty}
  (-1)^n\, {{ q_T^{r_1...r_nAB}(T_s,\vec \eta )}\over {n!}}
  &&{{\partial^n}\over
  {\partial \sigma^{r_1}...\partial \sigma^{r_n}}} \delta^3(\vec \sigma -
  \vec \eta (T_s)).
 \label{4.19}
\end{eqnarray}
\medskip

Note however that, as pointed out by Dixon \cite{28}, the
distributional equation (\ref{4.19}) is valid only if analytic
test functions are used defined on the support of the
energy-momentum tensor.

\bigskip

The quantities $q_T^{r_1...r_n\tau\tau}(T_s,\vec \eta )$,
$q_T^{r_1...r_n r\tau}(T_s,\vec \eta )=q_T^{r_1...r_n \tau
r}(T_s,\vec \eta )$, $q_T^{r_1...r_n uv}(T_s,\vec \eta )$ are the
{\it mass density, momentum density and stress tensor multi-poles}
with respect to the world-line $w^{\mu}(T_s)$  (barycentric for
$\vec \eta = {\vec {\tilde \eta}} = 0$).

\bigskip

\subsubsection{ Monopoles}

\medskip

The {\it monopoles}, corresponding to $n=0$, have the following
expression (they are $\vec \eta$-independent)
\medskip

\begin{eqnarray*}
q^{AB}_T(T_s,\vec \eta )&=&\delta^A_{\tau}\delta^B_{\tau} M
+\delta^A_u\delta^B_v\sum_{i=1}^N{{\kappa^u_i\kappa^v_i}\over
{\sqrt{m_i^2+{\vec \kappa}_i^2}}}+
(\delta^A_{\tau}\delta^B_u+\delta^A_u\delta^B_{\tau})
\kappa^u_{+},
 \end{eqnarray*}

\bea
 q^{\tau\tau}_T(T_s,\vec \eta )\, &{\rightarrow}_{c \rightarrow \infty}\,&
 \sum_{i=1}^Nm_ic^2+ H_{rel} +O(1/c),\nonumber \\
 &&{}\nonumber \\
 q_T^{r\tau}(T_s,\vec \eta ) &=&\kappa^r_{+} \approx 0,\quad rest-frame\, condition\,
 (also\, at\, the \, non-relativistic\, level),\nonumber \\
 &&{}\nonumber \\
 q^{uv}_T(T_s,\vec \eta )\, &{\rightarrow}_{c \rightarrow \infty}\,&
\sum_{ab}^{1..N-1} k^{-1}_{ab} \pi^u_{qa}\pi^v_{qb} +O(1/c), \nonumber \\
 &&{}\nonumber \\
 q^A_{T A}(T_s,\vec \eta )&=& t^{\mu}_{T \mu}(T_s,\vec \eta )=
 \sum_{i=1}^N {{m^2_i}\over {\sqrt{m_i^2+
 {\vec \kappa}_i^2}}}\nonumber \\
 &{\rightarrow}_{c \rightarrow \infty}\,&
 \sum_{i=1}^Nm_ic^2 -H_{rel} +O(1/c).
 \label{4.20}
\end{eqnarray}

\medskip

Therefore, independently of the choice of the world-line
$w^{\mu}(\tau )$, in the rest-frame instant form the {\it mass
monopole} $q^{\tau\tau}_T$ is the invariant mass $M_{sys} =
\sum_{i=1}^N\sqrt{m_i^2+{\vec \kappa}_i^2}$, while the {\it
momentum monopole} $q^{r\tau}_T$ vanishes and $q^{uv}_T$ is the
{\it stress tensor monopole}.
\bigskip

\subsubsection{ Dipoles}

\medskip

The mass, momentum and stress tensor {\it dipoles}, corresponding
to $n=1$, are

\medskip

\begin{eqnarray}
q_T^{rAB}(T_s,\vec \eta )&=&\delta^A_{\tau}\delta^B_{\tau} M
[R^r_{+}(T_s)-\eta^r(T_s)] +\delta^A_u\delta^B_v\Big[
\sum_{i=1}^N{{\eta_i^r\kappa^u_i\kappa^v_i}\over {\sqrt{m_i^2+
{\vec \kappa}_i^2}}}(T_s)-\eta^r(T_s)q_T^{uv}(T_s,\vec \eta )\Big]
+ \nonumber \\
&+&(\delta^A_{\tau}\delta^B_u+\delta^A_u\delta^B_{\tau}) \Big[
\sum_{i=1}^N[\eta^r_i\kappa_i^u](T_s)-\eta^r(T_s)\kappa^u_{+}\Big].
 \label{4.21}
\end{eqnarray}
\medskip

The vanishing of the {\it mass dipole} $q^{r\tau\tau}_T$ implies
$\vec \eta (\tau ) = {\vec {\tilde \eta}}(\tau ) -
\int_o^{\tau}d\tau_1\, \vec \lambda (\tau_1) = {\vec R}_{+}$ and
identifies the world-line $w^{\mu}(\tau ) = x^{({\vec q}_{+})
\mu}_s(\tau ) + \epsilon^{\mu}_r(u(p_s))\, \Big[ R^r_{+} +
\int_o^{\tau}d\tau_1\, \lambda_r(\tau_1)\Big]$. In the gauge
${\vec R}_{+} \approx {\vec q}_{+} \approx {\vec y}_{+} \approx
0$, where $\vec \lambda (\tau ) = 0$, this is the world-line
$w^{\mu}(\tau ) = x^{({\vec q}_{+}) \mu}_s(\tau )$ of the centroid
associated with the {\it rest-frame internal 3-center of mass}
${\vec q}_{+}$. We have, therefore, the implications following
from the vanishing of the barycentric (i.e. $\vec \lambda (\tau )
= 0$) mass dipole
\medskip

\begin{eqnarray}
q_T^{r\tau\tau}(T_s,\vec \eta )&=&\epsilon^{r_1}_{\mu_1}(u(p_s))
{\tilde t}_T^{\mu_1}(T_s,\vec \eta )= M\, \Big[ R^r_{+}(T_s) -
\eta^r(T_s)\Big] = 0,\quad and\,\, \vec \lambda (\tau ) =
0,\nonumber \\
 &&{}\nonumber \\
 &&\Rightarrow\quad \vec \eta (T_s)= {\vec {\tilde \eta}}(T_s) = {\vec R}_{+}
 \approx {\vec q}_{+} \approx {\vec y}_{+}.
 \label{4.22}
 \end{eqnarray}
\medskip

In the gauge ${\vec R}_{+} \approx {\vec q}_{+} \approx {\vec
y}_{+} \approx 0$,  Eq.(\ref{4.22}) with $\vec \eta = {\vec
{\tilde \eta}} = 0$ implies the vanishing of the time derivative
of the barycentric mass dipole: this identifies the {\it
center-of-mass momentum-velocity relation} (or constitutive
equation) for the system

\begin{equation}
{{d q_T^{r\tau\tau}(T_s,\vec \eta )}\over {dT_s}}\, {\buildrel
\circ \over =}\, \kappa^r_{+}-M {\dot R}^r_{+}\, = 0.
 \label{4.23}
\end{equation}

\medskip

The expression of the barycentric dipoles in terms of the internal
relative variables, when $\vec \eta = {\vec {\tilde \eta}} = {\vec
R}_{+} \approx {\vec q}_{+} \approx 0$ and $\vec{\kappa}_+\approx
0$, is obtained by using the Gartenhaus-Schwartz transformation.
\medskip

\begin{eqnarray}
q_T^{r\tau\tau}(T_s,{\vec R}_{+})&=&0,\nonumber \\
 &&{}\nonumber \\
 q_T^{ru\tau}(T_s,{\vec R}_{+})&=&\sum_{i=1}^N\eta_i^r\kappa_i^u-R_{+}^r\kappa_{+}^u=
 \sum_{a=1}^{N-1} \rho_a^r\pi^u_a+(\eta^r_{+}-R^r_{+})\kappa^u_{+}\nonumber \\
 &{\rightarrow}_{c \rightarrow \infty}\,& \sum_{a=1}^{N-1} \rho^r_a\pi^u_{qa} =
 \sum_{ab}^{1..N-1} k_{ab} \rho^r_a{\dot \rho}^u_b,\nonumber \\
 &&{}\nonumber \\
 q_T^{ruv}(T_s,{\vec R}_{+})&=& \sum_{i=1}^N \eta^r_i {{\kappa_i^u\kappa_i^v}\over {H_i}}-
 R^r_{+} \sum_{i=1}^N {{\kappa_i^u\kappa_i^v}\over {H_i}}=\nonumber \\
 &=&{1\over {\sqrt{N}}} \sum_{i=1}^N \sum_{a=1}^{N-1} \gamma_{ai} \rho^r_a
 {{\kappa^u_i\kappa_i^v}\over {H_i}}+(\eta^r_{+}-R^r_{+})\sum_{i=1}^N
 {{\kappa_i^u\kappa_i^v}\over {H_i}}\nonumber \\
&{\rightarrow}_{c \rightarrow \infty}\,& {1\over
{\sqrt{N}}}\sum_{abc}^{1..N-1}\Big[
N\sum_{i=1}^N{{\gamma_{ai}\gamma_{bi}\gamma_{ci}}\over
{m_i}}-{{\sum_{j=1}^Nm_j\gamma_{aj}}\over m}\Big]
\rho^r_a\pi^u_{qb}\pi^v_{qc}+O(1/c).\nonumber \\
 &&{}
 \label{4.24}
\end{eqnarray}

\medskip

The antisymmetric part of the related dipole
$p_T^{\mu_1\mu}(T_s,\vec \eta )$ identifies the {\it spin tensor}.
Indeed, the {\it spin dipole}  is
\medskip

\begin{eqnarray*}
S^{\mu\nu}_T(T_s)[\vec \eta]&=&2 p_T^{[\mu\nu ]}(T_s,\vec \eta ) =
2 \epsilon^{[\mu}_r(u(p_s))\, \epsilon^{\nu]}_A(u(p_s))\,
q_T^{rA\tau}(T_s, \vec \eta ) = \nonumber \\
 &=&M_{sys}\, [R^r_{+}(T_s)-\eta^r(T_s)]
\Big[
\epsilon^{\mu}_r(u(p_s))u^{\nu}(p_s)-\epsilon^{\nu}_r(u(p_s))u^{\mu}(p_s)
\Big] +\nonumber \\
&+&\sum_{i=1}^N[\eta^r_i(T_s)-\eta^r(T_s)]\kappa^s_i(T_s)
\Big[\epsilon^{\mu}_r(u(p_s))
\epsilon^{\nu}_s(u(p_s))-\epsilon^{\nu}_r(u(p_s))\epsilon^{\mu}_s(u(p_s))
\Big],\nonumber \\
 &&{}\nonumber \\
 m^{\nu}_{u(p_s)}(T_s,\vec \eta )&=& u_{\mu}(p_s)
S^{\mu\nu}_T(T_s)[\vec \eta ]=-\epsilon^{\nu}_r(u(p_s)) [{\bar
S}^{\tau r}_s - M_{sys}\, \eta^r(T_s)]= \nonumber \\
&=&-\epsilon^{\nu}_r(u(p_s)) M_{sys}\,
[R^r_{+}(T_s)-\eta^r(T_s)]=-
 \epsilon^{\nu}_r(u(p_s)) q_T^{r\tau\tau}(T_s,\vec \eta ),\nonumber \\
 &&{}\nonumber \\
 \Rightarrow&& u_{\mu}(p_s) S^{\mu\nu}_T(T_s)[\vec \eta ]=0,\quad\quad
 \Rightarrow \vec \eta =\vec{R}_+,
 \end{eqnarray*}

\begin{eqnarray}
 &&\Downarrow \quad {\it barycentric\, spin}\,\,
 for\,\, {\vec \kappa}_+ \approx 0,\,\,
 \vec \eta = {\vec {\tilde \eta}} = 0,\,\, see\, Eq(4.17),\nonumber \\
 &&{}\nonumber \\
S^{\mu\nu}_T(T_s)[\vec \eta =0]&=&S^{\mu\nu}_s\, {\buildrel \circ
\over =}\,  \epsilon^{rsu}\, {\bar S}^u_s\,
\epsilon^{\mu}_r(u(p_s)) \epsilon^{\nu}_s(u(p_s)).
 \label{4.25}
\end{eqnarray}
\medskip

This explains why $m^{\mu}_{u(p_s)}(T_s,\vec \eta )$ is also
called the {\it mass dipole moment}.
\medskip

We find, therefore, that in the gauge ${\vec R}_{+} \approx {\vec
q}_{+} \approx {\vec y}_{+} \approx 0$, with  $ P^{\mu}_T \approx
M_{sys}\, u^{\mu}(p_s) = M_{sys}\, {\dot x}_s^{({\vec q}
_{+})\mu}(T_s)$, the M$\o$ller and barycentric centroid $x^{({\vec
q}_{+})\mu}_s(T_s)$ is simultaneously the {\it Tulczyjew centroid}
\cite{48} (defined by $S^{\mu\nu}\, P_{\nu} =0$) and also the {\it
Pirani centroid} \cite{47} (defined by $S^{\mu\nu}\, {\dot
x}^{({\vec q} _{+})}_{s \nu} =0$). In general, {\it lacking a
relation between 4-momentum and 4-velocity}, they are different
centroids.

\subsubsection{ Quadrupoles\, and\, the\, barycentric\, tensor\,
of\, inertia}

\medskip

The {\it quadrupoles}, corresponding to $n=2$, are

\medskip

\begin{eqnarray}
q_T^{r_1r_2AB}(T_s,\vec \eta )&=&\delta^A_{\tau}\delta^B_{\tau}
\sum_{i=1}^N[\eta_i^{r_1}(T_s)-\eta^{r_1}(T_s)][\eta_i^{r_2}(T_s)-\eta^{r_2}(T_s)]
\sqrt{m_i^2+{\vec \kappa}_i^2(T_s)}+\nonumber \\
&+&\delta^A_u\delta^B_v
\sum_{i=1}^N[\eta_i^{r_1}(T_s)-\eta^{r_1}(T_s)][\eta_i^{r_2}(T_s)-\eta^{r_2}(T_s)]
{{\kappa_i^u\kappa_i^v}\over {\sqrt{m_i^2+{\vec
\kappa}_i^2}}}(T_s)+\nonumber \\
&+&(\delta^A_{\tau}\delta^B_u+\delta^A_u\delta^B_{\tau})
\sum_{i=1}^N[\eta_i^{r_1}(T_s)-\eta^{r_1}(T_s)][\eta_i^{r_2}(T_s)-\eta^{r_2}(T_s)]
\kappa_i^u(T_s),\nonumber \\
 &&{}.
 \label{4.26}
\end{eqnarray}

\noindent When the mass dipole vanishes, i.e.
$\vec{\eta}=\vec{R}_+ = \sum_i\, {\vec \eta}_i\, \sqrt{m_i^2 +
{\vec \kappa}_i^2} / M_{sys}$, we get

\begin{eqnarray*}
 q_T^{r_1r_2\tau\tau}(T_s,{\vec R}_{+})&=&\sum_{i=1}^N(\eta_i^{r_1}-R^{r_1}_{+})(\eta_i^{r_2}-
 R^{r_2}_{+}) \sqrt{m_i^2+{\vec \kappa}_i^2(T_s)},\nonumber \\
 &&{}\nonumber \\
  q_T^{r_1r_2u\tau}(T_s,{\vec R}_{+})&=&\sum_{i=1}^N (\eta_i^{r_1}-R^{r_1}_{+})
 (\eta_i^{r_2}-R^{r_2}_{+}) \kappa_i^u,
 \end{eqnarray*}

\bea
 q_T^{r_1r_2uv}(T_s,{\vec R}_{+})&=& \sum_{i=1}^N(\eta_i^{r_1}-R^{r_1}_{+})(\eta_i^{r_2}-
 R^{r_2}_{+}) {{\kappa_i^u\kappa_i^v}\over {\sqrt{m_i^2+{\vec \kappa}_i^2(T_s)}}}\nonumber \\
 &=&{1\over N} \sum_{ijk}^{1..N}\sum_{ab}^{1..N-1}(\gamma_{ai}-\gamma_{aj}).
 \label{4.27}
 \end{eqnarray}
\medskip

Following the non-relativistic pattern, Dixon starts from the {\it
mass quadrupole}
\medskip

\begin{equation}
q_T^{r_1r_2\tau\tau}(T_s,{\vec R}_{+} )=
\sum_{i=1}^N[\eta_i^{r_1}\eta_i^{r_2}\sqrt{m_i^2+{\vec
\kappa}_i^2}](T_s) - M_{sys}\, R_{+}^{r_1}\, R^{r_2}_{+},
 \label{4.28}
\end{equation}

\noindent and defines the following {\it barycentric tensor of
inertia}

\bea
 I_{dixon}^{r_1r_2}(T_s)&=&\delta^{r_1r_2} \sum_u
q_T^{uu\tau\tau}(T_s,\vec{R}_+)-
q_T^{r_1r_2\tau\tau}(T_s,\vec{R}_+)=\nonumber \\
&=&\sum_{i=1}^N[(\delta^{r_1r_2} ({\vec
\eta}_i-\vec{R}_+)^2-(\eta_i^{r_1}-R_+^{r_1})
(\eta_i^{r_2}-R_+^{r_2}))\sqrt{m_i^2+{\vec \kappa}_i^2}](T_s)
\nonumber \\
 &{\rightarrow}_{c \rightarrow \infty}\,&
 \sum_{ab}^{1..N-1} k_{ab} [ {\vec \rho}_{qa}\cdot {\vec \rho}_{qb}
 \delta^{r_1r_2}-\rho_{qa}^{r_1}\rho_{qb}^{r_2}] +O(1/c)=
  I^{r_1r_2}[{\vec q}_{nr}] + O(1/c).\nonumber \\
  &&{}
\label{4.29}
\end{eqnarray}

Note that in the non-relativistic limit we recover the {\it tensor
of inertia} of Eqs.(\ref{4.11}).

\bigskip
\noindent

On the other hand, Thorne's definition of {\it barycentric tensor
of inertia} \cite{30} is

\bea
 I_{thorne}^{r_1r_2}(T_s)&=&\delta^{r_1r_2} \sum_u
q_T^{uuA}{}_A(T_s,\vec{R}_+)-
q_T^{r_1r_2A}{}_A(T_s,\vec{R}_+)=\nonumber \\
 &=& \sum_{i=1}^N {{m_i^2 (\delta^{r_1r_2}{(\vec \eta}_i-\vec{R}_+)^2-(\eta_i^{r_1}-R_+^{r_1})
(\eta_i^{r_2}-R_+^{r_2}))}\over {\sqrt{m_i^2+{\vec
\kappa}_i^2}}}(T_s)\nonumber \\
 &{\rightarrow}_{c \rightarrow \infty}\,&
 \sum_{ab}^{1..N-1} k_{ab} [ {\vec \rho}_{qa}\cdot {\vec \rho}_{qb}
 \delta^{r_1r_2}-\rho_{qa}^{r_1}\rho_{qb}^{r_2}] +O(1/c)=
  I^{r_1r_2}[{\vec q}_{nr}] + O(1/c).\nonumber \\
  &&{}
\label{4.30}
\end{eqnarray}

In this case too we recover the {\it tensor of inertia} of
Eq.(\ref{4.11}). Note that the Dixon and Thorne barycentric
tensors of inertia differ at the post-Newtonian level

\medskip

\bea
 I^{r_1r_2}_{dixon}(T_s)-I^{r_1r_2}_{thorne}(T_s)&=& {1\over
c} \sum_{ab}^{1..N-1}\sum_{ijk}^{1..N}{{m_jm_k}\over {Nm^2}}
(\gamma_{ai}-\gamma_{aj})(\gamma_{bi}-\gamma_{bk}) \nonumber \\
 &&\Big[{\vec \rho}_{qa}\cdot {\vec \rho}_{qb}
 \delta^{r_1r_2}-\rho_{qa}^{r_1}\rho_{qb}^{r_2}\Big] {{N\sum_{cd}^{1..N-1}
 \gamma_{ci}\gamma_{di}{\vec \pi}_{qc}\cdot {\vec \pi}_{qd}}\over {m_i}}+O(1/c^2).
 \nonumber \\
 &&{}
 \label{4.31}
 \eea
\bigskip

\subsubsection{ The multipolar expansion}

\medskip

It can be shown that the multipolar expansion can be rearranged in
the form

\begin{eqnarray}
&&T^{\mu\nu}[ x_s^{({\vec q}_{+})
\beta}(T_s)+\epsilon^{\beta}_r(u(p_s))\sigma^r ]= T^{\mu\nu}[
w^{\beta}(T_s)+\epsilon^{\beta}_r(u(p_s)) (\sigma^r-\eta^r(T_s))
]=\nonumber \\
 &&{}\nonumber \\
 &=& u^{(\mu}(p_s) \epsilon^{\nu )}_A(u(p_s)) [\delta^A_{\tau}
  M_{sys} + \delta^A_u \kappa^u_{+}] \delta^3(\vec \sigma -\vec \eta
  (T_s))+\nonumber \\
  &&{}\nonumber \\
  &+&{1\over 2} S_T^{\rho (\mu}(T_s)[\vec \eta ] u^{\nu )}(p_s)
   \epsilon^r_{\rho}(u(p_s)) {{\partial}\over {\partial \sigma^r}}
   \delta^3(\vec \sigma -\vec \eta (T_s))+\nonumber \\
 &&{}\nonumber \\
   &+&\sum_{n=2}^{\infty} {{(-1)^n}\over {n!}} I_T^{\mu_1..\mu_n\mu\nu}(T_s,\vec \eta )
   \epsilon^{r_1}_{\mu_1}(u(p_s))..\epsilon^{r_n}_{\mu_n}(u(p_s))
   {{\partial^n}\over {\partial \sigma^{r_1}..\partial \sigma^{r_n}}}
   \delta^3(\vec \sigma -\vec \eta (T_s)),
\label{4.32}
\end{eqnarray}

\noindent where for $n\geq 2$ and $\vec \eta =0$, we have  $\quad
I_T^{\mu_1..\mu_n\mu\nu}(T_s)={{4(n-1)}\over {n+1}}
J_T^{(\mu_1..\mu_{n-1} | \mu | \mu_n)\nu}(T_s)$, with
$J_T^{\mu_1..\mu_n\mu\nu\rho\sigma}(T_s)$ being the generalized
Dixon {\it $2^{2+n}$-pole inertial moment tensors} given in
Ref.\cite{11}.

\bigskip

Note that, for an isolated system described by the multi-poles
appearing in Eq.(\ref{4.32}) [this is not true for those in
Eq.(\ref{4.19})] the equations $\partial_{\mu}T^{\mu\nu}\,
{\buildrel \circ \over =}\, 0$  imply no more than the following
{\it Papapetrou-Dixon-Souriau equations of motion} \cite{29} for
the total momentum $P^{\mu}_T(T_s)=\epsilon^{\mu}_A(u(p_s))
q_T^{A\tau}(T_s)= p^{\mu}_s$ and the spin tensor
$S^{\mu\nu}_T(T_s)[\vec \eta =0]$

\medskip

\begin{eqnarray}
{{d P^{\mu}_T(T_s)}\over {dT_s}}\, &{\buildrel \circ \over =}\,&
0,\nonumber \\
 {{d S^{\mu\nu}_T(T_s)[\vec \eta =0]}\over {dT_s}}\, &{\buildrel
 \circ \over =}\,& 2 P^{[\mu}_T(T_s) u^{\nu ]}(p_s)=2 \kappa^u_{+} \epsilon^{[\mu}_u(u(p_s))
 u^{\nu ]}(p_s) \approx 0,\nonumber \\
 &&{}\nonumber \\
 &&{}\nonumber \\
 or&& {{d M_{sys}}\over {dT_s}} \cir 0,\qquad {{d {\vec
 \kappa}_{+}}\over {dT_s}} \cir 0,\qquad {{d S_s^{\mu\nu}}\over
 {dT_s}} \cir 0.
\label{4.33}
\end{eqnarray}

\subsubsection{Cartesian Tensors}

In the applications to gravitational radiation, {\it irreducible
symmetric trace-free Cartesian tensors (STF tensors)} \cite{30,31}
are needed instead of {\it Cartesian tensors}. While a {\it
Cartesian multi-pole tensor of rank $l$} (like the rest-frame
Dixon multi-poles) on $R^3$ has $3^l$ components, ${1\over
2}(l+1)(l+2)$ of which are in general independent, a {\it
spherical multi-pole moment of order $l$} has only $2l+1$
independent components. Even if spherical multi-pole moments are
preferred in calculations of molecular interactions, spherical
harmonics have various disadvantages in numerical calculations:
for analytical and numerical calculations Cartesian moments are
often more convenient (see for instance Ref.\cite{54} for the case
of the electrostatic potential). It is therefore preferable using
the {irreducible Cartesian STF tensors} \cite{55} (having $2l+1$
independent components if of rank $l$), which are obtained by
using {\it Cartesian spherical (or solid) harmonic tensors} in
place of spherical harmonics.

Given an Euclidean tensor $A_{k_1...k_I}$ on $R^3$, one defines
the completely symmetrized tensor $S_{k_1..k_I} \equiv
A_{(k_1..k_I)} = {1\over {I!}} \sum_{\pi} A_{k_{\pi (1)}...k_{\pi
(I)}}$. Then, the associated STF tensor is obtained by removing
all traces ($[I/2]=$ largest integer $\leq I/2$)\medskip

\bea
 A_{k_1...k_I}^{(STF)} &=&\sum_{n=0}^{[I/2]} a_n\, \delta_{(k_1k_2} ...
\delta_{k_{2n-1}k_{2n}} S_{k_{2n+1}...k_I)
i_1i_1...j_nj_n},\nonumber \\
 &&{}\nonumber \\
 a_n &\equiv& (-1)^n {{ l! (2l-2n-1)!!}\over {(l-2n)! (2l-1)!!(2n)!!}}.
 \label{4.34}
 \eea

\noindent For instance $(T_{abc})^{STF} \equiv T_{(abc)}- {1\over
5}\Big[\delta_{ab} T_{(iic)}+
\delta_{ac}T_{(ibi)}+\delta_{bc}T_{(aii)}\Big]$.

\subsection{Open N-body systems.}

Consider now an open sub-system of the isolated system of $N$
charged positive-energy particles plus the electro-magnetic field
in the radiation gauge (see the second paper of Ref.\cite{6}). The
energy-momentum tensor and the Hamilton equations on the Wigner
hyper-plane are, respectively, [to avoid degenerations we assume
that all the masses $m_i$ are different; ${\vec \pi}_{\perp} =
{\vec E}_{\perp}$]

\begin{eqnarray*}
T^{\tau\tau}(\tau ,\vec \sigma )&=&
 \sum_{i=1}^N \delta^3(\vec
\sigma - {\vec \eta}_i(\tau )) \sqrt{m^2_i+[{ {\vec
\kappa}}_i(\tau ) -Q_i{{ \vec A}}_{\perp}(\tau ,{\vec \eta}_i(\tau
))]^2} + \nonumber \\
 &+& \sum_{i=1}^N\, Q_i\, {\vec \pi}_{\perp}(\tau ,\vec \sigma )\,
 \times {{\vec \partial}\over {\triangle}}\, \delta^3(\vec \sigma - {\vec
 \eta}_i(\tau )) + {1\over 2}\, [{\vec \pi}^2_{\perp} + {\vec
 B}^2](\tau ,\vec \sigma ) +\nonumber \\
 &+& {1\over 2}\, \sum_{i,k,i\not= k}^{1..N}\, Q_i\, Q_k\,
 {{\vec \partial}\over {\triangle}}\, \delta^3(\vec \sigma - {\vec
 \eta}_i(\tau )) \cdot  {{\vec \partial}\over {\triangle}}\, \delta^3(\vec \sigma - {\vec
 \eta}_k(\tau )),
 \end{eqnarray*}

\begin{eqnarray*}
 T^{r\tau}(\tau ,\vec \sigma )&=&\sum_{i=1}^N\delta^3(\vec \sigma -{\vec \eta}
_i(\tau )) [{\kappa}_i^r(\tau )-Q_i {A}_{\perp}^r(\tau ,{\vec
\eta}_i(\tau ))] +  \nonumber \\
 &+&[\Big( {{\vec \pi}}_{\perp}+\sum_{i=1}^NQ_i{\frac{{\vec \partial}}{
{\triangle}}}\delta^3(\vec \sigma -{\vec \eta}_i(\tau
))\Big)\times \vec B ](\tau ,\vec \sigma ),  \nonumber \\
 &&{}\nonumber \\
 T^{rs}(\tau ,\vec \sigma )&=& \sum_{i=1}^N \delta^3(\vec \sigma -{\vec \eta}
_i(\tau )) {\frac{{\ [{\kappa}_i^r(\tau )-Q_i{A} _{\perp}^r(\tau
,{\vec \eta}_i(\tau ))] [{\kappa}_i^s(\tau )-Q_i{
A}_{\perp}^s(\tau ,{\vec \eta}_i(\tau ))]}}{{\sqrt{m_i^2+[{ { \vec
\kappa}}_i(\tau )-Q_i{
 {\vec A}}_{\perp}(\tau ,{\vec \eta}_i(\tau ))]^2} }}} -  \nonumber \\
 &-&\Big[{\frac{1}{2}}\delta^{rs} [\Big( {{\vec \pi}}
_{\perp}+\sum_{i=1}^NQ_i{\frac{{\vec
\partial}}{{\triangle}}}\delta^3(\vec \sigma -{\vec \eta}_i(\tau
))\Big)^2+{\vec B}^2] -  \end{eqnarray*}

\bea
 &-&[\Big( {{\vec \pi}}_{\perp}+\sum_{i=1}^NQ_i{\frac{{\vec
\partial}}{ {\triangle}}}\delta^3(\vec \sigma -{\vec \eta}_i(\tau
))\Big)^r \Big( { {\vec \pi}}_{\perp}+\sum_{i=1}^NQ_i{\frac{{\vec
\partial}}{{\triangle} }}\delta^3(\vec \sigma -{\vec \eta}_i(\tau
))\Big)^s +  \nonumber \\
 &+&B^rB^s]\Big] (\tau ,\vec \sigma ).
 \label{4.35}
\end{eqnarray}
\medskip

\begin{eqnarray}
\dot{\vec{\eta}}_{i}(\tau )\, &\,\stackrel{\circ
}{=}&\,\frac{{\vec{ \kappa}}_{i}(\tau )-Q_{i}{\vec{A}}_{\perp
}(\tau ,\vec{\eta}_{i}(\tau
))}{\sqrt{m_{i}^{2}+({\vec{\kappa}}_{i}(\tau )-Q_{i}{\vec{A}}
_{\perp }(\tau ,\vec{\eta}_{i}(\tau )))^{2}}}, \nonumber \\
 \dot{{\vec{\kappa}}}_{i}(\tau )\,\stackrel{\circ
}{=} &&\, \sum_{k\neq i}\frac{Q_{i}Q_{k}({\vec{\eta}}_{i}(\tau
)-{\vec{\eta}}_{k}(\tau ))}{4\pi \mid \vec{\eta}_{i}(\tau
)-\vec{\eta}_{k}(\tau )\mid ^{3}}+
 Q_{i}\, \dot{\eta}_{i}^{u}(\tau )\, {\frac{{\partial }}{
\partial {{\vec{\eta}}_{i}}}}{{A}}_{\perp }^{u}(\tau ,\vec{\eta}
_{i}(\tau ))],  \nonumber \\
  &&{}\nonumber \\
  &&{}\nonumber \\
 &&\dot{{A}}_{\perp r}(\tau ,\vec{\sigma})\,\stackrel{\circ }{=}-{
\pi} _{\perp r}(\tau ,\vec{\sigma}),  \nonumber \\
 &&{}\nonumber \\
 \dot{{\pi}}_{\perp }^{r}(\tau ,\vec{\sigma})\, &\stackrel{\circ }{=}
&\,\Delta {A}_{\perp }^{r}(\tau ,\vec{\sigma}) -
\sum_{i}Q_{i}P_{\perp }^{rs}(\vec{\sigma})\dot{\eta}_{i}^{s}(\tau
)\delta ^{3}(\vec{\sigma}-\vec{\eta}_{i}(\tau )),\nonumber \\
 &&{}\nonumber \\
 &&{}\nonumber \\
 {{\vec{\kappa}}}_{+}(\tau )
&+&\int d^{3}\sigma \lbrack {{\vec{ \pi}}}_{\perp }\times
{{\vec{B}}}](\tau ,\vec{\sigma})\approx 0\,\, (rest-frame\,
condition).
 \label{4.36}
\end{eqnarray}
\medskip

Let us note that in this reduced phase space there are only either
particle-field interactions or action-at-a-distance 2-body
interactions. The particle world-lines are $x^{\mu}_i(\tau ) =
x^{\mu}_o + u^{\mu}(p_s)\, \tau + \epsilon^{\mu}_r(u(p_s))\,
\eta^r_i(\tau )$, while their 4-momenta are $p^{\mu}_i(\tau ) =
\sqrt{m^2_i + [{\vec \kappa}_i - Q_i\, {\vec A}_{\perp}(\tau
,{\vec \eta}_i) ]^2}\, u^{\mu}(p_s) + \epsilon^{\mu}_r(u(p_s))\,
[\kappa^r_i - Q_i\, A^r_{\perp}(\tau ,{\vec \eta}_i)]$.
\bigskip

The generators of the internal Poincar\'{e} group are

\begin{eqnarray*}
 {\cal P}_{(int)}^{\tau } &=&M=\sum_{i=1}^{N}\sqrt{m_{i}^{2}+({\vec{
\kappa}}_{i}(\tau )-Q_{i}{\vec{A}}_{\perp }(\tau ,\vec{\eta}
_{i}(\tau )))^{2}}+  \nonumber \\
 &+&{1\over 2}\, \sum_{i\neq j}\frac{Q_{i}Q_{j}}{4\pi \mid \vec{\eta}_{i}(\tau )-\vec{\eta}
_{j}(\tau )\mid }+\int d^{3}\sigma
{\frac{1}{2}}[{\vec{\pi}}_{\perp }^{2}+{\vec{B}}^{2}](\tau
,\vec{\sigma}),  \nonumber \\
 &&{}\nonumber \\
 {\cal \vec{P}}_{(int)} &=& {{\vec{\kappa}}}
_{+}(\tau )+\int d^{3}\sigma \lbrack {{\vec{\pi}}}_{\perp }\times
{ {\vec{B}}}](\tau ,\vec{\sigma})\approx 0, \nonumber \\
 &&{} \nonumber \\
 {\cal J}_{(int)}^{r} &=& \sum_{i=1}^{N}( \vec{\eta}_{i}(\tau )\times
{{\vec{\kappa}}}_{i}(\tau ))^{r}+\int d^{3}\sigma
\,(\vec{\sigma}\times \,{[{{\vec{\pi}}}}_{\perp }{\times {
{\vec{B}}]}}^{r}{(\tau ,\vec{\sigma})},  \nonumber \\
 {\cal K}_{(int)}^{r} &=& -\sum_{i=1}^{N}\vec{\eta}_{i}(\tau
)\sqrt{m_{i}^{2}+[{{{\vec{\kappa}} }}_{i}(\tau
)-Q_{i}{{\vec{A}}}_{\perp }(\tau ,{\ \vec{\eta}}_{i}(\tau
))]{}^{2}}+\end{eqnarray*}

\bea
 &+&{1\over 2}\, \Big[ Q_i\, \sum_{i=1}^{N} \sum_{j\not=i}^{1..N}\,
 Q_j\, \int d^3\sigma\, \sigma^r\, \vec c(\vec \sigma -
 {\vec \eta}_i(\tau )) \cdot \vec c(\vec \sigma - {\vec
 \eta}_j(\tau )) + \nonumber \\
 &+& Q_{i}\, \int d^{3}\sigma {{\pi}}_{\perp }^{r}(\tau ,\vec{\sigma})c(
\vec{\sigma} - {\ \vec{\eta}}_{i}(\tau )) \Big]-{\frac{1}{2}}\int
d^{3}\sigma \sigma ^{r}\,({{{\vec{\pi}}}}_{\perp }^{2}+{{{\vec{B}
}}}^{2})(\tau ,\vec{\sigma}),
 \label{4.37}
\end{eqnarray}

\noindent with $c({\vec{\eta}}_{i}-{\vec{\eta}}_{j})= 1/(4\pi
|\vec{\eta}_{j}-\vec{ \eta}_{i}|)$ [$\triangle\, c(\vec \sigma ) =
\delta^3(\vec \sigma )$, $\triangle = - {\vec \partial}^2$, $\vec
c(\vec \sigma ) = \vec \partial\, c(\vec \sigma ) = \vec \sigma /
(4\pi\, |\vec \sigma |^3)$]. \medskip

${\cal P}^{\tau}_{(int)} = q^{\tau\tau}$ and ${\cal P}^r_{(int)} =
q^{r\tau}$ are the mass and momentum monopoles, respectively.

\bigskip

For the sake of simplicity, consider the sub-system formed by the
two particles of mass $m_1$ and $m_2$. Our considerations may be
extended to any cluster of particles. This sub-system is {\it
open}: besides their mutual interaction the two particles have
Coulomb interaction with the other $N - 2$ particles and are
affected by the transverse electric and magnetic fields.
\medskip

Exploiting the multi-poles we will select a set of {\it effective
parameters} (mass, 3-center of motion, 3-momentum, spin)
describing the two-particle cluster as a global entity subject to
external forces in the global rest-frame instant form. This was
indeed the original motivation of the multipolar expansion in
general relativity: replacing an extended object (an open system
due to the presence of the gravitational field) by a set of
multi-poles concentrated on a center of motion. Now, in the
rest-frame instant form it is possible to show that there is no
preferred centroid for open system so that, unlike the case of
isolated systems where, in the rest frame ${\vec \kappa}_{+}
\approx 0$, all possible conventions identify the same centroid,
different centers of motion can be selected according to different
conventions. We will see, however, that one specific choice exists
showing preferable properties.
\medskip

Given the energy-momentum tensor $T^{AB}(\tau ,\vec \sigma )$
(\ref{4.35}) of the isolated system,  it would seem natural to
define {\it the energy-momentum tensor $T^{AB}_{c(n)}(\tau ,\vec
\sigma )$ of an open sub-system composed by a cluster of $n \leq
N$ particles} as the sum of all the terms in Eq.(\ref{4.35})
containing a dependence on the variables ${\vec \eta}_i$, ${\vec
\kappa}_i$, of the particles of the cluster. Besides kinetic
terms, this tensor would contain  internal mutual interactions as
well as external interactions of the cluster particles with the
environment composed by the other $N-n$ particles and by the
transverse electro-magnetic field. There is an ambiguity, however.
While there is no problem in attributing to the cluster the whole
interaction with the electro-magnetic field, why should we
attribute to it just {\it all the external interactions with the
other $N-n$ particles}?  Since we have 2-body interactions, it
seems more reasonable to attribute only {\it half} of these
external interactions to the cluster and consider the other half
as a property of the remaining $N-n$ particles. Let us remark that
considering, e.g., two clusters composed by two non-overlapping
sets of $n_1$ and $n_2$ particles, respectively, since the mutual
Coulomb interactions between the clusters are present in both
$T^{AB}_{c(n_1)}$ and $T^{AB}_{c(n_2)}$, according to the first
choice we would get $T^{AB}_{c(n_1+n_2)} \not= T^{AB}_{c(n_1)} +
T^{AB}_{c(n_2)}$. On the other hand, according to the second
choice we get $T^{AB}_{c(n_1+n_2)} = T^{AB}_{c(n_1)} +
T^{AB}_{c(n_2)}$. Since this property is important for studying
the mutual relative motion of two clusters in actual cases, we
will adopt {\it the convention that the energy-momentum tensor of
a $n$ particle cluster contains only half of the external
interaction with the other $N-n$ particles}.

\bigskip

Let us remark that, in the case of $k$-body forces, this
convention should be replaced by the following rule: i) for each
particle $m_i$ of the cluster and each $k$-body term in the
energy-momentum tensor involving this particle, $k = h_i + (k -
h_i)$, where $h_i$ is the number of particles of the cluster
participating to this particular $k$-body interaction; ii) only
the fraction $h_i/k$ of this particular $k$-body interaction term
containing $m_i$ must be attributed to the cluster.

\bigskip

Let us consider the cluster composed by the two particles with
mass $m_1$ and $m_2$. The knowledge of $T^{AB}_c\, {\buildrel
{def}\over =}\, T^{AB}_{c(2)}$ on the Wigner hyper-plane of the
global rest-frame instant form allows us to find the following 10
{\it non conserved} charges [due to $Q^2_i =0$ we have
$\sqrt{m^2_i + [{\vec \kappa}_i - Q_i\, {\vec A}_{\perp}(\tau
,{\vec \eta}_i)]^2} = \sqrt{m^2_i + {\vec \kappa}_1^2} - Q_i\,
{{{\vec \kappa}_i \cdot {\vec A}_{\perp}(\tau ,{\vec
\eta}_i)}\over {\sqrt{m^2_i + {\vec \kappa}_i^2 }}} $] \bigskip

\begin{eqnarray*}
 M_c &=& \int d^3\sigma\, T^{\tau\tau}_c(\tau ,\vec \sigma ) =
 \sum_{i=1}^2\, \sqrt{m^2_i + [{\vec \kappa}_i(\tau ) - Q_i\,
  {\vec A}_{\perp}(\tau ,{\vec \eta}_i(\tau ))]^2} +\nonumber \\
  &+& {{Q_1\, Q_2}\over {4\pi\, |{\vec \eta}_1(\tau ) - {\vec \eta}_2(\tau )|^2}}
  + {1\over 2}\, \sum_{i=1}^2\, \sum_{k\not= 1,2}\, {{Q_i\, Q_k}\over {4\pi\,
  |{\vec \eta}_i(\tau ) - {\vec \eta}_k(\tau )|^2}} =\nonumber \\
  &=& M_{c(int)} + M_{c(ext)},\nonumber \\
  &&M_{c(int)} = \sum_{i=1}^2\, \sqrt{m_i^2 + {\vec \kappa}_i^2} -
  {{Q_1\, Q_2}\over {4\pi\, |{\vec \eta}_1(\tau ) - {\vec \eta}_2(\tau
  )|^2}},\nonumber \\
  &&{}\nonumber \\
  {\vec {\cal P}}_c &=& \{ \int d^3\sigma\, T^{r\tau}_c(\tau ,\vec \sigma ) \} =
  {\vec \kappa}_1(\tau ) + {\vec \kappa}_2(\tau
  ),\nonumber \\
  &&{}\nonumber \\
  {\vec {\cal J}}_c &=& \{ \epsilon^{ruv}\, \int d^3\sigma\, [\sigma^u\, T^{v\tau}_c
  - \sigma^v\, T^{u\tau}_c](\tau ,\vec \sigma ) \} =
   {\vec \eta}_i(\tau ) \times {\vec
  \kappa}_1(\tau ) + {\vec \eta}_2(\tau ) \times {\vec
  \kappa}_2(\tau ),
  \end{eqnarray*}

\begin{eqnarray*}
  {\vec {\cal K}}_c &=& - \int d^3\sigma\, \vec \sigma\,
  T^{\tau\tau}_c(\tau ,\vec \sigma ) =
   - \sum_{i=1}^2\, {\vec \eta}_i(\tau )\, \sqrt{m^2_i + [{\vec \kappa}_i(\tau ) - Q_i\,
  {\vec A}_{\perp}(\tau ,{\vec \eta}_i(\tau ))]^2} -\nonumber \\
  &-&  \sum_{i=1}^2\, Q_i\, \int d^3\sigma\, {\vec
  \pi}_{\perp}(\tau ,\vec \sigma )\, c(\vec \sigma - {\vec
  \eta}_i(\tau )) -
  \end{eqnarray*}

\bea
 &-&  Q_1\, Q_2\, \int d^3\sigma\, \vec \sigma\, \vec c(\vec
 \sigma - {\vec \eta}_1(\tau )) \cdot \vec c(\vec \sigma - {\vec
 \eta}_2(\tau )) -\nonumber \\
 &-& {1\over 2}\, \sum_{i=1}^2\, Q_i\, \sum_{k\not= 1,2}\, Q_k\,
\int d^3\sigma\, \vec \sigma\, \vec c(\vec
 \sigma - {\vec \eta}_i(\tau )) \cdot \vec c(\vec \sigma - {\vec
 \eta}_k(\tau )) =\nonumber \\
  &=& {\vec {\cal K}}_{c(int)} + {\vec {\cal
  K}}_{c(ext)},\nonumber \\
  &&{\vec {\cal K}}_{c(int)} = - \sum_{i=1}^2\, {\vec \eta}_i(\tau
  )\, \sqrt{m^2_i + {\vec \kappa}_i^2} - Q_1\, Q_2\,
 \int d^3\sigma\, \vec \sigma\, \vec c(\vec
 \sigma - {\vec \eta}_1(\tau )) \cdot \vec c(\vec \sigma - {\vec
 \eta}_2(\tau )).\nonumber \\
 &&{}
 \label{4.38}
 \eea

\noindent Such charges do not satisfy the algebra of an internal
Poincare' group just because of the openness of the system.
Working in an instant form of dynamics, only the cluster internal
energy and boosts depend on the (internal and external)
interactions. Again, $M_c = q_c^{\tau\tau}$ and ${\cal P}^r_c =
q_c^{r\tau}$ are the mass and momentum monopoles of the cluster.
\bigskip

Another quantity to be considered is the momentum dipole

\bea
 p^{ru}_c &=& \int d^3\sigma\, \sigma^r\, T^{u\tau}_c(\tau ,\vec
 \sigma ) =\nonumber \\
 &=& \sum_{i=1}^2\, \eta^r_i(\tau )\, \kappa^u_i(\tau ) -
  \sum_{i=1}^2\, Q_i\, \int d^3\sigma\, c(\vec
 \sigma - {\vec \eta}_i(\tau ))\, [\partial^r\, A^s_{\perp} + \partial^s\, A^r_{\perp}]
 (\tau ,\vec \sigma),\nonumber \\
 &&{}\nonumber \\
 &&p^{ru}_c + p^{ur}_c = \sum_{i=1}^2\, [\eta^r_i(\tau )\,
 \kappa^u_i(\tau ) + \eta^u_i(\tau )\, \kappa^r_i(\tau
 )] -\nonumber \\
 &-& 2\, \sum_{i=1}^2\, Q_i\, \int d^3\sigma\, c(\vec
 \sigma - {\vec \eta}_i(\tau ))\, [\partial^r\, A^s_{\perp} + \partial^s\, A^r_{\perp}]
 (\tau ,\vec \sigma ),\nonumber \\
 &&{}\nonumber \\
 &&p^{ru}_c - p^{ur}_c = \epsilon^{ruv}\, {\cal J}^v_c.
 \label{4.39}
 \eea
\medskip

The time variation of the 10 charges (\ref{4.38}) can be evaluated
by using the equations of motion (\ref{4.36}) \bigskip

\begin{eqnarray*}
 {{d M_c}\over {d\tau}} &=& \sum_{i=1}^2\, Q_i\, \Big( {{{\vec
 \kappa}_i(\tau ) \cdot {\vec \pi}_{\perp}(\tau ,{\vec
 \eta}_i(\tau ))}\over {\sqrt{m^2_i + {\vec \kappa}_i^2}}}
 +\nonumber \\
 &+&  {1\over 2}\, \sum_{k\not= 1,2}\, Q_k\, \Big[{{{\vec \kappa}_i(\tau
)}\over {\sqrt{m^2_i + {\vec \kappa}_i^2}}} + {{{\vec
\kappa}_k(\tau )}\over {\sqrt{m^2_k + {\vec \kappa}_k^2}}}\Big]
\cdot {{{\vec \eta}_i(\tau ) - {\vec \eta}_k(\tau ) }\over {4\pi\,
 |{\vec \eta}_i(\tau ) - {\vec \eta}_k(\tau ) |^3}}\Big),\nonumber \\
 &&{}\nonumber \\
 {{d {\cal P}^r_c}\over {d\tau }} &=& \sum_{i=1}^2\, Q_i\, \Big( {{{\vec \kappa}_i(\tau )}\over
{\sqrt{m^2_i + {\vec \kappa}_i^2}}} \cdot {{\partial
A^r_{\perp}(\tau ,{\vec \eta}_i(\tau ))}\over {\partial {\vec
\eta}_i}} + \sum_{k\not= 1,2}\, Q_k\, {{\eta^r_i(\tau ) -
\eta^r_k(\tau ) }\over {4\pi\, |{\vec \eta}_i(\tau ) - {\vec
\eta}_k(\tau ) |^3}} \Big),
 \end{eqnarray*}

\begin{eqnarray*}
 {{d {\vec {\cal J}}_c}\over {d\tau }} &=&
 \sum_{i=1}^2\, Q_i\, \Big( {{{\vec \kappa}_i(\tau )}\over
{\sqrt{m^2_i + {\vec \kappa}_i^2}}} \times {\vec A}_{\perp}(\tau
,{\vec \eta}_i(\tau )) + {\vec \eta}_i(\tau ) \times \Big[
 {{{\vec \kappa}_i(\tau )}\over
{\sqrt{m^2_i + {\vec \kappa}_i^2}}} \cdot {{\partial}\over
{\partial {\vec \eta}_i}}\Big]\, {\vec A}_{\perp}(\tau ,{\vec
\eta}_i(\tau )) -\nonumber \\
 &-& \sum_{k \not= i}\, Q_k\, {{{\vec \eta}_i(\tau ) \times {\vec \eta}_k(\tau )}
 \over {4\pi\, |{\vec \eta}_i(\tau ) - {\vec \eta}_k(\tau )|^3}}\Big),
 \end{eqnarray*}

\begin{eqnarray*}
 {{d {\cal K}^r_c}\over {d\tau}} &=&
  Q_1Q_2\, \int d^3\sigma\, \vec \sigma\, \Big( \Big[ \Big( {{{\vec
 \kappa}_1(\tau )}\over {\sqrt{m^2_1 + {\vec \kappa}_1^2(\tau )}}}
\cdot \vec \partial\Big)\, \vec c(\vec \sigma - {\vec \eta}_1(\tau
))\Big] \cdot \vec c(\vec \sigma - {\vec \eta}_2(\tau ))
+\nonumber \\
 &+& \vec c(\vec \sigma - {\vec \eta}_1(\tau )) \cdot \Big[ \Big( {{{\vec
 \kappa}_2(\tau )}\over {\sqrt{m^2_2 + {\vec \kappa}_2^2(\tau )}}}
 \cdot \vec \partial \Big)\, \vec c(\vec \sigma - {\vec
 \eta}_2(\tau ))\Big] \Big) -\end{eqnarray*}

\bea
 &-& {1\over 2}\, \sum_{i=1}^2\, Q_i\, \sum_{k\not= 1,2}\, Q_k\,
  \int d^3\sigma\, \vec \sigma\, \Big( \Big[ \Big( {{{\vec
 \kappa}_i(\tau )}\over {\sqrt{m^2_i + {\vec \kappa}_i^2(\tau )}}}
\cdot \vec \partial\Big)\, \vec c(\vec \sigma - {\vec \eta}_i(\tau
))\Big] \cdot \vec c(\vec \sigma - {\vec \eta}_k(\tau ))
+\nonumber \\
 &+& \vec c(\vec \sigma - {\vec \eta}_i(\tau )) \cdot \Big[ \Big( {{{\vec
 \kappa}_k(\tau )}\over {\sqrt{m^2_k + {\vec \kappa}_k^2(\tau )}}}
 \cdot \vec \partial \Big)\, \vec c(\vec \sigma - {\vec
 \eta}_k(\tau ))\Big] \Big).
 \label{4.40}
 \eea

\medskip

Note that, if we have two clusters of $n_1$ and $n_2$ particles
respectively, our definition of cluster energy-momentum tensor
implies

\bea
 M_{c(n_1+n_2)} &=& M_{c(n_1)} + M_{c(n_2)},\nonumber \\
 {\vec {\cal P}}_{c(n_1+n_2)} &=& {\vec {\cal P}}_{c(n_1)} + {\vec
 {\cal P}}_{c(n_2)},\nonumber\\
 {\vec {\cal J}}_{c(n_1+n_2)} &=& {\vec {\cal J}}_{c(n_1)} + {\vec
 {\cal J}}_{c(n_2)},\nonumber \\
 {\vec {\cal K}}_{c(n_1+n_2)} &=& {\vec {\cal K}}_{c(n_1)} + {\vec
 {\cal K}}_{c(n_2)}.
  \label{4.41}
 \eea
\bigskip

The main problem is now the determination of an {\it effective
center of motion} $\zeta^r_c(\tau )$ with world-line
$w^{\mu}_c(\tau ) = x^{\mu}_o + u^{\mu}(p_s)\, \tau +
\epsilon^{\mu}_r(u(p_s))\, \zeta^r_c(\tau )$ in the gauge $T_s
\equiv \tau$, ${\vec q}_{+} = {\vec R}_{+} = {\vec y}_{+} \equiv
0$ of the isolated system. The unit 4-velocity of this center of
motion is $u^{\mu}_c(\tau ) = {\dot w}_c^{\mu}(\tau ) / \sqrt{1 -
{\dot {\vec \zeta}}^2_c(\tau )}$ with ${\dot w}_c^{\mu}(\tau ) =
u^{\mu}(p_s) + \epsilon^{\mu}_r(u(p_s))\, {\dot \zeta}^r_c(\tau
)$. By using $\delta\, z^{\mu}(\tau ,\vec \sigma ) =
\epsilon^{\mu}_r(u(p_s))\, (\sigma^r - \zeta^r(\tau ))$ we can
define the multipoles of the cluster with respect to the
world-line $w^{\mu}_c(\tau )$

\begin{equation} q_c^{r_1..r_nAB}(\tau ) = \int d^3\sigma \, [\sigma^{r_1} -
 \zeta_c^{r_1}(\tau )] .. [\sigma^{r_n} - \zeta_c^{r_n}(\tau )]\,
 T^{AB}_c(\tau ,\vec \sigma ).
 \label{4.42}
 \end{equation}
\medskip

The mass and momentum monopoles, and the mass, momentum and spin
dipoles are, respectively
\medskip

\begin{eqnarray*}
 q^{\tau\tau}_c &=& M_c,\qquad q_c^{r\tau} = {\cal
 P}_c^r,\nonumber \\
 q_c^{r\tau\tau} &=& - {\cal K}^r_c - M_c\, \zeta^r_c(\tau ) = M_c\,
 (R^r_c(\tau ) - \zeta^r_c(\tau )),\qquad
 q_c^{ru\tau} = p^{ru}_c(\tau ) - \zeta^r_c(\tau ) \, {\cal
 P}^u_c,
 \end{eqnarray*}

\bea
 S^{\mu\nu}_c &=& [\epsilon^{\mu}_r(u(p_s))\, u^{\nu}(p_s) -
 \epsilon^{\nu}_r(u(p_s))\, u^{\mu}(p_s)]\, q_c^{r\tau\tau} +
 \epsilon^{\mu}_r(u(p_s))\, \epsilon^{\nu}_u(u(p_s))\,
 (q^{ru\tau}_c - q_c^{ur\tau}) =\nonumber \\
 &=& [\epsilon^{\mu}_r(u(p_s))\, u^{\nu}(p_s) -
 \epsilon^{\nu}_r(u(p_s))\, u^{\mu}(p_s)]\, M_c\, (R^r_c -
 \zeta^r_c) +\nonumber \\
 &+&  \epsilon^{\mu}_r(u(p_s))\, \epsilon^{\nu}_u(u(p_s))\,
 \Big[ \epsilon^{ruv}\, {\cal J}^v_c - (\zeta^r_c\, {\cal P}_c^u -
 \zeta^u_c\, {\cal P}_c^r)\Big],\nonumber \\
 &&{}\nonumber \\
 &&\Rightarrow \, m^{\mu}_{c(p_s)} = - S^{\mu\nu}_c\, u_{\nu}(p_s)
 = - \epsilon^{\mu}_r(u(p_s))\, q_c^{r\tau\tau}.
 \label{4.43}
 \eea
\medskip

Then, consider the following possible definitions of effective
centers of motion (clearly, many other possibilities
exist)\bigskip

1) {\it Center of energy as center of motion}, ${\vec
\zeta}_{c(E)}(\tau ) = {\vec R}_c(\tau )$, where ${\vec R}_c(\tau
)$ is {\it a 3-center of energy} for the cluster, built by means
of the standard definition

\begin{equation} {\vec R}_c = - {{ {\vec {\cal K}}_c}\over {M_c}}.
 \label{4.44}
 \end{equation}
\medskip

It is determined by the requirement that either the mass dipole
vanishes, $q_c^{r\tau\tau} = 0$, or the mass dipole moment with
respect to $u^{\mu}(p_s)$ vanishes, $m^{\mu}_{c(p_s)} = 0$.
\medskip

The center of energy seems to be the only center of motion
enjoying the simple composition rule

\begin{equation} {\vec R}_{c(n_1+n_2)} = {{M_{c(n_1)}\, {\vec R}_{c(n_1)} +
 M_{c(n_2)}\, {\vec R}_{c(n_2)}}\over {M_{c(n_1+n_2)}}}.
 \label{4.45}
 \end{equation}
\medskip

The constitutive relation between ${\vec {\cal P}}_c$ and ${\dot
{\vec R}}_c(\tau )$, see Eq.(\ref{4.23}), is

\bea
 0 &=& {{d q^{r\tau\tau}_c}\over {d\tau}} = - {\dot {\cal K}}^r_c -
{\dot M}_c\, R^r_c - M_c\, {\dot R}^r_c,\nonumber \\
 &&{}\nonumber \\
 &&\Downarrow\nonumber \\
 &&{}\nonumber \\
 {\vec {\cal P}}_c &=&
  Q_1Q_2\, \int d^3\sigma\, \vec \sigma\, \Big( \Big[ \Big( {{{\vec
 \kappa}_1(\tau )}\over {\sqrt{m^2_1 + {\vec \kappa}_1^2(\tau )}}}
\cdot \vec \partial\Big)\, \vec c(\vec \sigma - {\vec \eta}_1(\tau
))\Big] \cdot \vec c(\vec \sigma - {\vec \eta}_2(\tau ))
+\nonumber \\
 &+& \vec c(\vec \sigma - {\vec \eta}_1(\tau )) \cdot \Big[ \Big( {{{\vec
 \kappa}_2(\tau )}\over {\sqrt{m^2_2 + {\vec \kappa}_2^2(\tau )}}}
 \cdot \vec \partial \Big)\, \vec c(\vec \sigma - {\vec
 \eta}_2(\tau ))\Big] \Big) -\nonumber \\
 &-& {1\over 2}\, \sum_{i=1}^2\, Q_i\, \sum_{k\not= 1,2}\, Q_k\,
  \int d^3\sigma\, \vec \sigma\, \Big( \Big[ \Big( {{{\vec
 \kappa}_i(\tau )}\over {\sqrt{m^2_i + {\vec \kappa}_i^2(\tau )}}}
\cdot \vec \partial\Big)\, \vec c(\vec \sigma - {\vec \eta}_i(\tau
))\Big] \cdot \vec c(\vec \sigma - {\vec \eta}_k(\tau ))
+\nonumber \\
 &+& \vec c(\vec \sigma - {\vec \eta}_i(\tau )) \cdot \Big[ \Big( {{{\vec
 \kappa}_k(\tau )}\over {\sqrt{m^2_k + {\vec \kappa}_k^2(\tau )}}}
 \cdot \vec \partial \Big)\, \vec c(\vec \sigma - {\vec
 \eta}_k(\tau ))\Big] \Big).
 \label{4.46}
 \eea

\medskip

>From Eq.(\ref{4.25}), it follows that the associated cluster spin
tensor is

\bea
  S^{\mu\nu}_c &=& \epsilon^{\mu}_r(u(p_s))\, \epsilon^{\nu}_u(u(p_s))\,
  [q_c^{ru\tau} - q_c^{ur\tau}] =\nonumber \\
  &=& \epsilon^{\mu}_r(u(p_s))\, \epsilon^{\nu}_u(u(p_s))\,
 \epsilon^{ruv}\, \Big[ {\cal J}^v_c - ({\vec R}_c \times {\vec {\cal P}}_c)^v\Big].
 \label{4.47}
 \eea

\bigskip

2)  {\it Pirani centroid ${\vec \zeta}_{c(P)}(\tau )$ as center of
motion}. It is determined by the requirement that the mass dipole
moment with respect to 4-velocity ${\dot w}^{\mu}_c(\tau )$
vanishes (it involves the anti-symmetric part of $p^{ur}_c$)

\bea
  m^{\mu}_{c({\dot w}_c)} &=& - S^{\mu\nu}_c\, {\dot w}_{c\nu} = 0,\quad
 \Rightarrow {\dot {\vec \zeta}}_{c(P)} \cdot {\vec \zeta}_{c(P)} = {\dot {\vec
 \zeta}}_{c(P)} \cdot {\vec R}_c,\nonumber \\
 &&\Downarrow \nonumber \\
 &&{}\nonumber \\
 {\vec \zeta}_{c(P)}(\tau ) &=& {1\over {M_c - {\vec {\cal P}}_c
 \cdot {\dot {\vec \zeta}}_{c(P)}(\tau )}}\, \Big[ M_c\, {\vec R}_c -
 {\vec R}_c \cdot {\dot {\vec \zeta}}_{c(P)}(\tau )\, {\vec {\cal P}}_c
 - {\dot {\vec \zeta}}_{c(P)}(\tau ) \times  {\vec {\cal J}}_c\Big].
 \label{4.48}
 \eea

Therefore this centroid is implicitly defined as the solution of
these three coupled first order ordinary differential equations.

\bigskip

3) {\it Tulczyjew centroid ${\vec \zeta}_{c(T)}(\tau )$ as center
of motion}. If we define the cluster 4-momentum  $P^{\mu}_c =
M_c\, u^{\mu}(p_s) + {\cal P}^s_c\, \epsilon^{\mu}_s(u(p_s))$
[$P^2_c = M^2_c - {\vec {\cal P}}_c^2 \, {\buildrel {def}\over
=}\, {\cal M}_c^2$], its definition is the requirement that the
mass dipole moment with respect to $P^{\mu}_c$ vanishes (it
involves the anti-symmetric part of $p^{ur}_c$)

\bea
 m^{\mu}_{c(P_c)} &=& - S^{\mu\nu}_c\, P_{c\nu} = 0,\quad
 \Rightarrow {\vec {\cal P}}_c \cdot {\vec \zeta}_{c(T)} = {\vec
 {\cal P}}_c \cdot {\vec R}_c,\nonumber \\
 &&\Downarrow \nonumber \\
 &&{}\nonumber \\
 {\vec \zeta}_{c(T)}(\tau ) &=& {1\over {M^2_c - {\vec {\cal
 P}}_c^2}}\, \Big[ M^2_c\, {\vec R}_c - {\vec {\cal P}}_c \cdot
 {\vec R}_c\, {\vec {\cal P}}_c -
 {\vec {\cal P}}_c \times  {\vec {\cal J}}_c\Big].
 \label{4.49}
 \eea
\medskip

Let us show that this centroid satisfies the free particle
relation as constitutive relation

\bea
 {\vec {\cal P}}_c &=& M_c\, {\dot {\vec \zeta}}_{c(T)},\quad
 \Rightarrow \quad
P^{\mu}_c = M_c\, \Big[ u^{\mu}(p_s) + {\dot \zeta}^s_{c(T)}\,
  \epsilon^{\mu}_s(u(p_s))\Big],\nonumber \\
 &&{}\nonumber \\
 &&q^{r\tau\tau}_{c(T)} = {{M_c}\over {M^2_c - {\vec {\cal
 P}}_c^2}}\, \Big[ {\vec {\cal P}}^2_c\, {\vec R}_c + {\vec {\cal
 P}}_c \cdot {\vec R}_c\, {\vec {\cal P}}_c + {\vec {\cal P}}_c
 \times {\vec {\cal J}}_c\Big],\nonumber \\
 &&S^{\mu\nu}_c = [\epsilon^{\mu}_r(u(p_s))\, u^{\nu}(p_s) -
 \epsilon^{\nu}_r(u(p_s))\, u^{\mu}(p_s)]\, q_{c(T)}^{r\tau\tau}
 +\nonumber \\
 &&\quad +  \epsilon^{\mu}_r(u(p_s))\, \epsilon^{\nu}_u(u(p_s))\,
 \epsilon^{ruv}\, \Big[ {\cal J}^v_c - ({\vec \zeta}_{c(T)} \times {\vec {\cal P}}_c)^v\Big].
 \label{4.50}
 \eea

If we use Eq.(\ref{4.48}) to find a Pirani centroid such that
${\dot {\vec \zeta}}_c = {\vec {\cal P}}_c / M_c$, it turns out
that the condition (\ref{4.48}) becomes Eq.(\ref{4.49}) and this
implies Eq.(\ref{4.50}).

\medskip

The equations of motion

\begin{equation} M_c(\tau )\, {\ddot {\vec \zeta}}_{c(T)}(\tau ) =
{\dot {\vec {\cal P}}}_c(\tau ) - {\dot M}_c(\tau )\, {\dot {\vec
\zeta}}_{c(T)}(\tau ),
 \label{4.51}
  \end{equation}\medskip

\noindent  contain both internal and external forces. In spite of
the nice properties (\ref{4.50}) and (\ref{4.51}) of the Tulczyjew
centroid, this effective center of motion fails to satisfy a
simple composition property. The relation among the Tulczyjew
centroids of clusters with $n_1$, $n_2$ and $n_1+n_2$ particles
respectively is much more complicated of the composition
(\ref{4.45}) of the centers of energy.

\medskip

All the previous centroids coincide for an isolated system in the
rest-frame instant form with ${\vec {\cal P}}_c = {\vec
\kappa}_{+} \approx 0$ in the gauge ${\vec q}_{+} \approx {\vec
R}_{+} \approx {\vec y}_{+} \approx 0$.
\medskip

4) The {\it Corinaldesi-Papapetrou centroid with respect to a
time-like observer with 4-velocity $v^{\mu}(\tau )$, ${\vec
\zeta}^{(v)}_{c(CP)}(\tau )$ as center of motion}.

\begin{equation}
 m^{\mu}_{c(v)} = - S^{\mu\nu}_c\, v_{\nu} = 0.
 \label{4.52}
 \end{equation}
Clearly these centroids are unrelated to the previous ones being
dependent on the choice of an arbitrary observer.

\medskip

5) The {\it Pryce center of spin or classical canonical
Newton-Wigner centroid} ${\vec \zeta}_{c(NW)}$. It defined as the
solution of the differential equations implied by the requirement
$\{ \zeta^r_{c(NW)}, \zeta^s_{c(NW)} \} = 0$, $\{ \zeta^r_{c(NW)},
{\cal P}_c^s \} = \delta^{rs}$. Let us remark that, being in an
instant form of dynamics, we have $\{ {\cal P}_c^r, {\cal P}_c^s
\} = 0$ also for an open system.

\bigskip

The two effective centers of motion which appear to be more useful
for applications are the center of energy ${\vec
\zeta}_{c(E)}(\tau )$ and Tulczyjew's centroid ${\vec
\zeta}_{c(T)}(\tau )$, with ${\vec \zeta}_{c(E)}(\tau )$ preferred
for the study of the mutual motion of clusters due to
Eq.(\ref{4.45}).

\bigskip

Therefore, in the spirit of the multipolar expansion, our two-body
cluster may be described by an effective non-conserved internal
energy (or mass) $M_c(\tau ) $, by the world-line $w^{\mu}_c =
x^{\mu}_o + u^{\mu}(p_s)\, \tau + \epsilon^{\mu}_r(u(p_s))\,
\zeta_{c(E\, or\, T)}^r(\tau )$ associated with the effective
center of motion ${\vec \zeta}_{c(E\, or\, T)}(\tau )$ and by the
effective 3-momentum ${\vec {\cal P}}_c(\tau )$, with ${\vec
\zeta}_{c(E\, or\, T)}(\tau )$ and ${\vec {\cal P}}_c(\tau )$
forming a non-canonical basis for the collective variables of the
cluster. A non-canonical effective spin for the cluster in the 1)
and 3) cases is defined by\medskip

\noindent a) case of the center of energy:

\begin{eqnarray*}
  &&{}\nonumber \\
 {\vec {\cal S}}_{c(E)}(\tau ) &=& {\vec {\cal J}}_c(\tau ) - {\vec R}_c(\tau )
 \times {\vec {\cal P}}_c(\tau ),\nonumber \\
 &&{}\nonumber \\
 {{d {\vec \zeta}_{c(E)}(\tau )}\over {d\tau}} &=& {{d {\vec {\cal
 J}}_c(\tau )}\over {d\tau }} - {{d {\vec R}_c(\tau )}\over {d\tau}}\, \times {\vec
 {\cal P}}_c(\tau ) - {\vec R}_c(\tau ) \times
 {{d {\vec {\cal P}}_c(\tau )}\over {d\tau}},
 \end{eqnarray*}

\noindent b) case of the Tulczyjew centroid:

\bea
 &&{}\nonumber \\
 {\vec {\cal S}}_{c(T)}(\tau ) &=& {\vec {\cal J}}_c(\tau ) - {\vec \zeta}_{c(T)}(\tau )
 \times {\vec {\cal P}}_c(\tau )= \nonumber \\
 &=& {{M^2_c(\tau )\, {\vec {\cal S}}_{c(E)}(\tau ) - {\vec {\cal
 P}}_c(\tau ) \cdot {\vec {\cal J}}_c(\tau )\, {\vec {\cal
 P}}_c(\tau )}\over {M^2_c(\tau ) - {\vec {\cal P}}_c^2(\tau
 )}},\nonumber \\
 &&{}\nonumber \\
 {{d {\vec \zeta}_{c(T)}(\tau )}\over {d\tau}} &=& {{d {\vec {\cal
 J}}_c(\tau )}\over {d\tau }} - {\vec \zeta}_{c(T)}(\tau ) \times
 {{d {\vec {\cal P}}_c(\tau )}\over {d\tau}}.
 \label{4.53}
 \eea

Since our cluster contains only two particles, this {\it
pole-dipole description} concentrated on the world-line
$w^{\mu}_c(\tau )$ is equivalent to the original description in
terms of the canonical variables ${\vec \eta}_i(\tau )$, ${\vec
\kappa}_i(\tau )$ (all the higher multipoles are not independent
quantities in this case).
\medskip

Finally, in Ref.\cite{12} there is an attempt to replace the
description of the two body system as an {\it effective
pole-dipole system} with a description as an {\it effective
extended two-body system} by introducing two non-canonical
relative variables ${\vec \rho}_{c(E\, or\, T)}(\tau )$, ${\vec
\pi}_{c(E\, or\, T)}(\tau )$ with the following definitions
\medskip

\bea
 {\vec \eta}_1 &{\buildrel {def}\over =}& {\vec \zeta}_{c(E\, or\, T)} + {1\over
 2}\, {\vec \rho}_{c(E\, or\, T)},\qquad {\vec \zeta}_{c(E\, or\, T)} = {1\over 2}\, ({\vec
 \eta}_1 + {\vec \eta}_2),\nonumber \\
 {\vec \eta}_2  &{\buildrel {def}\over =}& {\vec \zeta}_{c(E\, or\, T)} - {1\over
 2}\, {\vec \rho}_{c(E\, or\, T)},\qquad {\vec \rho}_{c(E\, or\, T)} = {\vec \eta}_1 - {\vec
 \eta}_2,\nonumber \\
 &&{}\nonumber \\
 {\vec \kappa}_1  &{\buildrel {def}\over =}& {1\over 2}\, {\vec
 {\cal P}}_c + {\vec \pi}_{c(E\, or\, T)},\qquad {\vec {\cal P}}_c = {\vec
 \kappa}_1 + {\vec \kappa}_2,\nonumber \\
 {\vec \kappa}_2  &{\buildrel {def}\over =}& {1\over 2}\, {\vec
 {\cal P}}_c - {\vec \pi}_{c(E\, or\, T)},\qquad {\vec \pi}_{c(E\, or\, T)} = {1\over 2}\,
 ({\vec \kappa}_1 - {\vec \kappa}_2),\nonumber \\
 &&{}\nonumber \\
 {\vec {\cal J}}_c &=& {\vec \eta}_1 \times {\vec \kappa}_1 +
 {\vec \eta}_2 \times {\vec \kappa}_2 = {\vec \zeta}_{c(E\, or\, T)} \times {\vec
 {\cal P}}_c + {\vec \rho}_{c(E\, or\, T)} \times {\vec \pi}_{c(E\, or\, T)},\nonumber \\
 &&{}\nonumber \\
 &&\Rightarrow\,\, {\vec {\cal S}}_{c(E\, or\, T)} = {\vec \rho}_{c(E\, or\, T)} \times {\vec
 \pi}_{c(E\, or\, T)}.
 \label{4.54}
 \eea
\medskip

\noindent Even if suggested by a canonical transformation, it is
{\it not} a canonical transformation and it only exists because we
are working in an instant form of dynamics in which both ${\vec
{\cal P}}_c$ and ${\vec {\cal J}}_c$ do not depend on the
interactions. Note that we know everything about this new basis
except for the unit vector ${\vec \rho}_{c(E\, or\, T)}/|{\vec
\rho}_{c(E\, or\, T)}|$ and the momentum ${\vec \pi}_{c(E\, or\,
T)}$. The relevant lacking information can be extracted from the
symmetrized momentum dipole $p^{ru}_c + p^{ur}_c$, which  is a
known effective quantity due to Eq.(\ref{4.40}). However, strictly
speaking, this type of attempt fails, because $p_c^{ru} +
p_c^{ur}$ {\it does not depend only} on the cluster properties but
{\it also} on the external electro-magnetic transverse vector
potential at the particle positions, as shown by Eq.(\ref{4.39}).
Consequently, the spin frame, or equivalently the 3 Euler angles
associated with the internal spin, depend upon the external
fields.

\subsection{The Multipoles of the Real Klein-Gordon Field.}

In the rest-frame instant form we have the following expression
\cite{26} for the energy-momentum of the real Klein-Gordon field

\begin{eqnarray*}
T^{\mu\nu}[x^{\mu}_s(T_s)+\epsilon^{\mu}_u(u(p_s))\sigma^u][\phi
]&=&
T^{\mu\nu}[x^{\mu}_s(T_s)+\epsilon^{\mu}_u(u(p_s))\sigma^u][X^A_{\phi},
P^A_{\phi},{\bf H},{\bf K}]=\nonumber \\
&=&{1\over 2}u^{\mu}(p_s)u^{\nu}(p_s)[\pi^2+(\vec \partial \phi
)^2+
m^2\phi^2](T_s,\vec \sigma )+\nonumber \\
&+&\epsilon^{\mu}_r(u(p_s))\epsilon^{\nu}_s(u(p_s))[-{1\over
2}\delta_{rs} [\pi^2-(\vec \partial \phi )^2-m^2\phi^2]+
 \end{eqnarray*}

\bea
 &+&\partial_r\phi\partial_s\phi ](T_s,
\vec \sigma )-\nonumber \\
&-&[u^{\mu}(p_s)\epsilon^{\nu}_r(u(p_s))+u^{\nu}(p_s)\epsilon^{\mu}_r(u(p_s))]
[\pi \partial_r\phi ](T_s,\vec \sigma )=\nonumber \\
&=& \Big[ \rho [\phi ,\pi ] u^{\mu}(p_s)u^{\nu}(p_s)+{\cal P}[\phi
,\pi ] [\eta
^{\mu\nu}-u^{\mu}(p_s)u^{\nu}(p_s)]+\nonumber \\
&+&u^{\mu}(p_s)q^{\nu}[\phi ,\pi ]+
u^{\nu}(p_s)q^{\mu}[\phi ,\pi ]+\nonumber \\
&+&T^{rs}_{an\, stress}[\phi ,\pi ]
\epsilon^{\mu}_r(u(p_s))\epsilon^{\nu}
_s(u(p_s))\Big] (T_s,\vec \sigma ),\nonumber \\
 &&{}\nonumber \\
 \rho [\phi ,\pi ]&=&{1\over 2}[\pi^2+(\vec \partial \phi
)^2+m^2\phi^2],\nonumber \\
 {\cal P}[\phi ,\pi ]&=&{1\over 2}
[\pi^2-{5\over 3}(\vec \partial \phi )^2- m^2\phi^2],\nonumber \\
 q^{\mu}[\phi ,\pi ]&=&-\pi \partial_r\phi
\epsilon^{\mu}_r(u(p_s)),\nonumber \\
 T^{rs}_{an\, stress}[\phi ,\pi
]&=&-[\partial^r\phi \partial^s\phi -{1\over 3}
\delta^{rs}(\vec \partial \phi )^2],\nonumber \\
 &&{}{}{} \delta_{uv}T^{uv}_{an\,
stress}[\phi ,\pi ]=0,\nonumber \\
&&{}\nonumber \\
T^{rs}_{stress}(T_s,\vec \sigma )[\phi
]&=&\epsilon^r_{\mu}(u(p_s))\epsilon^s
_{\nu}(u(p_s))T^{\mu\nu}[x^{\mu}_s(T_s)+\epsilon^{\mu}
_u(u(p_s))\sigma^u][\phi ]=\nonumber \\
&=&[\partial^r\phi \partial^s\phi ](T_s,\vec \sigma )- {1\over
2}\delta^{rs}
[\pi^2-(\vec \partial \phi )^2-m^2\phi^2](T_s,\vec \sigma ),\nonumber \\
&&{}\nonumber \\
 P^{\mu}_T[\phi]&=&\int d^3\sigma
T^{\mu\nu}[x^{\mu}_s(T_s)+\epsilon^{\mu}
_u(u(p_s))\sigma^u][\phi ]u_{\nu}(p_s)=\nonumber \\
&=&P^{\tau}_{\phi}
u^{\mu}(p_s)+P^r_{\phi}\epsilon^{\mu}_r(u(p_s))\approx
P^{\tau}_{\phi} u^{\mu}(p_s) \approx p^{\mu}_s,\nonumber \\
M_{\phi}&=&P^{\mu}_T[\phi ] u_{\mu}(p_s) =P^{\tau}_{\phi}.
 \label{4.55}
\end{eqnarray}

\noindent The stress tensor $T^{rs}_{stress}(T_s,\vec \sigma
)[\phi ]$ of the Klein-Gordon field on the Wigner hyper-planes
acquires a form reminiscent of the energy-momentum tensor of an
ideal relativistic fluid as seen from a local observer at rest
(Eckart decomposition; see Ref.\cite{27} ): i) the constant normal
$u^{\mu}(p_s)$ to the Wigner hyper-planes replaces the
hydrodynamic velocity field of the fluid; ii) $\rho [\phi ,\pi
](T_s,\vec \sigma ]$ is the energy density; iii) ${\cal P}[\phi
,\pi ](T_s,\vec \sigma )$ is the analogue of the pressure (sum of
the thermodynamical pressure and of the non-equilibrium bulk
stress or viscous pressure); iv) $q^{\mu}[\phi ,\pi ](T_s, \vec
\sigma )$ is the analogue of the heat flow; v) $T^{rs}_{an\,
stress}[\phi ,\pi ](T_s,\vec \sigma )$ is the shear (or
anisotropic) stress tensor.

\medskip

Given the Hamiltonian version of the energy-momentum tensor,
Eqs.(\ref{4.18}) allow us to find all Dixon multi-poles of the
Klein-Gordon field \cite{26}, with respect to a natural center of
motion identified by the collective variable defined in Subsection
IIIH. Finally, Eqs.(\ref{3.70}), conjoined with the assumption
that we can interchange the sums with the integrals, allow us to
define another class of multi-poles \cite{26} with respect to the
centroid $X^{\mu}_s(\tau )$, origin of the 3-coordinates $\vec
\sigma$, in the gauge ${\vec X}_{\phi} \approx 0$, which, when the
fields have a compact support in momentum space, form a closed
algebra with a generalized Kronecker symbol, that could be
quantized instead of the Fourier coefficients.

\section{Final comments and open problems.}

In this article we have shown how the traditional Jacobi technique
based on the clustering of centers of mass for an N-body system,
can be profitably replaced by a technique based on the {\it
clustering of spins} in order to develop a theory of the
relativistic, rotational, and multi-pole, kinematics for
deformable systems. More generally, the relativistic extension is
also made possible by a systematic use of the so-called {\it
rest-frame instant form} of relativistic dynamics on Wigner
hyper-planes, orthogonal to the conserved 4-momentum of the
isolated system. This is a form of dynamics characterized by a
typical {\it doubling} of the Poincar\'e canonical realizations
into so-called {\it external} and {\it internal} realizations. In
fact, this framework appears to be the most natural theoretical
background for the description of isolated systems (particles,
strings, fields, fluids) in special relativity.

The rest-frame instant form of the N-body problem has Newton
mechanics in the center-of-mass frame as non-relativistic limit.
At the same time it is  the special relativistic limit of the
rest-frame instant form of canonical metric and tetrad gravity
\cite{7}, when the Newton constant G is turned off. Both in
relativistic and non-relativistic theories there is a SO(3) left
action on the $(6N-6)$-dimensional phase space of the canonical
{\it relative variables} with respect to the 3-center of mass,
generated by the non-Abelian Noether constants for the angular
momentum. Correspondingly, the notion of {\it clustering of the
spins of sub-clusters} do exist at both levels, while the Jacobi
clustering of the sub-cluster 3-centers of mass is not extendible
to special relativity. Previously introduced concepts, like {\it
dynamical body frames} (replacing the standard notion of {\it body
frames} valid for rigid systems), {\it spin frames}, and {\it
canonical spin bases}, proper of the Galilean group-theoretical
and Hamiltonian description of N-body or deformable systems, are
likewise directly extended to special relativity. The canonical
spin frames and a finite number of dynamical body frames (SO(3)
{\it right actions} on the $(6N-6)$-dimensional phase space) can
be introduced for every N with well defined non-point canonical
transformations (when the total angular momentum does not vanish).
At the non-relativistic level, our treatment generalizes the
orientation-shape bundle approach of Ref.\cite{1}.

The main results obtained can be summarized as follows: Our
procedure leads to: i) The relativistic separation of the center
of mass of the isolated system and the characterization of all the
relevant notions of external and internal 4- and 3- centers of
mass. ii) The construction of $6N-6$ canonical relative variables
with respect to the internal canonical 3-center of mass for the
N-body systems (unlike the non-relativistic case, they are defined
by a canonical transformation which is {\it point} in the momenta
in absence of interactions and becomes interaction-dependent when
interactions are turned on). iii) The simplest construction of
{\it Dixon multi-poles} of an isolated system with respect to a
center of motion, naturally identified with the internal canonical
3-center of mass. This provides, in particular, the only way for
introducing relativistic tensors of inertia.

Such results are intermediate steps in view of future
developments. Actually, they furnish: iv) The natural theoretical
background for a future relativistic {\it theory of orbits} for a
N-body system, by taking into account the fact that in every
instant form of relativistic dynamics, action-at-a-distance
potentials appear both in the Hamiltonian and in the Lorentz
boosts. v) A new definition of canonical relative variables with
respect to the internal canonical 3-center of mass for every {\it
field configuration} admitting a collective 4-vector conjugate to
the conserved field 4-momentum. vi) The analysis of the relevant
notions of centers of motion and associated Dixon multi-poles for
an {\it open sub-system} of an isolated system, in a way which can
be easily extended to general relativity. vii) The theoretical
background for a post-Minkowskian approximation of binary systems
in general relativity. The relativistic theory of orbits should
provide the relativistic counterparts of the post-Keplerian
parameters used in the post-Newtonian approximation \cite{56}.
viii) The theoretical background for the characterization of a
{\it relativistic rest-frame micro-canonical mean-field
thermodynamics} of N-body systems with long range interactions
(Coulomb or Darwin potentials \cite{6}) as it has already been
done in Ref.\cite{57} for non-relativistic self-gravitating and
rotating systems. ix) A theoretical background that could be
extended to the weak-field general relativistic N-body problem
after a suitable regularization of self-energies. Actually, this
framework has already been extended to charged Klein-Gordon fields
interacting with the electro-magnetic field \cite{26}, Dirac
fields \cite{58} and relativistic perfect fluids \cite{27}.
Finally: x) Parametrized Minkowski theories in {\it non-inertial
frames} are prepared for a systematic study \cite{9} of the
allowed conventions for clock synchronizations, the influence of
relativistic inertial forces and the time-delays (to order
$1/c^3$) for the one-way propagation of light rays (for instance
between an Earth station and a satellite).

\bigskip

The extension of our new rotational kinematics to continuous
systems like the Klein-Gordon field, an extension which should be
instrumental to atomic and molecular physics at least at the
classical level, is under investigation \cite{52}. When applied to
the electro-magnetic field, these methods could lead to
interesting results for the problem of {\it phases} \cite{53} in
optics and laser physics. Also, relativistic perfect fluids have
been studied in the rest-frame instant form \cite{50} and a future
application of the new rotational kinematics might give new
insights for their description.

\medskip

Let us conclude by noting the the non-point nature of the
canonical transformations will make the quantization more
difficult than in the {\it orientation-shape bundle} approach,
where a separation of rotations from vibrations in the
Schr\"odinger equation is reviewed in Ref.\cite{1}. The
quantizations of the original canonical relative variables and of
the canonical spin bases will give equivalent quantum theories
only if the non-point canonical transformations could be unitarily
implementable. Up to now, these problems are completely
unexplored.

\vfill\eject


\begin{thebibliography}{}
\bibitem{4}This is due to the topological complexity of
the shape space generated by the singular configurations \cite{3},
which are dispersed among the generic configurations for $N \geq
4$.
\bibitem{5}We shall put $c=1$ everywhere and use the convention
$\eta^{\mu\nu}=\epsilon (+---)$ for the Minkowski metric (with
$\epsilon = \pm 1$ according to the particle physics or general
relativity convention).
\bibitem{23}We follow the convention of
Ref.\cite{1}; note that this action is usually denoted as a {\it
right} action in mathematical texts. The $S^r$'s are the
Hamiltonians, associated with the momentum map from the symplectic
manifold to $so(3)^{*}$ [the dual of the Lie algebra $so(3)$],
which allow the implementation of the symplectic action through
Hamiltonian vector fields.
\bibitem{33}With ${\check S}^r=\vec S\cdot
{\hat e}_r$, the conditions $\{ {\check S}^r,{\check S}^s \} =
-\epsilon^{rsu} {\check S}^u$ imply the equations $\vec S\cdot
{\hat e}_r\times {\hat e}_s +S^iS^j \{ {\hat e}^i_r,{\hat e}^j_s\}
= \epsilon_{rsu} S^k{\hat e}_k^u$, hence the quoted result.
\bibitem{34}Let us remark that any pair of
orthonormal vectors $\vec N$, $\vec \chi$ function only of the
relative coordinates can be used to build a body frame. This
freedom is connected to the possibility of redefining a body frame
by using a configuration-dependent arbitrary rotation, which
leaves $\vec N$ in the $\vec S$-$\hat R$ plane.
\end{thebibliography}
\end{document}